\documentclass[a4paper,11pt]{article}
\usepackage{jheppub} 
\usepackage{hyperref}
\usepackage{xcolor,graphicx}
\usepackage{amsmath}
\usepackage{multicol,multirow}
\usepackage{rotating}
\usepackage{etoolbox}

\newcommand{\ie}{{\it i.e.}\ }

\newcommand{\be}{\begin{eqnarray}}
\newcommand{\ee}{\end{eqnarray}}

\allowdisplaybreaks
\robustify\bfseries

\begin{document}

\title{\boldmath The Euclidean Adler Function and its \\ Interplay with $\Delta\alpha^{\mathrm{had}}_{\mathrm{QED}}$ and $\alpha_s$}

\author[a]{M.~Davier,}
\author[b]{D.~Díaz-Calderón,}
\author[c]{B.~Malaescu,}
\author[b]{A.~Pich,}
\author[c,d,e,f]{A.~Rodríguez-Sánchez,}
\author[a]{Z.~Zhang}

\affiliation[a]{Universit\'{e} Paris-Saclay, CNRS/IN2P3, IJCLab, 91405 Orsay, France}

\affiliation[b]{Departament de Física Teòrica, IFIC, Universitat de València – CSIC\\
\hskip .6cm Parque Científico, Catedrático José Beltrán 2, E-46980 Paterna, Spain}

\affiliation[c]{LPNHE, Sorbonne Université, Université Paris Cité, CNRS/IN2P3, Paris, France}

\affiliation[d]{Sorbonne Universit\'e, CNRS, Laboratoire de Physique Th\'eorique et Hautes Energies (LPTHE), F-75252 Paris, France}

\affiliation[e]{SISSA International School for Advanced Studies, Via Bonomea 265, 34136, Trieste, Italy}

\affiliation[f]{INFN, Sezione di Trieste, SISSA, Via Bonomea 265, 34136, Trieste, Italy}

\abstract{
Three different approaches to precisely describe the Adler function in the Euclidean regime at around $2\, \mathrm{GeVs}$ are available:
dispersion relations based on the hadronic production data in $e^+e^-$ annihilation, lattice simulations and perturbative QCD (pQCD).
We make a comprehensive study of the perturbative approach, supplemented with the leading power corrections in the operator product expansion. All known contributions are included, with a careful assessment of uncertainties.
The pQCD predictions are compared with the Adler functions extracted from $\Delta\alpha^{\mathrm{had}}_{\mathrm{QED}}(Q^2)$, using both the DHMZ compilation of $e^+e^-$ data and published lattice results. Taking as input the FLAG value of $\alpha_s$, the pQCD Adler function turns out to be in good agreement with the lattice data, while the dispersive results lie systematically below them. Finally, we explore the sensitivity to $\alpha_s$ of the direct comparison between the data-driven, lattice and QCD Euclidean Adler functions. The precision with which the renormalisation group equation can be tested is also evaluated.}

\maketitle

\section{Introduction} 

Two-point functions are among the most basic objects that one can define within a Quantum Field Theory (QFT). Every possible action encodes associated distributions for them. From the phenomenological point of view, a particularly interesting one is the two-point correlation function of two vector neutral quark currents,
\begin{equation}
\Pi_{ij}^{\mu\nu}(q)\equiv   i \int d^{4}x \, e^{-iqx} \langle 0 | T\left(\bar{q}_i(x) \gamma^{\mu} q_i(x) \,\bar{q}_j(0) \gamma^{\nu} q_j(0)\right)  | 0 \rangle=(q^{\mu}q^{\nu}-g^{\mu\nu}q^2)\,\Pi_{ij}(s\equiv q^2)  \, .
\end{equation}
Even in the partonic approximation, this is in the absence of quantum corrections and neglecting quark masses, these two-point correlation functions are divergent and depend on an arbitrary subtraction prescription. This approximation makes sense as the leading term of an Operator Product Expansion (OPE)~\cite{Wilson:1969zs,Shifman:1978bx,Shifman:1978by} which is well defined for large Euclidean momenta, $Q^2=-q^2$ and leads to\footnote{In the rest of the complex plane $\Pi_{ij}(Q^2)$ can be obtained by analytic continuation. We take $\mathrm{Arg}(Q^2)\in[-\pi,\pi)$ and $q^2=e^{i\pi}Q^2$, so that $\mathrm{Arg}(q^2)\in[0,2\pi)$. In this way, 
$\mathrm{Im}\Pi(|Q^2| e^{-i\pi})=\frac{1}{2i}[\Pi(|Q^2| e^{-i\pi})-\Pi(|Q^2| e^{i\pi})]$.}

\begin{equation}\label{eq:partonic}
\Pi^{\mathrm{part}}_{ij}(Q^2)=-\frac{N_C}{12\pi^2}\delta_{ij}
\left[\log(Q^2)+C_{\mathrm{subtraction}}\right] \, ,
\end{equation}
where $N_C$ is the number of quark colours.
In order to avoid unphysical subtraction ambiguities, it is convenient to define the object of study in this work, the Euclidean Adler function~\cite{Adler:1974gd}:
\begin{equation}
\label{eq:AdlerDef}
D_{ij}(Q^2)\equiv -12\pi^2 Q^2 \frac{d\Pi_{ij}(Q^2)}{dQ^2} \, ,
\end{equation}
which gives in the partonic limit
\begin{equation}
D^{\mathrm{part}}_{ij}(Q^2)=N_{C}\,\delta_{ij} \, .
\end{equation}

The most important corrections in the Standard Model (SM) come from strong interactions. At large Euclidean momenta, far enough from the non-analytic behaviour in the distributions induced by hadrons, deviations from asymptotic freedom are described by perturbative QCD (pQCD), so that the leading corrections are simply given by $\delta D_{ij}(Q^2)=N_C\,\delta_{ij}\frac{\alpha_s(Q^2)}{\pi}$. For light quarks $i,j\leq 3$, $D_{ij}^{L}(Q^2)$, the perturbative QCD description breaks down in the infrared region: 
the low-energy hadron dynamics is not well described by approximately-free quarks and gluons. Our knowledge about it at very low energies (there are no physical singularities at $Q^2=0$) implies that $D_{ij}(Q^2\rightarrow 0)=0$.\footnote{Let us note that if one defines $\alpha^{\mathrm{eff}}_s(Q^2)\equiv \pi\left(\frac{D_{ii}(Q^2)}{N_{C}}-1 \right)$, then $\alpha^{\mathrm{eff}}_s(0)=-\pi$, which per-se does not tell us any new fundamental knowledge about strong interactions.} 
Low-energy Effective Field Theories (EFTs), such as Chiral Perturbation Theory \cite{Gasser:1983yg,Gasser:1984gg,Amoros:1999dp}, give some nontrivial information about
the infrared behaviour of $D^{L}_{ij}(Q^2)$, but their predictive power is limited, especially at intermediate energies.  For massive quarks, $i,j>3$, $D^{\mathrm{H}}_{ij}(Q^2)$, a perturbative QCD description is known to give a precise description of the Adler function even in the neighbourhood of $Q^2\rightarrow 0$ \cite{Novikov:1977dq,Vainshtein:1978wd,Chetyrkin:2010ic,Chetyrkin:2017lif}, since quark masses regularize, at least up to a certain extent, the gluon singularities associated to infrared propagators. 

Other powerful nonperturbative methods can also be used to obtain $D_{ij}(Q^2)$. 
Numerical simulations in a discretized space-time lattice \cite{Wilson:1974sk} allow for a precise computation of the two-point functions at Euclidean momenta without relying on perturbation theory. Indeed huge efforts are recently being  made to compute the electromagnetic correlator,
\begin{equation}
\Pi(Q^2)\equiv \sum_{i,j} Q_{i}Q_{j}\, \Pi_{ij}(Q^2) \, ,
\end{equation}
where $Q_{i}$ is the electromagnetic charge of the associated quark in units of $e$ (e.g. $Q_1=\frac{2}{3}$), since it plays a fundamental role in our understanding of the anomalous magnetic moment of the muon and in the so-called hadronic running of the QED coupling \cite{Borsanyi:2020mff,Ce:2022eix,Chakraborty:2017tqp,Borsanyi:2017zdw,Blum:2018mom,Giusti:2019xct,Shintani:2019wai,FermilabLattice:2019ugu,Gerardin:2019rua,Aubin:2019usy,Giusti:2019hkz}. Currently the predictive power of lattice methods becomes severely limited as one goes above $Q \sim 2-3\, \mathrm{GeVs}$ due to discretization effects, leading to an interesting complementarity with respect to pQCD.

Similar motivations have increased the knowledge of $\Pi(Q^2)$ obtained from another powerful nonperturbative method, the dispersive data-driven approach, which mainly uses electron-positron data to determine $\Pi(Q^2)$
\cite{Davier:2017zfy,Keshavarzi:2018mgv,Colangelo:2018mtw,Hoferichter:2019mqg,Davier:2019can,Keshavarzi:2019abf}. A well-known limitation in the current precision comes from a series of tensions involving electron-positron data. 
Besides the long-established discrepancy between $e^+e^-\to\pi^+\pi^-$ and $\tau^-\to\pi^-\pi^0\nu_\tau$ data (invoking an isospin rotation)~\cite{Davier:2010fmf}, there are significant tensions
among different $e^{+}e^{-}$ data sets (mainly KLOE vs BABAR, for the same $2\pi$ channel). Additionally, there is a clear discrepancy between the experimental value of $(g-2)_{\mu}$ and the theoretical SM prediction obtained
when using $e^+e^-$ data to evaluate
the Hadronic Vacuum Polarization (HVP) contribution (even after inflating uncertainties to account for the KLOE-BABAR tension). Finally, further tensions arise between $e^{+}e^{-}$ data and lattice evaluations of both the hadronic running of the QED coupling and  again the HVP contribution to $(g-2)_{\mu}$~\cite{Borsanyi:2020mff,Ce:2022eix,Ce:2022kxy,Alexandrou:2022amy,Blum:2023qou,Bazavov:2023has}.

In this work we study analytically the Euclidean Adler function, in the $Q^2$ region where perturbation theory is expected to be valid, with the aim of comparing it to both the dispersive Adler function $D(Q^2)$ obtained with the DHMZ compilation of data~\cite{Davier:2019can} and the one obtained from recently published lattice results for the hadronic running of the QED coupling~\cite{Ce:2022eix}.
 
On the one hand, assuming the validity of pQCD at a certain Euclidean momentum $Q$, one can check whether the description is consistent with the other approaches. On the other hand, assuming that the other approaches are correct, we have a uniquely clean window to learn about the onset of the asymptotic regime and the value of the associated QCD coupling. We start by introducing the overall theoretical framework to connect the different descriptions of the Euclidean Adler functions and the HVP. This is done in Sec.~\ref{sec:theo}. In Sec.~\ref{sec:pqcdadler} we perform a comprehensive study of the perturbative Adler function in the regime we are interested in, combining many existing results. Then, in Sec.~\ref{sec:otheradlers}, we explain in detail how the data-driven and the lattice Adler functions are obtained, and in Sec.~\ref{sec:comparison} we compare them with the perturbative expression. Power corrections are discussed in Sec.~\ref{sec:nonpert}.
Finally an exploration to the sensitivity of the comparison to the strong coupling and discussion about some possible fitting strategies can be found in Sec.~\ref{sec:alphas}. Conclusions and final remarks are presented in Sec.~\ref{sec:conclusions}.

\section{Theoretical framework}\label{sec:theo}
Let us start by connecting the needed  descriptions and observables related to the HVP. The hadronic running of the QED coupling can be defined in terms of the electromagnetic correlator $\Pi(Q^2)$ as follows,
\begin{equation}\label{eq:defalphad}
\Delta \alpha_{\mathrm{had}}(Q^2)\equiv 4\pi \alpha\, \bar{\Pi}(Q^2) \, , 
\qquad\qquad   
\bar{\Pi}(Q^2)=\Pi(0)-\Pi(Q^2) \, ,
\end{equation}
with $\alpha = \alpha (0) = 1/137.035\, 999\, 084\; (21)$ \cite{Workman:2022ynf}.
Since both the hadronic running of $\alpha (Q^2)$ and the Adler function are defined in terms of the electromagnetic correlator (\emph{cf.} Eq.~(\ref{eq:AdlerDef})), it is straightforward to relate both of them,
\begin{equation}\label{eq:defqedadler}
D(Q^2)
\equiv\sum_{i,j} Q_i Q_j\, D_{ij}(Q^2)
=3\pi Q^2\,\frac{d\Delta\alpha_{\mathrm{had}}(Q^2)}{\alpha\, dQ^2} \, .
\end{equation}
Alternatively,
\begin{equation}
\Delta\alpha_{\mathrm{had}}(Q^2)-\Delta\alpha_{\mathrm{had}}(Q_0^2)=\frac{\alpha}{3\pi}\int_{Q_0^2}^{Q^2}\frac{dQ'^2}{Q'^2} D(Q'^2) \, .
\end{equation}

On the other hand, it can be shown that the ratio of hadronic and muonic production cross sections in $e^+e^-$ annihilation is directly related to the imaginary part of $\Pi(Q^2)$. Formally, it is defined as
\begin{equation}
R(s)\equiv\frac{3s}{4\pi\alpha}\;\sigma^{0}(e^{+}e^{-}\rightarrow \mathrm{hadrons}\, (+\gamma))=12\pi\; \mathrm{Im} \Pi(Q^2=s\, e^{-i\pi}) \,  , \label{eq:Rratio}
\end{equation}
where $\sigma^{0}(e^{+}e^{-}\rightarrow \mathrm{hadrons}\, (+\gamma))$ refers to the so-called ``bare'' hadronic cross section in electron-positron annihilation, subtracting the vacuum polarization contribution to the photon propagator and the Initial State Radiation (and ISR-FSR interference) but including all Final State Radiation.

\begin{figure}
    \centering
    \includegraphics[width=0.5\textwidth]{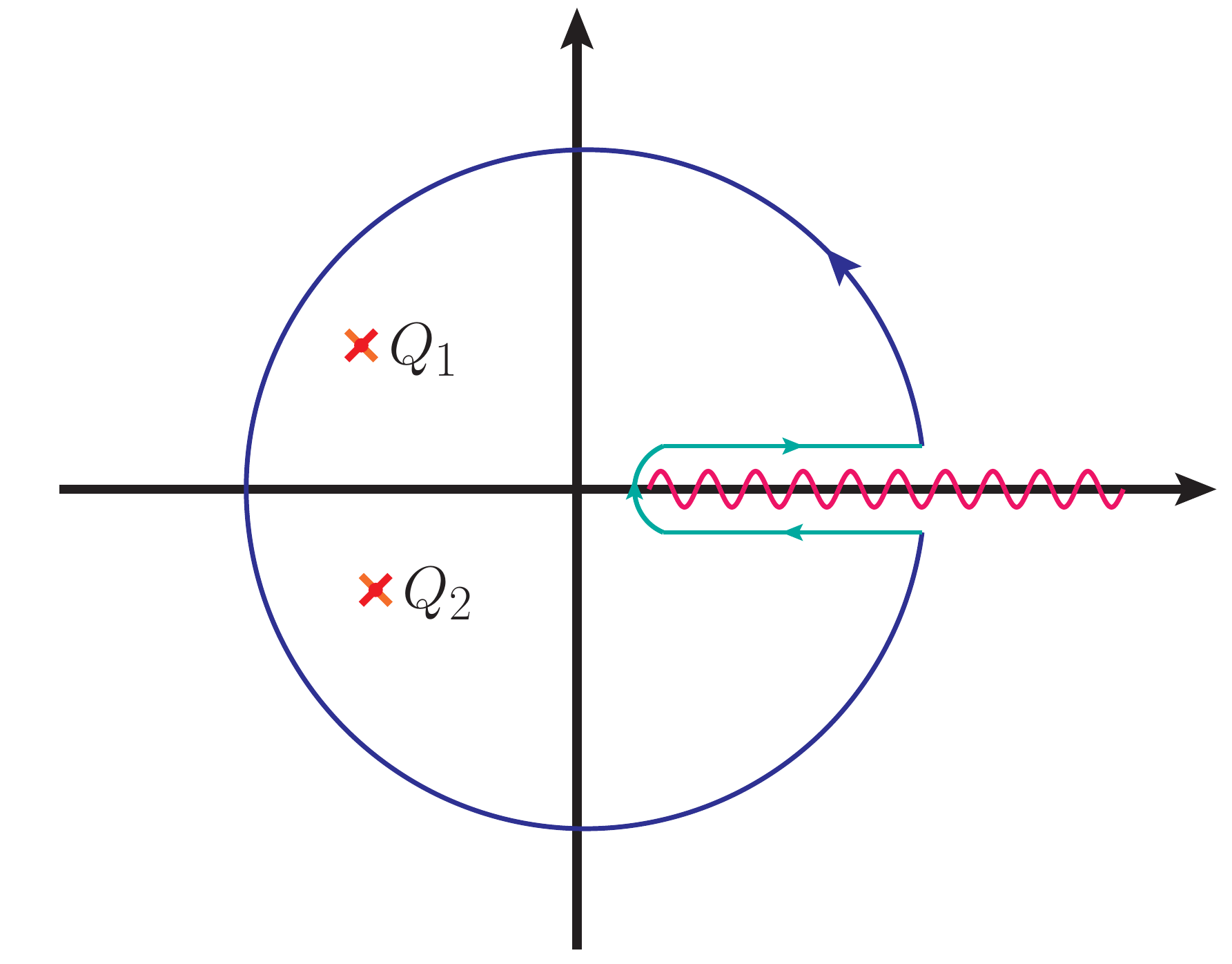}
    \caption{Circuit of integration of $\frac{1}{2\pi i}W(Q^2; Q_1^2,Q_2^2)\cdot\Pi_{ij}(Q^2)$
    in the $q^2=-Q^2$ complex plane.} 
    \label{fig:circuit}
\end{figure}

Thus, in order to relate $R(s)$ with $D(Q^2)$ and $\Delta \alpha_{\mathrm{had}}(Q^2)$, we have to relate the imaginary part of the analytic continuation of $\Pi(Q^2)$ with $\Pi(Q^2)$ itself. This is accomplished by using dispersion relations, which combine our knowledge of the analytic structure of the electromagnetic correlator, $\Pi(Q^2)$, with its known asymptotic behaviour. Indeed, the partonic picture, see Eq.~(\ref{eq:partonic}), is valid at very high energies and $\Pi(Q^2)$ is an analytic function in the whole complex plane except for a branch cut starting at the hadronic threshold,\footnote{If QED effects are included, the threshold becomes $Q^2_{\mathrm{th}}= - m_{\pi}^2$, corresponding to the $\pi^{0}\gamma$ channel.} $Q_{\mathrm{th}}^2=-4m_{\pi}^2$. As a consequence, we can integrate $\Pi(Q^2)$ in the complex plane along any contour as long as we avoid this cut. We may also weight the integral with some function
\begin{equation}
W(Q^2; Q_1^2,Q_2^2)=\frac{1}{(Q^2-Q_1^2)(Q^2-Q_2^2)} \, ,
\end{equation}
which is analytic in the complex plane except for the two poles. Then, integrating along the contour in Fig.~\ref{fig:circuit} and using the Residue Theorem, we get
\begin{equation}
   \frac{1}{2\pi i} \int_{\mathcal{C}} dQ^2\, \frac{\Pi_{ij}(Q^2)}{(Q^2-Q_1^2)(Q^2-Q_2^2)}=\frac{\Pi_{ij}(Q_2^2)}{Q_2^2-Q_1^2}+\frac{\Pi_{ij}(Q_1^2)}{Q_1^2-Q_2^2}.
\end{equation}
We may separate the different contributions to the contour integral as
\begin{align}\nonumber
&\frac{1}{2\pi i}\oint_{|Q^2|=Q_0^2} dQ^2 \frac{\Pi_{ij}(Q^2)}{(Q^2-Q_1^2)(Q^2-Q_2^2)}+\frac{1}{2\pi i}\left(\int_{|Q_{th}^2| e^{-i\pi}}^{|Q_{0}^2| e^{-i\pi}}- \int_{|Q_{th}^2| e^{i\pi}}^{|Q_{0}^2| e^{i\pi}}\right)dQ^2\frac{\Pi_{ij}(Q^2)}{(Q^2-Q_1^2)(Q^2-Q_2^2)}\\
&=\frac{\Pi_{ij}(Q_2^2)}{Q_2^2-Q_1^2}+\frac{\Pi_{ij}(Q_1^2)}{Q_1^2-Q_2^2} \, .
\end{align}
For $|Q_0^2|\rightarrow \infty$ one can easily show that the first term in the first line goes to zero by using the partonic description. Performing a change of variables and using the Schwarz reflection principle in the second term, one arrives at
\begin{equation}
\frac{1}{\pi}\int_{|Q_{\mathrm{th}}^2| }^{\infty} dQ^2\,\frac{\mathrm{Im}\Pi_{ij}(Q^2e^{-i\pi})}{(Q^2+Q_1^2)(Q^2+Q_2^2)}=-\frac{\Pi_{ij}(Q_2^2)-\Pi_{ij}(Q_1^2)}{Q_2^2-Q_1^2}  \, .
\end{equation}
Starting from this expression, one can obtain dispersion relations for different objects by choosing different 
values of $Q_1^2$ and $Q_2^2$. 
Choosing either $Q_1^2$ or $Q_2^2$ to be $0$ leads to a dispersion relation for the correlator $\Pi(Q^2)$,
\begin{equation}
\label{eq:CorrDispRel}
\frac{\Pi(Q^2)-\Pi(0)}{Q^2}=-\frac{1}{\pi}\int_{|Q_{\mathrm{th}}|^2 }^{\infty} dQ'^2\,\frac{\mathrm{Im}\Pi(Q'^2e^{-i\pi})}{Q'^2(Q'^2+Q^2)}  \, .
\end{equation}
A dispersion relation for the Adler function can be obtained by differentiating $\Pi(Q^2)$ (\emph{cf}. Eq.~(\ref{eq:AdlerDef})) in Eq.~(\ref{eq:CorrDispRel}) or by choosing $Q_1^2=Q_2^2+\delta$ with $\delta \rightarrow 0$ in the weight function. Either way, one obtains\footnote{Let us note that the result holds for complex momenta, which promotes the analytic continuation of the Adler function to an observable.}
\begin{equation}
\label{eq:AdlerDispRel}
D_{ij}(Q^2)=12\pi Q^2\int^{\infty}_{|Q_{th}^2|}dQ'^2\, \frac{\mathrm{Im}\Pi_{ij}(Q'^2e^{-i\pi})}{(Q'^2+Q^2)^2}.
\end{equation}
It is then straightforward to write an equation for the data-driven determination of the Euclidean Adler function using Eq.~(\ref{eq:AdlerDispRel}) and Eq.~(\ref{eq:Rratio}),
\begin{equation}
D(Q^2)=Q^{2}\int_{s_{th}}^{\infty}ds\,\frac{R(s)}{(s+Q^2)^2} \, .
\end{equation}
Finally, $\Delta \alpha_{\mathrm{had}}(Q^2)$ is also related to $R(s)$ through the dispersion relation of Eq.~(\ref{eq:CorrDispRel}) and Eq.~(\ref{eq:Rratio}),
\begin{equation}
\Delta\alpha_{\mathrm{had}}(Q^2)=\frac{\alpha Q^{2}}{3\pi}\int_{s_{th}}^{\infty}ds\,\frac{R(s)}{s(s+Q^2)} \, .
\end{equation}

\section{The perturbative Adler function}\label{sec:pqcdadler}

The perturbative Adler function has been the subject of many studies in different energy regimes, since it is directly linked to different precisely known inclusive observables~\cite{Pich:2020gzz}. At order $\alpha_s^2$ and beyond, the coefficients in the expansion depend on the renormalization scheme. For practical purposes, 
the $\overline{\mathrm{MS}}$ scheme is usually adopted
due to its computational simplicity.
In the limit of $n_f$ massless quarks and no massive ones, the corresponding Adler functions are known up to (and including) five loops ({\ie order $\alpha_s^4$).

The real world contains six quark flavours with a striking hierarchy of quark masses:
\begin{equation}
m_{u,d} \ll m_s \ll \Lambda_{\mathrm{QCD}} \ll m_c \ll m_b \ll m_t  \, ,
\end{equation}
where $\Lambda_{\mathrm{QCD}}$ is the QCD scale, so that perturbative QCD does not make sense below it. 
In mass-independent renormalization schemes such as the $\overline{\mathrm{MS}}$,
the perturbative series with six quark flavours does not give a very accurate approximation to the QCD Adler function at $|Q^2| \ll m_t^2$.  This is a consequence of the lack of decoupling associated with this type of schemes, which leads to $\log{(m_t^2/Q^2)}$ factors that slow down and eventually break the perturbative series ({\it e.g.} see~\cite{Pich:1998xt} for a pedagogical introduction to the problem). If one wants to keep track of the full quark-mass dependence of the Adler function~\cite{Eidelman:1998vc}, useful for example for the perturbative running of $\Delta\alpha^{\mathrm{had}}(Q^2)$ from $Q\sim 2.5 \, \mathrm{GeV}$ to $Q=M_Z$~\cite{Jegerlehner:2019lxt}, one possibility is employing a renormalization scheme that automatically performs the decoupling of heavy masses, such as MOM, at the cost of more complex calculations and less known perturbative corrections. 
Alternatively, 
one can still work in the $\overline{\mathrm{MS}}$ scheme by introducing a series of QCD effective field theories with different number $n_{f}$ of massless quark flavours, which need to be matched at the corresponding quark-mass thresholds. The decoupling of heavy masses is then implemented by hand and the massless Adler function can be supplemented with the corresponding power-suppressed corrections from the heavy quark masses. The small contributions from the non-zero light-quark masses can be taken into account through perturbative expansions in powers of $m_q^2/Q^2$.

Since the other methods analyzed in this work are also more powerful below $|Q^2|\sim 5 \, \mathrm{GeV}^2$, we will focus on the Adler function $D(Q^2)$ at $|Q^2|<4 m_c^2$. In the next subsections we study the different perturbative Adler functions $D_{ij}(Q^2)$ separately, first focusing on those with $i,j<4$ (light-quark contributions) and then on
the heavy ones, $i,j>3$ (heavy-quark contributions). Eventually we add the leading QED corrections, put them together and also explain why the mixed ($i<4$, $j>3$ and vice-versa) contributions to $D(Q^2)$ are very suppressed.

\subsection{Light-quark contributions}
\subsubsection{Leading massless contributions}
For light flavours $i,j<4$, the $\overline{\mathrm{MS}}$ scheme with $n_f=3$ massless
quarks gives an accurate description of the perturbative Adler function in the energy range $\Lambda_{\mathrm{QCD}}^2 \ll Q^2 \ll 4 \, m_{c}^2$. 
The flavour-diagonal ($i=j$) correlators contribute to $D(Q^2)$ through the so-called non-singlet topology with the two electromagnetic currents connected by one quark loop. The result can be written as
\be\label{eq:Adler}
D^{L,(0)}_{ii}(Q^2)&\!\! = &\!\! 
N_C\;\left\{
1 + \sum_{n=1}\sum_{p=0}^{n-1}\; K_{n,p}\,
\left( {\alpha_s(\mu^2)\over \pi}\right)^n\,\log^p{(Q^2/\mu^2)}\right\} \, ,
\ee
with $\mu$ the renormalization scale.
Additional disconnected diagrams with each current in a separate quark loop (singlet topology), which are also present for $i \neq j$, start to contribute at order $\alpha^3_s$, but with three massless quarks the flavour trace of both quark loops cancels in the sum ($Q_u+Q_d+Q_s=0$), leading to a completely negligible effect of $\mathcal{O}\left(\alpha_s^3\, \frac{m_s^4}{Q^4}\right)$ for $D(Q^2)$ once the non-zero masses are taken into account.

The $(0)$ superscript indicates that we have not yet incorporated any quark mass correction. Since the Adler function is independent of the renormalization scale, one can trivially reconstruct the coefficients $K_{n,p}$ with $p>0$ from those with $p=0$, simply taking into account the Renormalization Group Equation (RGE) satisfied by the strong coupling:
\begin{equation}\label{eq:beta-rge}
\mu\,\frac{d\alpha_s}{d\mu}\; =\; \alpha_s\;\beta(\alpha_s)\, ,
\qquad\qquad\qquad\qquad
\beta(\alpha_s)\; =\;\sum_{n=1}\, \beta_n\, 
\left(\frac{\alpha_s}{\pi}\right)^n\, .
\end{equation}
The coefficients of the Adler function are known up to five loops, \ie at $\mathcal{O}(\alpha_s^4)$, while the QCD $\beta$ function has been already computed to $\mathcal{O}(\alpha_s^5)$. We collect the values
of $K_{n\le 4,0}$, $\beta_{n\le 5}$
and the rest of perturbative coefficients in App.~\ref{app:pertcoef}. Since we need to truncate the series, a residual scale dependence (of the first unaccounted order) remains. In order 
to avoid higher-order corrections enhanced by large logarithms of the renormalization scale, one should set  $\mu^2=\xi^2 Q^2$ with $\xi^2$ a number of order $1$. The exact choice is however ambiguous and a priori arbitrary, just as it is the exact scheme choice of how to minimally subtract when renormalizing. Modifying the residual scale dependence through the variation of $\xi^2$ in a reasonable interval around unity can then be used to estimate perturbative uncertainties~\cite{LeDiberder:1992jjr}. One has
\be
D^{L,(0)}_{ii}(Q^2)=N_C\left(1+\sum_{n=1} K_{n}(\xi^2)\left( {\alpha_s(\xi^2 Q^2)\over \pi}\right)^n\right) \, ,
\ee
with
\begin{equation}
K_{n}(\xi^2)=\sum_{p=0}^{n-1} K_{n,p} \; \log^{p}(1/\xi^2) \, .
\end{equation}

In order to numerically evaluate the Adler functions below the charm threshold, we then need $\alpha_s(\mu^2)$ with $n_f=3$. The standard input is however $\alpha_{s}(M_{Z}^2)$ with $n_f=5$.  One can translate one into another by supplementing the RGE given above with the decoupling relations,
\be
\alpha_s^{(n_f-1)}(\mu^2) & = & \alpha_s^{(n_f)}(\mu^2)\,\left\{
1 + \sum_{k=1} \sum_{n=0}^k d_{kn}\; \left[a_s^{(n_f)}(\mu^2)\right]^k\, 
\log^n{(\mu^2/M_q^2)}\right\} ,
\\
m_q^{(n_f-1)}(\mu^2) & = & m_q^{(n_f)}(\mu^2)\,\left\{
1 + \sum_{k=2} \sum_{n=0}^k h_{kn}\; \left[a_s^{(n_f)}(\mu^2)\right]^k\, 
\log^n{(\mu^2/M_q^2)}\right\} ,
\ee
where $M_q\equiv m_q^{(n_f)}(\mu^2)$ is the running  mass of the heavy quark that has been integrated out and $a_s^{(n_f)}\equiv \alpha_s^{(n_f)}/\pi$, which is also needed as input. The running masses satisfy the following renormalization group equation
\be \label{eq:runningMass}
\mu\,\frac{d m_q}{d\mu}\; =\; -m_q \;\gamma(\alpha_s)\, ,
\qquad\qquad\qquad\qquad
\gamma(\alpha_s)\; =\;\sum_{n=1}\, \gamma_n \,
\left(\frac{\alpha_s}{\pi}\right)^n\, ,
\ee
where the perturbative $\gamma_n$ coefficients are known up to $n=5$. We will take as extra input $m_c(m_c^2)=1.275\, (5)\, \mathrm{GeV}$ and $m_b(m_b^2)=4.171\, (20)\, \mathrm{GeV}$ from the FLAG lattice review \cite{FlavourLatticeAveragingGroupFLAG:2021npn,McNeile:2010ji,Yang:2014sea,Nakayama:2016atf,Petreczky:2019ozv,EuropeanTwistedMass:2014osg,Chakraborty:2014aca,Alexandrou:2014sha,FermilabLattice:2018est,Hatton:2020qhk,Hatton:2021syc,Colquhoun:2014ica,ETM:2016nbo,Gambino:2017vkx} and perform the decoupling at $\mu=m_q(m_q^2)$. Quark-mass uncertainties are small enough to be negligible. Our results for $\alpha_{s}(Q^2)\equiv\alpha_{s}^{(n_f=3)}(Q^2)$ and $D_{ii}^{L,(0)}(Q^2)$ are given in Table~\ref{tab:massless} for several choices of $\alpha_s^{(n_f=5)}(M_Z^2)$  and $Q$, at different orders in the strong coupling. We have checked that the $\alpha_s(Q^2)$ values fully agree with the corresponding values obtained with {\sc RUNDEC} \cite{Herren:2017osy}. In Fig.~\ref{fig:masslessD} we show the corresponding $Q^2$ dependence for $\alpha_s^{(n_f=5)}(M_Z^2)=0.115,0.120$, at $\mathcal{O}(\alpha_s^5)$,
adding as perturbative uncertainties the quadratic sum of variations due to changing $K_5$ in a conservative interval $(-125,675)$ and $\xi^2 \in (0.5,2)$ \cite{Pich:2020gzz}. The range chosen for $K_5$ includes the values  advocated by renormalon models, Padé approximants, effective charges and conformal mappings \cite{Baikov:2008jh,Beneke:2008ad,Boito:2018rwt,Caprini:2019kwp,Jamin:2021qxb,Goriachuk:2021ayq,Ayala:2022cxo}, but also allows for a correction of opposite sign. The interval of variation for the renormalization scale 
is conventional for low-energy analyses, e.g.~see \cite{Pich:2016bdg,Salam:2017qdl}. This is partially justified by the fact that taking a too small value for $\xi^2$, one would be using an ill-defined expansion parameter $\alpha_{s}(\xi^2 Q^2)$ without any real justification, leading to unreasonable uncertainties.\footnote{In this sense the observed blow-up in the uncertainty at $Q^2\approx 1\, \mathrm{GeV}^2$ when $\alpha_s$ is increased, it is a consequence of the variation of $\xi^2$ towards too small values.} Alternative prescriptions to circumvent this issue, such as asymmetric scale variations, can be found in the literature. See for example Ref.~\cite{DelDebbio:2021ryq}.

\begin{table}[tbh]\centering\renewcommand{\arraystretch}{1.2}
{\begin{tabular}{|c|c|c||c|c|c|c|c|c|}\hline 
\multicolumn{3}{|c||}{} & \multicolumn{6}{c|}{$D_{ii}^{L,(0)}(Q^2)$}
\\ \hline 
$\alpha_{s}^{(n_f=5)}(M_Z^2)$ & $Q$  & $\alpha_s(Q^2)$ & 0 &1&2&3&4&5 \\ \hline
\multirow{3}{*}{$0.115$}  &
$1.0$ & $0.4227$ & $3$ & $3.4036$   & $3.4927$ &  $3.5392$ & $3.5874$ & $3.6238$
\\
& $1.5$ & $0.3197$ & $3$ & $3.3053$ & $3.3562$  & $3.3764$ & $3.3921$ & $3.4011$
\\
& $2.0$ & $0.2751$ & $3$ & $3.2627$ & $3.3005$  & $3.3133$ & $3.3219$ & $3.3262$
\\ \hline
\multirow{3}{*}{$0.120$} &
$1.0$  & $0.5277$ & $3$  & $3.5039$ & $3.6427$ &  $3.7332$ & $3.8504$ & $3.9606$
\\
& $1.5$  & $0.3681$ & $3$ & $3.3515$ &$3.4191$ &  $3.4498$ & $3.4776$ & $3.4958$
\\
& $2.0$  & $0.3085$ & $3$ & $3.2946$ &$3.3420$ &  $3.3601$ & $3.3738$ & $3.3813$
\\ \hline
\end{tabular}}
\caption{Values of $\alpha_s(Q^2)$ with $n_f=3$ (left)
at different scales $Q$ (in $\mathrm{GeV}$ units)
for two input choices
of $\alpha_s^{(n_f=5)}(M_Z^2)$.
The corresponding values of $D^{L,(0)}_{ii}(Q^2)$ are given in the right columns
at different orders in $\alpha_s(Q^2)$. Our central value for the fifth-order coefficient, $K_5=275$, has been adopted in the last column.}
\label{tab:massless}
\end{table}

\begin{figure}[tbh]
    \centering
    \includegraphics[width=0.45\textwidth]{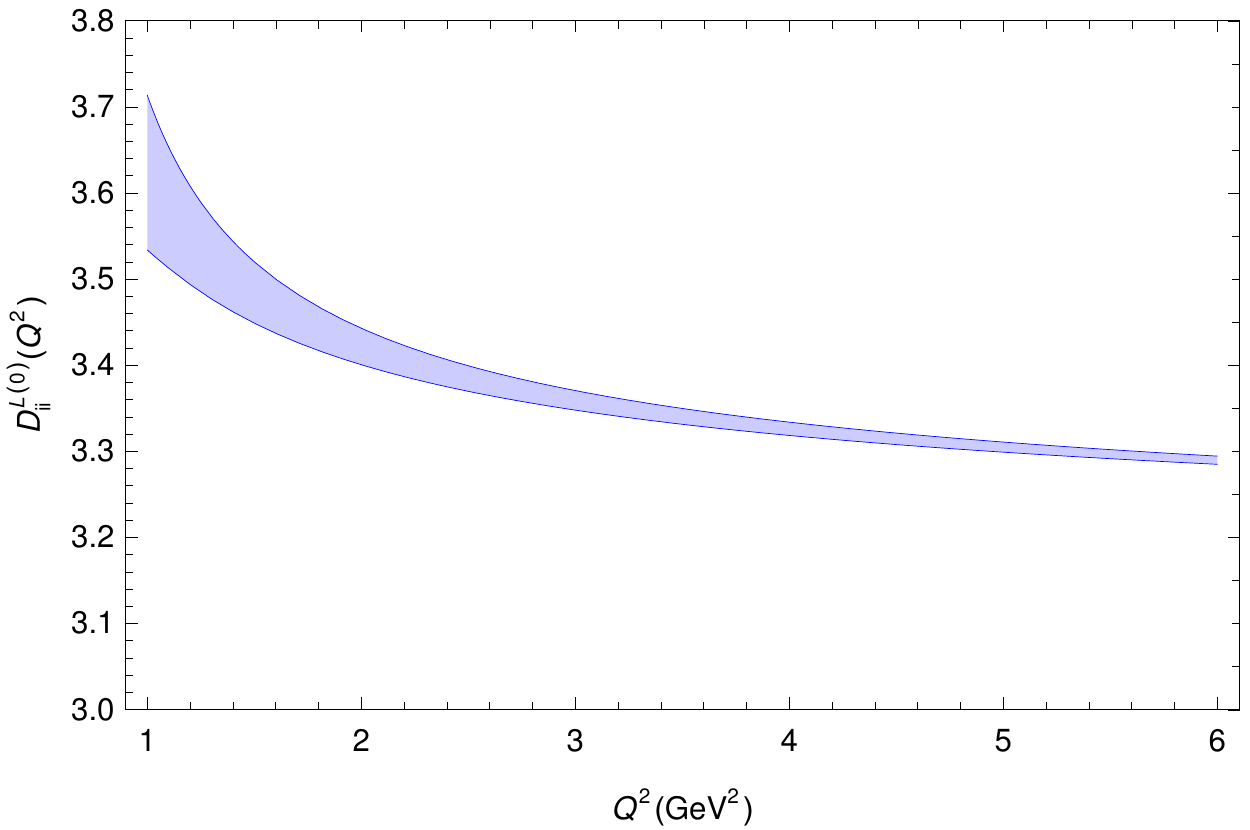}\hskip .5cm
        \includegraphics[width=0.45\textwidth]{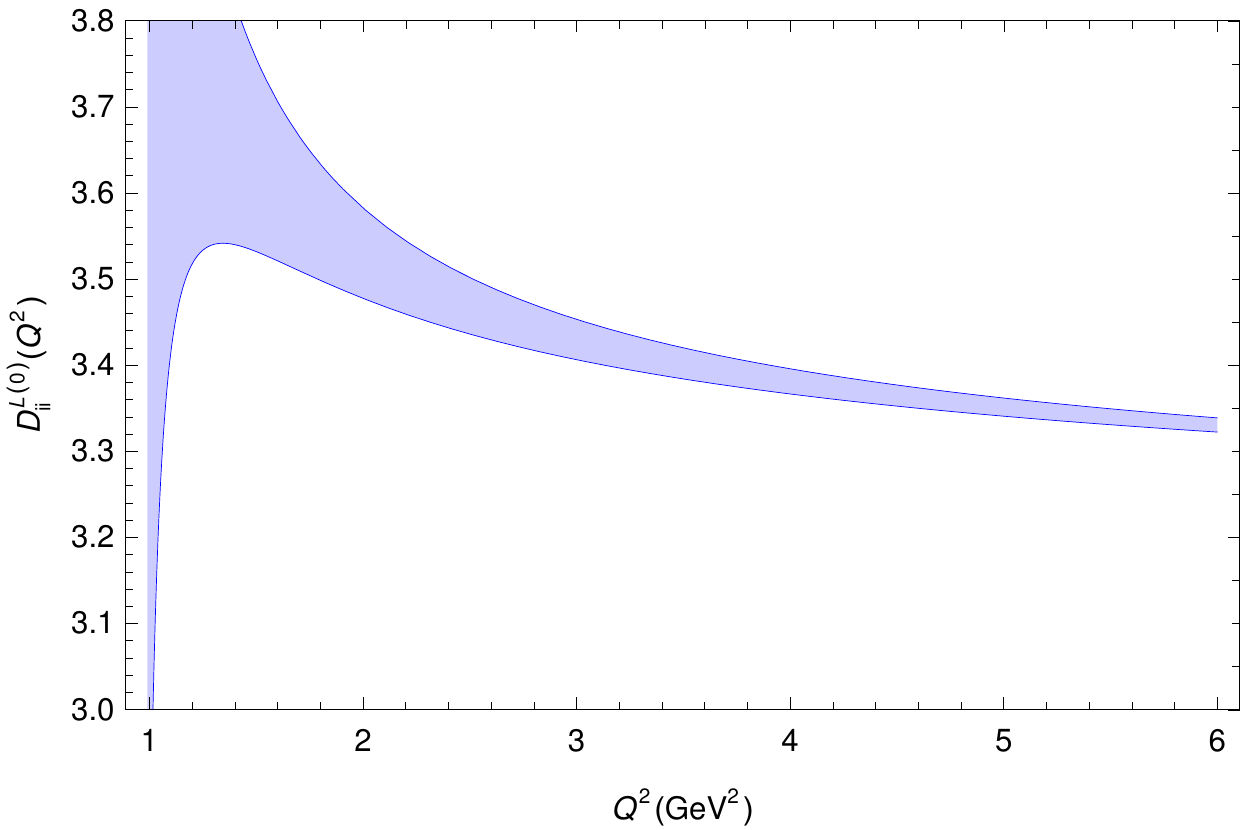}
    \caption{\label{fig:masslessD}
    $Q^2$ ($\mathrm{GeV}^2$ units) dependence
    of $D^{L,(0)}_{ii}(Q^2)$ 
      in perturbative $n_f=3$ massless QCD, at $\mathcal{O}(\alpha_s^5)$,
    for $\alpha_s^{(n_f=5)}(M_Z^2)=0.115$ (left) and $0.120$
    (right), including perturbative uncertainties.}
\end{figure}

\subsubsection{Strange mass corrections}\label{sec:strange}
The light quark masses are not exactly zero, and this leaves a small imprint, fully dominated by the strange quark mass. Let us then safely neglect the tiny effects suppressed by $\mathcal{O}(\frac{m_{u,d}^2}{Q^2})$. Taking $\mu^2=Q^2$ one has \cite{Pich:2020gzz}
\be\label{eq:strangecorr}
\Delta_{m_s} D^{L}_{33}(Q^2)=-3 N_C \frac{m_s^2(Q^2)}{Q^2}\sum_{n} (2c_n^{L+T}+e_n^{L+T}+f_{n}^{L+T})\left( {\alpha_s(Q^2)\over \pi}\right)^n +\mathcal{O}\left( \frac{m_s^4}{Q^4}\right) ,\;
\ee
where $m_s(Q^2)\equiv m_s^{(n_f=3)}(Q^2)$ and
the coefficients are once again shown in App.~\ref{app:pertcoef}. Numerical values for the associated corrections at different $Q$, taking as input $m_s(\mu_0^2)=(92.03 \pm 0.88)\, \mathrm{MeV}$ at $\mu_0=2\, \mathrm{GeV}$
\cite{FlavourLatticeAveragingGroupFLAG:2021npn,MILC:2009ltw,Durr:2010vn,Durr:2010aw,McNeile:2010ji,RBC:2014ntl,FermilabLattice:2018est,Lytle:2018evc,EuropeanTwistedMass:2014osg,Chakraborty:2014aca},
can be found in Table~\ref{tab:strange}.\footnote{Once again, we have cross-checked that both the strong coupling values and the strange quark mass ones fully agree with the corresponding ones obtained by using instead the {\sc RUNDEC} package~\cite{Herren:2017osy}.} Let us note the very bad behaviour of the perturbative series (\ref{eq:strangecorr}), which appears to show its asymptotic behaviour from the first terms.
Fortunately, the whole mass correction is very suppressed by the small value of the strange quark mass and its electric charge. We will take as central value the average between truncating at $\mathcal{O}(\alpha_s^2)$ and $\mathcal{O}(\alpha_s^3)$ and half their difference as an additional perturbative uncertainty. The associated $Q^2$ dependence, together with our estimated error bars are shown in Fig.~\ref{fig:strange}.
\begin{table}[bth]\centering\renewcommand{\arraystretch}{1.2}
{\begin{tabular}{|c|c|c|c||c|c|c|c|}\hline 
\multicolumn{4}{|c||}{} & \multicolumn{4}{c|}{$\Delta_{m_s} D^{L}_{33}(Q^2)$}
\\ \hline 
$\alpha_{s}^{(n_f=5)}(M_Z^2)$ & $Q$  & $\alpha_s(Q^2)$&$m_s(Q^2)$ & 0 &1&2&3 \\ \hline
\multirow{3}{*}{$0.115$}  &
$1.0$ & $0.4227$ & $0.1177$ & $-0.2495$ & $-0.4062$   & $-0.6007$ & $-0.8821$
\\
& $1.5$ & $0.3197$ & $0.09997$ & $-0.07994$ & $-0.1179$ & $-0.1536$ & $-0.1926$
\\
& $2.0$ & $0.2751$ & $0.09203$ & $-0.03811$ & $-0.0537$ & $-0.0663$ & $-0.0781$
\\ \hline
\multirow{3}{*}{$0.120$} &
$1.0$  & $0.5277$ & $0.1276$ &  $-0.2932$  & $-0.5229$ & $-0.8790$  & $-1.5223$
\\
& $1.5$  & $0.3681$ & $0.1018$ & $-0.08289$ & $-0.1282$ &$-0.1772$  & $-0.2390$
\\
& $2.0$  & $0.3085$ & $0.09203$ & $-0.03811$ & $-0.0556$ &$-0.0714$ & $-0.0881$
\\ \hline
\end{tabular}}
\caption{Values of $\alpha_s(Q^2)$ and $m_s(Q^2)$ with $n_f=3$ (left) for several choices of $Q$ (GeV units) and $\alpha_s^{(n_f=5)}(M_Z^2)$. The corresponding values of $\Delta_{m_s} D^{L}_{33}(Q^2)$ are shown in the right columns at different perturbative orders.
}
\label{tab:strange}
\end{table}
\begin{figure}[tbh]
    \centering
    \includegraphics[width=0.45\textwidth]{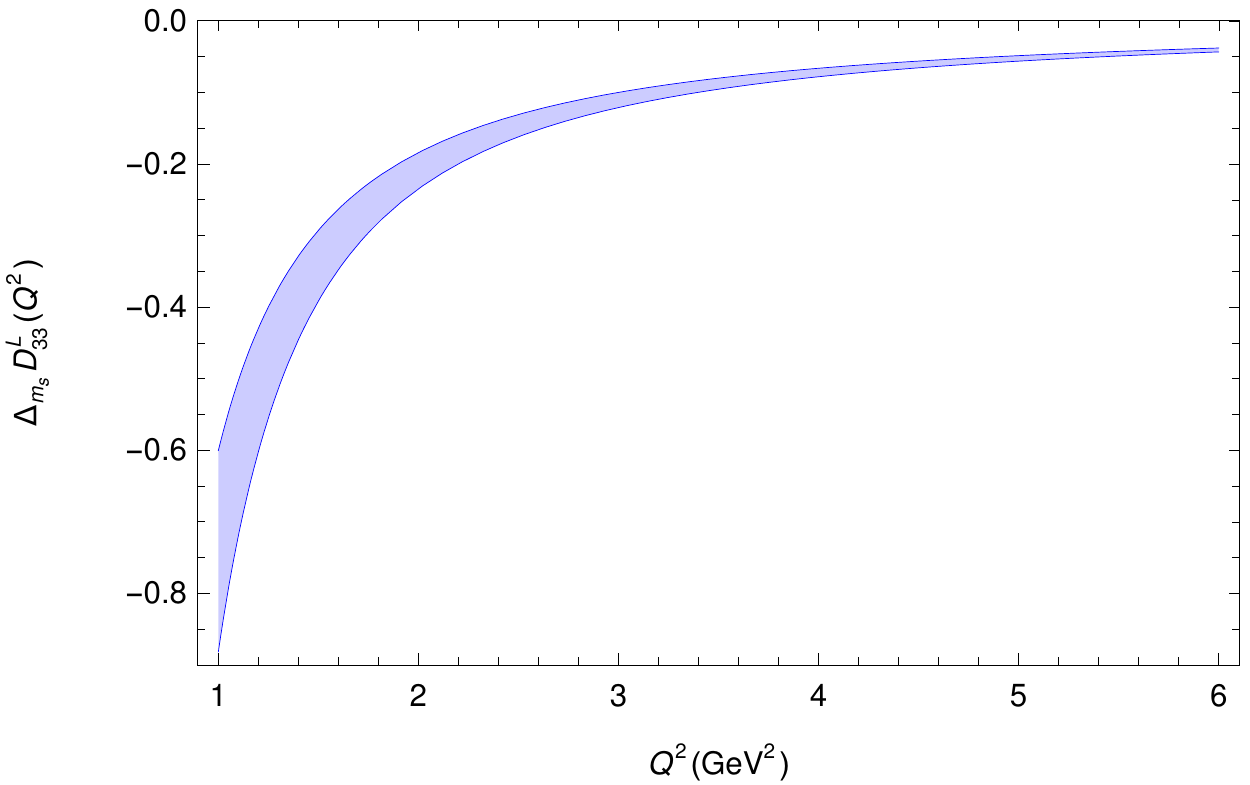}\hskip .5cm
        \includegraphics[width=0.45\textwidth]{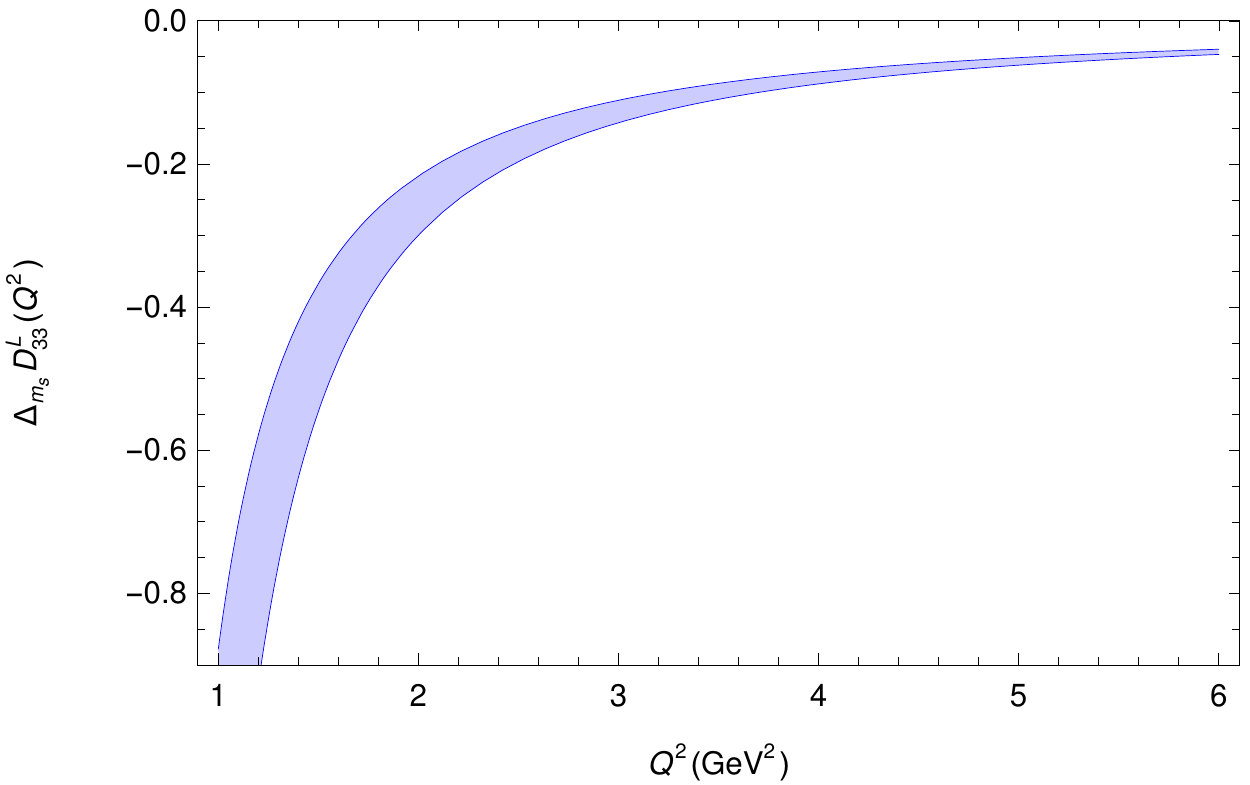}
    \caption{\label{fig:strange}
    $Q^2$ 
    dependence of $\Delta_{m_s} D^{L}_{33}(Q^2)$ for $\alpha_s^{(n_f=5)}(M_Z^2)=0.115$ (left) and $0.120$ (right).}
\end{figure}

\subsubsection{Heavy-quark mass corrections}
Internal heavy-quark loops induce charm-mass corrections into the light-quark correlators, which are suppressed by powers of $\frac{Q^2}{4m_{c}^2}$ and start to contribute at $\mathcal{O}(\alpha_s^2)$. In order to take into account these small, but sizable when 
approaching the charm threshold, contributions to the Euclidean Adler function, we can use the known results at order $\alpha_s^2$ for the associated contributions to $R(s)$. Two distinct topologies appear there. One corresponds to the four-quark cut, $\rho_R$, which starts at $s=4\, m_{c}^2$. The second corresponds to the vertex correction $\rho_V$, which, in the chiral limit, starts at $s=0$. The exact expressions can be found in~\cite{Hoang:1994it}. Taking into account that these are the only cuts induced by those topologies, one can reconstruct the associated contribution to the Adler function by using the same kind of dispersion relation as above:
\begin{align}\nonumber
D_{ii}^{L,m_c}(Q^2)&=Q^{2}\int_{s_{th}}^{\infty}ds\,\frac{\delta R_q(s)}{(s+Q^2)^2}\\&=N_C C_FT_F\,Q^2\left(\int^{4 m_c^2}_0 ds\, \frac{\rho_V
(s)}{(s+Q^2)^2}+\int^{\infty}_{4 m_c^2} ds\, \frac{\rho_R
(s)+\rho_V(s)}{(s+Q^2)^2} \right)\left(\frac{\alpha_s(\mu^2)}{\pi} \right)^2\, ,
\end{align}
where $C_FT_F=2/3$. As expected, for small enough values of $Q^2$ both integrals admit expansions in powers of $\frac{Q^2}{4m_c^2}$. For the former, the $Q^2\ll4m_c^2$ expansion can only be performed after integration, since $s \in (0,4m_c^2)$. However, one can first expand $\rho_V(s)$ in 
powers of $\frac{s}{4m_c^2}$ whose leading, next-to-leading  and next-to-next-to-leading terms can also be found in Refs.~\cite{Chetyrkin:1993tt,Larin:1994va}:
\begin{align}\nonumber
\label{eq:rhov}
\rho_{V}(s)&=
\frac{1}{45}\left[\frac{22}{5}+\log{\left(\frac{m_c^2}{s}\right)}\right]\frac{s}{m_c^2}-\frac{1}{1680}\left[\frac{1303}{420}+\log{\left(\frac{m_c^2}{s}\right)} \right]\left(\frac{s}{m_c^2}\right)^2\\
&+\frac{1}{28350}\left[\frac{1643}{630}+\log{\left(\frac{m_c^2}{s}\right)} \right]\left(\frac{s}{m_c^2}\right)^3-\frac{1}{332640}\left[\frac{32429}{13860}+\log{\left(\frac{m_c^2}{s}\right)} \right]\left(\frac{s}{m_c^2}\right)^4 \nonumber
\\ & + \mathcal{O}\left(\frac{s}{m_c^2}\right)^5\, .
\end{align}
In the left panel of Fig.~\ref{fig:rhov} we show how truncating at this order is already an excellent approximation in the needed interval. Using that expanded version, it is straightforward to analytically integrate in $s$ and then expand in powers of $\frac{Q^2}{4m_c^2}$. One finds
\begin{figure}
    \centering
    \includegraphics[width=0.48\textwidth]{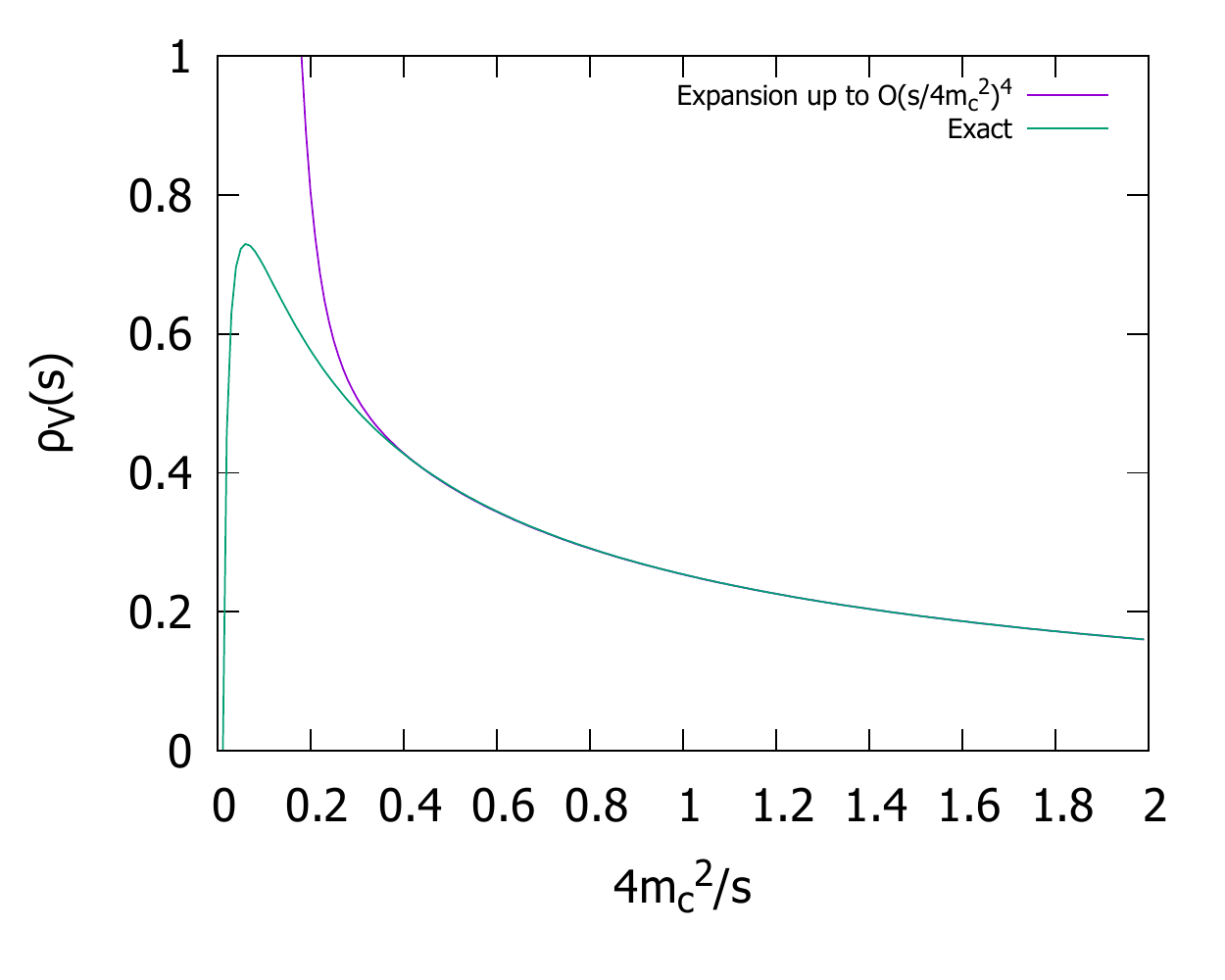}
        \includegraphics[width=0.48\textwidth]{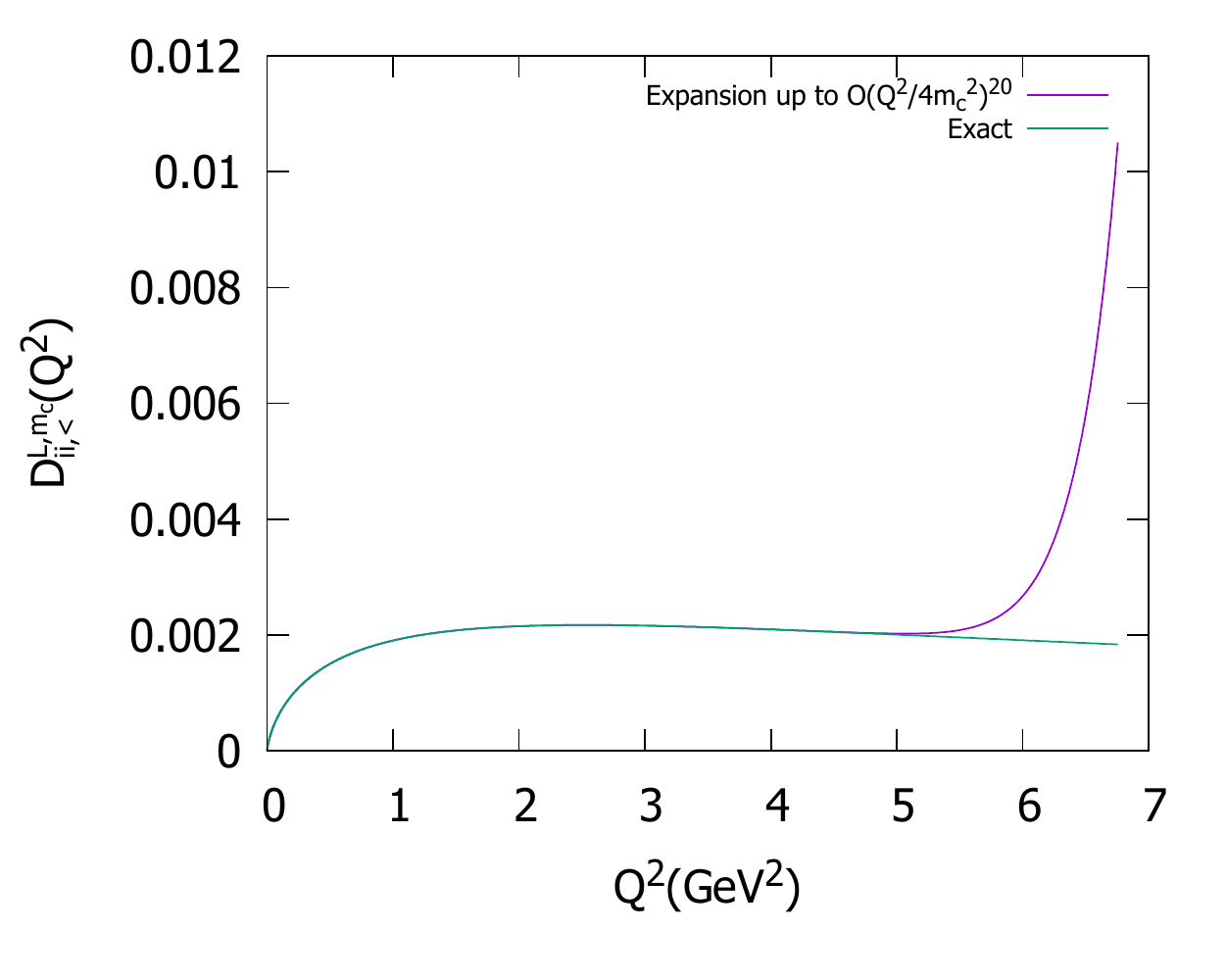}
    \caption{\emph{Left}: Comparison between the exact form of $\rho_V(s)$ and its expansion in the limit $4m_c^2/s \rightarrow \infty$. As can be checked, they are equal for $4m_c^2/s>1$. \emph{Right}: Comparison between $D_{ii,<}^{L,m_c}(Q^2)$ computed numerically and computed analytically, using the expansion~(\ref{eq:rhov}) for $\rho_V(s)$, and performing an expansion in $Q^2/4m_c^2$ afterwards. The input value $\alpha_s^{(n_f=5)}(M_Z^2) = 0.1184$ has been adopted.}
    \label{fig:rhov}
\end{figure}

\begin{align}\nonumber
    D_{ii,<}^{L,m_c}(Q^2)&\, =\, N_C C_F T_F
    \left(\frac{\alpha_s(\mu^2)}{\pi}\right)^2
    \left\{\left[-\frac{923574439 }{2161120500}+\frac{4 }{45}\zeta_2+\frac{124666 }{155925}\log (2) -\frac{8}{45}  \log ^2(2)\right. \right. \\ \nonumber
    & \left. \left.+\frac{8  }{45}\log^2\left(\frac{m_c}{Q}\right)+\frac{136  }{225}\log\left(\frac{m_c}{Q}\right)\right]\frac{Q^2}{4m_c^2} + \left[\frac{1211942621 }{2161120500}+\frac{2}{105}\zeta_2-\frac{250907 }{1091475}\log (2)\right. \right. \\ \nonumber
    & \left. \left. -\frac{4}{105} \log ^2(2)+\frac{1093 }{11025} \log\left(\frac{m_c}{Q}\right)+\frac{4  }{105}\log^2\left(\frac{m_c}{Q}\right) \right]\left(\frac{Q^2}{4m_c^2}\right)^2 + \left[-\frac{1023355847
   }{2161120500} \right. \right. \\ \nonumber
    & \left. \left. +\frac{32 }{4725}\zeta_2+\frac{4084286  }{16372125}\log (2)-\frac{64  }{4725}\log ^2(2) +\frac{45856 }{1488375} \log\left(\frac{m_c}{Q}\right) \right. \right.\\
   & \left. \biggl.+\frac{64 }{4725} \log^2\left(\frac{m_c}{Q}\right)  \right]\left(\frac{Q^2}{4m_c^2}\right)^3+ \cdots \biggr\}
   +\mathcal{O}\left(\frac{Q^2}{4 m_c^2}\alpha_s^3
   \right) \, .
\end{align}
In the right panel of Fig.~\ref{fig:rhov}, the agreement between this expansion and the result obtained computing the integral numerically, using the full expression for $\rho_V(s)$, can be seen. Following an analogous criteria as for the light quarks, we will estimate uncertainties from higher orders by changing $\mu$ between $\frac{m_c(m_c^2)}{\sqrt{2}}$ and $\sqrt{2}\, m_c(m_c^2)$.\footnote{This is just one possible choice to circumvent the overestimated uncertainty due to an ill-defined expansion parameter that one would have if $\mu=\frac{m_c(m_c^2)}{2}$ were taken. For heavy quarks an alternative way of achieving this, based on using different scale choices for $m_c(\mu^2)$ and $\alpha_{s}(\mu^2)$, which is useful to avoid underestimating uncertainties in fits with combined moments, can be found in Refs.~\cite{Dehnadi:2011gc,Boito:2019pqp}.}

Similar arguments can be used for the second term. In this case, the expansion in powers of $Q^2$ can be performed before or after integration. 
In the first case, the integrals can be computed numerically using the full expressions for $\rho_V(s)$ and $\rho_R(s)$. On the other hand, to perform the $Q^2$ expansion after integration, we have expanded $\rho^V(s)+\rho^R(s)$ first at large $s$,
\begin{align}\nonumber
\label{eq:rhov+rhor}
\rho_{R}(s)+\rho_{V}(s)&=\left[-\frac{1}{4}\log{\left(\frac{m_c^2}{s}\right)} +\zeta_3-\frac{11}{8} \right] +\left(\frac{m_c^2}{s}\right)^2\left[-\frac{3}{2}\log{\left(\frac{m_c^2}{s}\right)}-6\zeta_3+\frac{13}{2} \right]\\
&+\left(\frac{m_c^2}{s}\right)^3\left[-\frac{4}{9}\log^2{\left(\frac{m_c^2}{s}\right)}+\frac{28}{27}\log{\left(\frac{m_c^2}{s}\right)}+\frac{8}{9}\zeta(2)+\frac{68}{81} \right]+\mathcal{O}\left(\frac{m_c^2}{s}\right)^4 \, ,
\end{align}
which agrees with Ref.~\cite{Hoang:1994it}, in order to integrate them analytically. Both ways yield very similar results since, as can be seen in Fig.~\ref{fig:rhovpr}, the expansion of $\rho_V(s)+\rho_R(s)$ agrees very well with the full expression. The contribution to the Adler function then is:
\begin{align}\nonumber
    D_{ii,>}^{L,m_c}(Q^2)=& N_CC_FT_F\left\{\left[-\frac{40523 }{41472}+\frac{1}{288}\zeta_2+\frac{7}{8}\zeta_3+\frac{955 }{1728}\log (2)-\frac{1}{144} \log ^2(2) \right]\frac{Q^2}{4 m_c^2} \right. \\ \nonumber
    & \left. +\left[\frac{2673461}{2592000} -\frac{1}{180}\zeta_2-\frac{13 }{16}\zeta_3-\frac{12497 }{21600}\log (2)+\frac{1}{90}\log ^2(2) \right]\left(\frac{Q^2}{4 m_c^2}\right)^2 \right.\\ \nonumber
    &\left.+\left[-\frac{66851}{64800}+\frac{1}{144}\zeta_2+\frac{31 }{40}\zeta_3+\frac{1283 }{2160}\log (2)-\frac{1}{72}\log ^2(2)\right]\left(\frac{Q^2}{4 m_c^2}\right)^3 \right.\\
    &\biggl.+ \,\cdots\,\biggr\}\left(\frac{\alpha_s(\mu^2)}{\pi}\right)^2+\,\mathcal{O}\left(\frac{Q^2}{4 m_c^2}\alpha_s^3
    \right)\,. 
\end{align}
The comparison between this expansion and the exact result is plotted on Fig.~\ref{fig:rhovpr}. As expected, the expansion in powers of $Q^2/4m_c^2$ diverges from the exact result near threshold.

\begin{figure}
    \centering
    \includegraphics[width=0.48\textwidth]{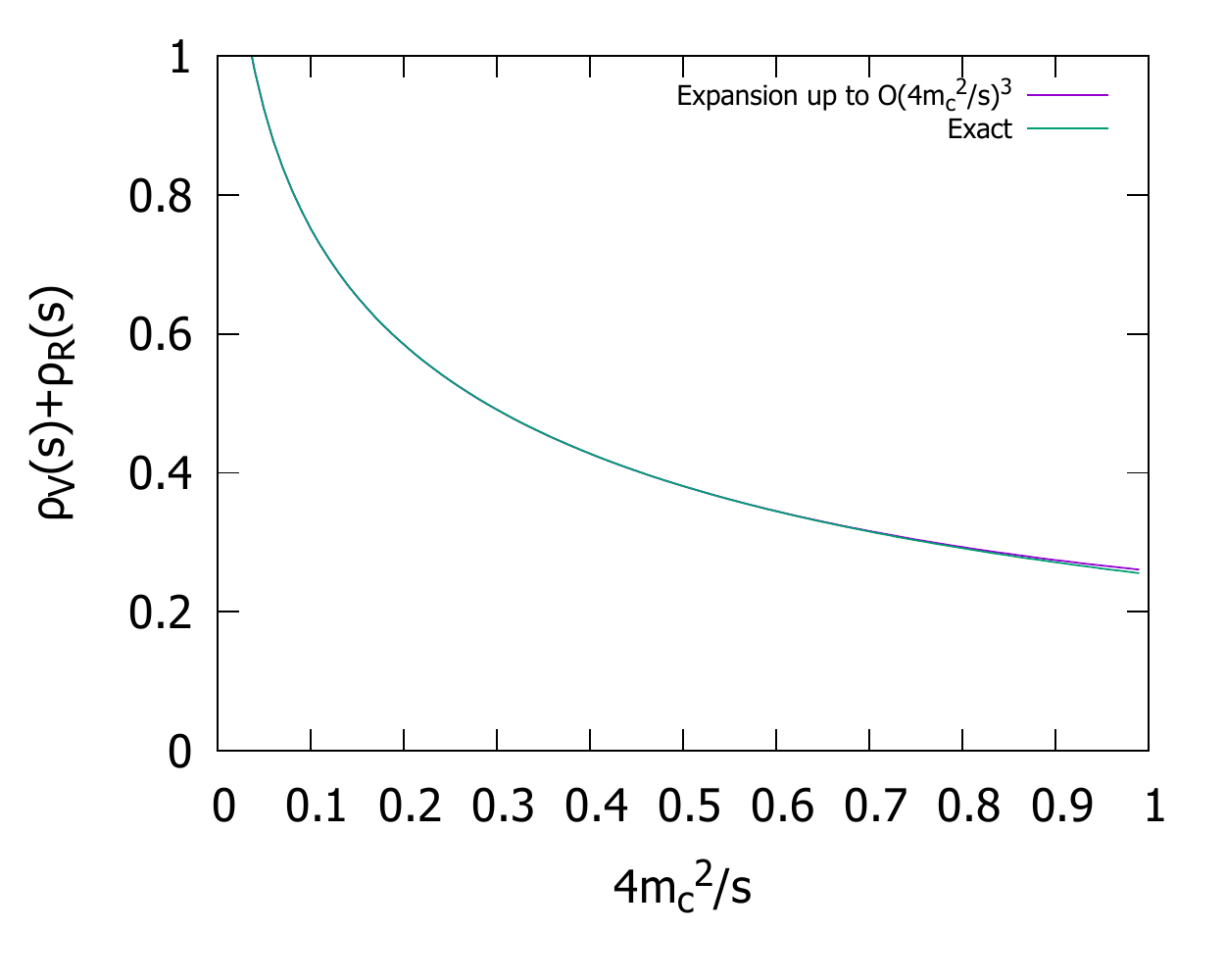}
        \includegraphics[width=0.48\textwidth]{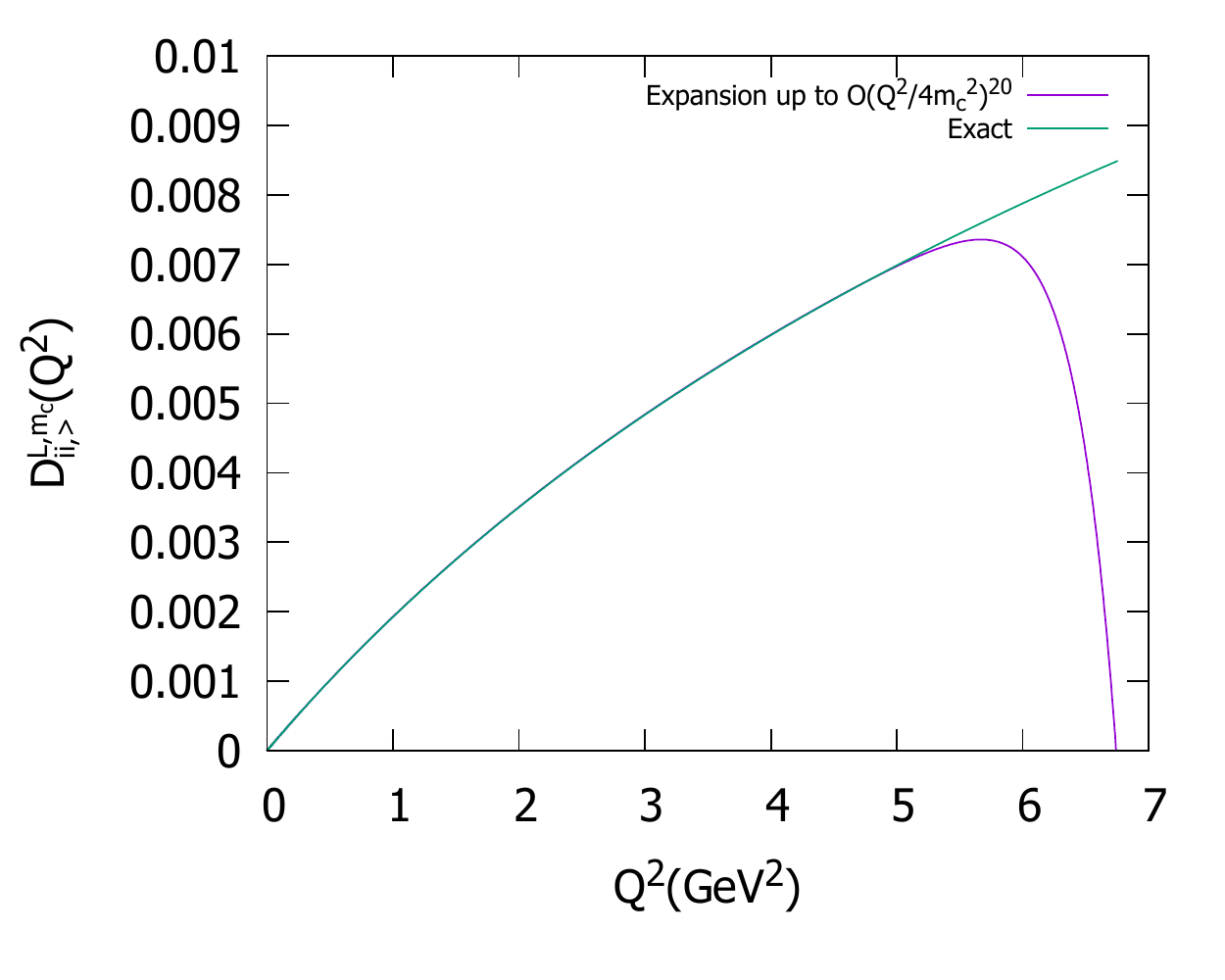}
    \caption{\emph{Left}: Comparison between the exact form of $\rho_V(s)+\rho_R(s)$ and its expansion in the limit $4m_c^2/s \rightarrow 0$. As can be seen, they are equal for $0<4m_c^2/s<1$. \emph{Right}: Comparison between $D_{ii,>}^{L,m_c}(Q^2)$ computed numerically and computed analytically, using the expansion~(\ref{eq:rhov+rhor}) for $\rho_V(s)+\rho_R(s)$ and performing an expansion in $Q^2/4m_c^2$ afterwards. The input value $\alpha_s^{(n_f=5)}(M_Z^2) = 0.1184$ has been adopted.}
    \label{fig:rhovpr}
\end{figure}

It is worth making a parenthesis to note how this second integral fully dominates if one takes instead $Q^2\gg 4 m_c^2$. In that case one may 
use the expansion (\ref{eq:rhov+rhor}) of $\rho_R(s)+\rho_V(s)$ at $s \gg m_c^2$. The leading term diverges as $s\rightarrow \infty$,
generating a large $\alpha_s^2$ contribution to $D(Q^2)$ that grows logarithmically
at large $Q^2$:
\begin{equation}
    N_C C_FT_F \;
    Q^2\int_{4m_c^2}^{\infty} ds\; \frac{-\frac{1}{4}\log\frac{m_c^2}{s} +\zeta_3-\frac{11}{8}}{(s+Q^2)^2}
    \,\approx\, N_C C_FT_F \left(-\frac{1}{4}\log\frac{m_c^2}{Q^2} +\zeta_3-\frac{11}{8} \right)\, .
\end{equation}
However one should keep in mind that the strong coupling with $n_f=3$ flavors should not be used as expansion parameter far above the charm threshold. In this limit, at order $\alpha_s^2$, one then has
\begin{align}\nonumber
D_{ii}^{L}(Q^2)&=D_{ii}^{L,(n_f=3)}\left(\alpha_{s}^{(n_f=3)}(\mu^2),Q^2\right)+D_{ii}^{L,m_c}(Q^2)\\ \nonumber
&=N_C\left[1+\frac{\alpha_s^{(n_f=3)}(\mu^2)}{\pi}+\left(\frac{\alpha_s^{(n_f=3)}(\mu^2)}{\pi}\right)^2\left(K_{2,0}^{(n_f=3)}-\frac{\beta_{1}^{(n_f=3)}}{2}\log\frac{\mu^2}{Q^2}\right)\right. \\ \nonumber
&\left.\hskip 1.cm +\left(\frac{\alpha_{s}^{(n_f=3)}(\mu^2)}{\pi}\right)^2 C_F T_F \left(-\frac{1}{4}\log\frac{m_c^2}{Q^2} +\zeta_3-\frac{11}{8} \right) +\mathcal{O}(\alpha_s^3)\right]\\ \nonumber
&=N_C\left[1+\frac{\alpha_s^{(n_f=4)}(\mu^2)}{\pi}+\left(\frac{\alpha_s^{(n_f=4)}(\mu^2)}{\pi}\right)^2\left(K_{2,0}^{(n_f=4)}-\frac{\beta_{1}^{(n_f=4)}}{2}\log\frac{\mu^2}{Q^2}\right)\right]+\mathcal{O}(\alpha_s^3)\nonumber\\
&=D_{ii}^{L,(n_f=4)}\left(\alpha_s^{(n_f=4)}(\mu^2),Q^2\right)+\mathcal{O}(\alpha_s^3) \, ,
\end{align}
which shows how the logarithm of the charm mass is properly reabsorbed into the $n_f=4$ strong coupling, through the QCD matching conditions, while the constant $\alpha_s^2$ term reproduces the known $n_f$ dependence of $K_{2,0}$.

The total heavy-quark contribution to the light Adler function,
$D_{ii}^{L,m_c}(Q^2)=D_{ii,<}^{L,m_c}(Q^2)+D_{ii,>}^{L,m_c}(Q^2)$,
is plotted in Fig.~\ref{fig:DHeavyLight}. The expanded expression in powers of $Q^2/(4 m_c^2)$ turns out to provide an excellent approximation to the exact numerical result in the full range of $Q^2$ values analysed.

\begin{figure}
    \centering
    \includegraphics[width=0.75\textwidth]{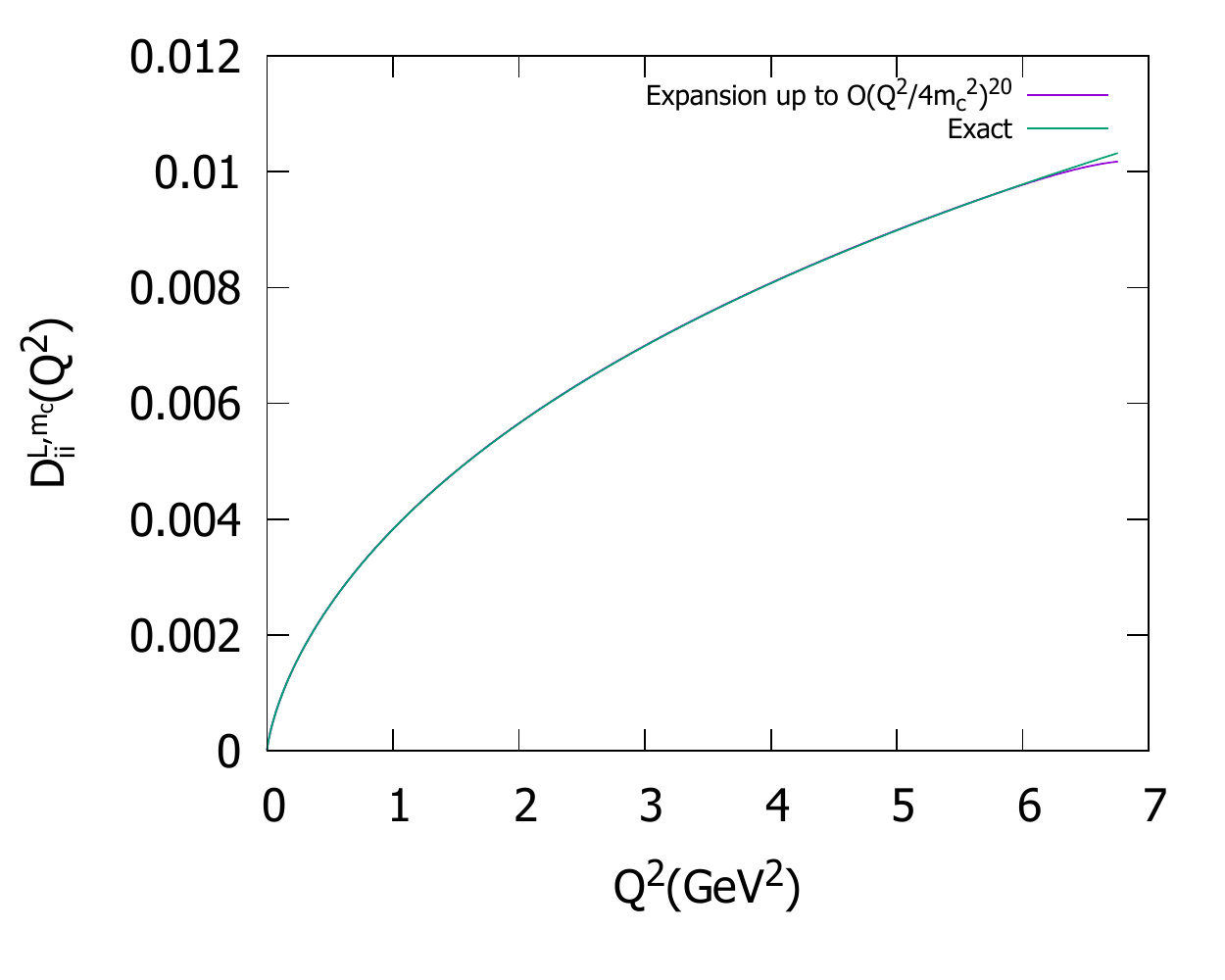}
    \caption{Comparison between $D_{ii}^{L,m_c}(Q^2)=D_{ii,<}^{L,m_c}(Q^2)+D_{ii,>}^{L,m_c}(Q^2)$ computed numerically and computed analytically, using the expansions and performing an expansion in $Q^2/4m_c^2$ afterwards. The input value $\alpha_s^{(n_f=5)}(M_Z^2) = 0.1184$ has been adopted.}
    \label{fig:DHeavyLight}
\end{figure}

\subsection{Heavy-quark contributions}
In this section we benefit from the many works devoted to computing the needed coefficients for the Adler function induced by heavy quarks~\cite{Novikov:1977dq,Chetyrkin:1995ii,Chetyrkin:1996cf,Chetyrkin:1997mb,Maier:2007yn,Boughezal:2006uu,Chetyrkin:2006xg,Boughezal:2006px,Maier:2008he,Maier:2009fz,Maier:2017ypu,Hoang:2008qy,Kiyo:2009gb}. One may write the low-energy expansion of the heavy-quark loops as
\begin{equation}\label{eq:PiHeavy}
\Pi_{ii}=\frac{3}{16\pi^2}\sum_j \overline{C}_{j}(\mu)\, z^j(\mu) \, ,
\end{equation}
where 
\begin{equation}\label{eq:CjCoefficients}
\overline{C}_{j}(\mu)=\sum_{n}\overline{C}_{j}^{(n)}(\mu)\left(\frac{\alpha_s(\mu^2)}{\pi}\right)^n  \, ,
\end{equation}
and
\begin{equation}
z(\mu)=-\frac{Q^2}{4 m_i^2(\mu^2)} \, .
\end{equation}
The associated Adler function is then
\begin{equation}
D_{ii}(Q^2)=-\frac{9}{4}\,\sum_j\, (-1)^j\, j \, \overline{C}_j(\mu)\left(\frac{Q^2}{4 m^2_{i}(\mu^2)}\right)^j \, .
\end{equation}
The $\overline{C}_{j}^{(0,1,2)}$ coefficients are known up to  
$j=30$ \cite{Maier:2007yn,Boughezal:2006uu} and $\overline{C}_{j}^{(3)}$ up to $j=3$ \cite{Chetyrkin:2006xg,Boughezal:2006px,Maier:2008he,Maier:2009fz}. The contribution to $\overline{C}_{4}^{(3)}$ from topologies associated to quark-connected (non-singlet) contributions has also been computed \cite{Maier:2017ypu}, and very good approximations to the coefficients $\overline{C}_{j}^{(3)}$ from $j=5$ to $j=10$ are also known \cite{Hoang:2008qy,Kiyo:2009gb}. They are typically given at the renormalization scale $\mu=m_c(m_c^2)$, but one can then trivially recover them at arbitrary scales by using RGEs. We compile them in App.~\ref{app:pertcoef}. In Fig.~\ref{fig:HeavyCharm-Diff-Order-alpha} we show the associated Adler function up to different orders in $\alpha_s$, cutting the series at $j=10$.\footnote{For the fourth loop and $j>4$ we take the approximate coefficients obtained in \cite{Kiyo:2009gb}, which succeeded in giving an excellent prediction for the nowadays exactly known $j=4$ \cite{Maier:2017ypu}.} The series is observed to stabilize after including the two-loop corrections. We also show in Fig.~\ref{fig:HeavyCharm-Diff-Order-Q} the convergence of the energy expansion at three loops, truncating at several values of $j$. As expected, the energy series breaks down slightly below $Q^2\sim 4 m_c^2$. We find that one actually needs to keep quite high orders. Nevertheless, for practical purposes $j=10$ is high enough to safely neglect higher orders below $Q^2\sim 5.5\, \mathrm{GeV^2}$. In order to estimate perturbative uncertainties we will vary the renormalization scale in the interval $\mu=(\frac{1}{\sqrt{2}},1)\, m_c(m_c^2)$.\footnote{
The rationale behind this asymmetric choice is that $j=10$ is not precise enough near $Q^2\sim 5.5\, \mathrm{GeV^2}$ if $\mu=\sqrt{2}\, m_c(m_c^2)$ is taken, which has nothing to do with the perturbative uncertainties of the Taylor coefficients that we want to estimate.} This, together with the uncertainty coming from the
input value $m_c(m_c^2)=1.275\, (5)\, \mathrm{GeV}$, are the main sources of error from this contribution.

\begin{figure}
    \centering
    \includegraphics[width=0.75\textwidth]{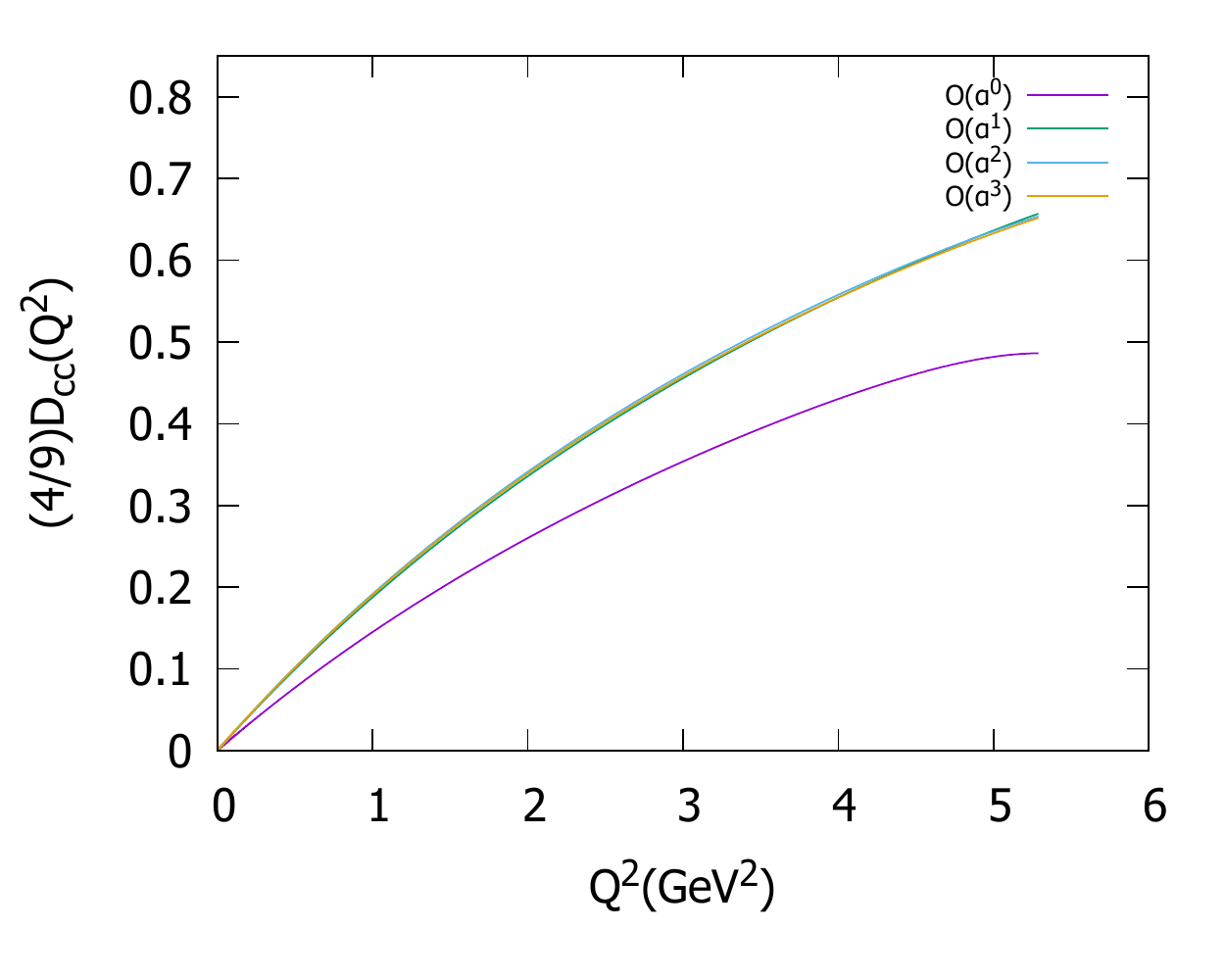}
    \caption{$D_{cc}(Q^2)$ at different orders in $\alpha_s$.
    The input value $\alpha_s^{(n_f=5)}(M_Z^2) = 0.1184$ has been adopted.}
    \label{fig:HeavyCharm-Diff-Order-alpha}
\end{figure}
\begin{figure}
    \centering
    \includegraphics[width=0.75\textwidth]{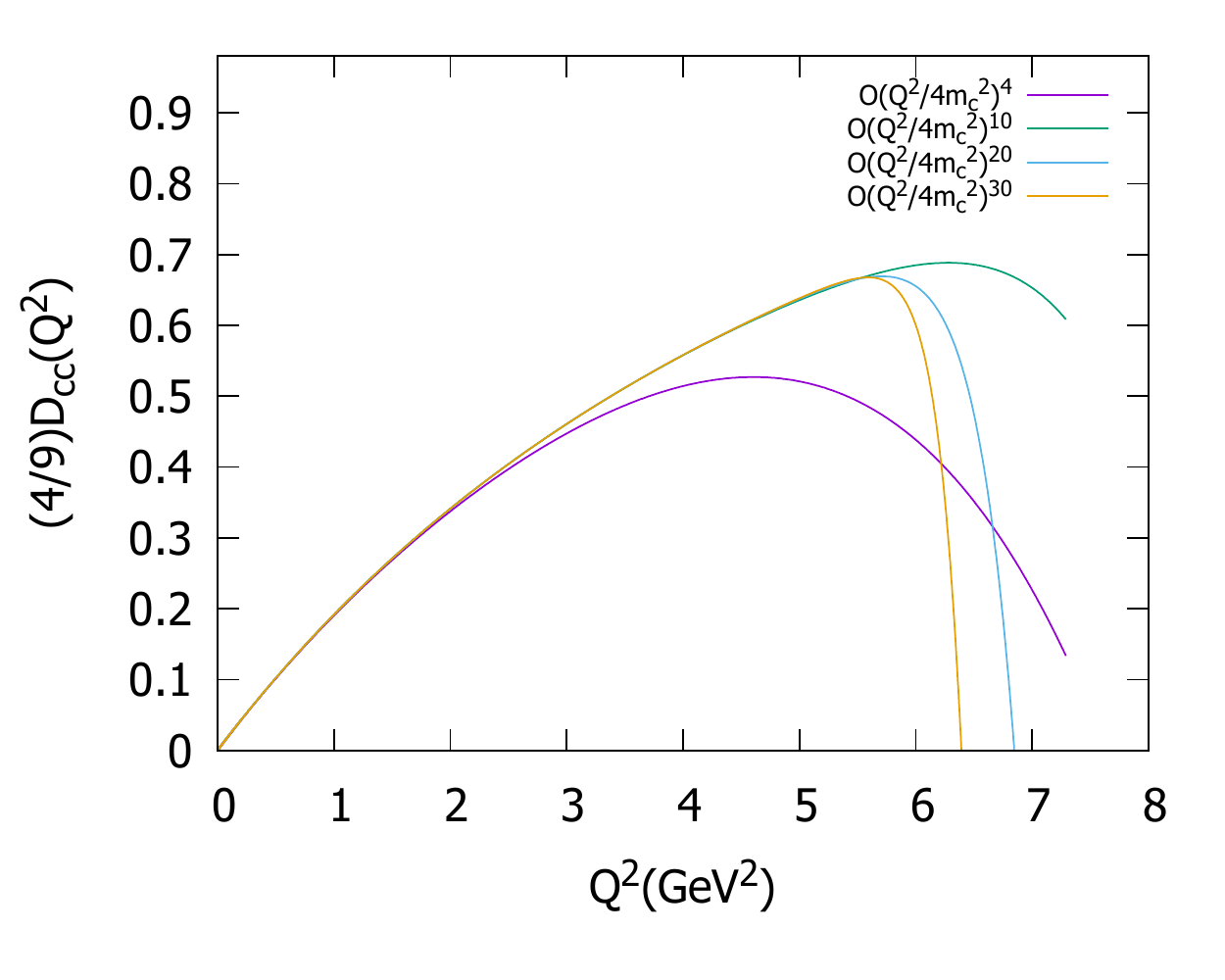}
    \caption{$D_{cc}(Q^2)$ at order $\alpha_s^2$ and at different orders in the expansion of $Q^2/4m_c^2$. As expected, the convergence is better the higher the order of the expansion is. At the same time, when going beyond the radius of convergence, the higher the order the faster it goes to infinity.}
    \label{fig:HeavyCharm-Diff-Order-Q}
\end{figure}

Finally, from $j=4$ a singlet topology with a massless (three-gluon) cut also appears at four loops, whose leading contribution at $Q^2\ll m_c^2$ is driven by a known logarithm~\cite{Groote:2001py}
\begin{equation}\label{eq:SingletHeavy}
\Pi_{cc}^{s}(Q^2)=-\frac{17 d_{abc}d_{abc}}{243000}\left(\frac{Q^2}{4m_c^2}\right)^4\left(\log\frac{Q^2}{m_c^2}+C \right)\left(\frac{\alpha_s(\mu^2)}{\pi} \right)^3 \, ,
\end{equation}
with $d_{abc}d_{abc}=40/3$. While this leads to a logarithmic divergence in the associated coefficient, the limit $Q\rightarrow 0$ does not give any problems for the Adler function itself.  One finds
\begin{equation}
D_{cc}^{s}(Q^2)\approx \frac{17\pi^2 d_{abc}d_{abc}}{20250}\left(\frac{Q^2}{4m_c^2}\right)^4\left(4\log\frac{Q^2}{m_c^2}+1+4C \right)\left(\frac{\alpha_s(\mu^2)}{\pi} \right)^3 \, .
\end{equation}
Based on the known relative values of the constant coefficients in the analogous axial-current singlet contributions (starting at $\alpha_s^2$) \cite{Maier:2007yn}, one expects $C=0\pm 3$, so the predictive power for this tiny correction is very limited. Taking into account this uncertainty and the one from changing the residual scale dependence of $\mathcal{O} (\alpha_s^4)$ in the interval $\mu=\sqrt{2}\left(\frac{1}{2},1 \right) m_c(m_c^2)$, one finds the result shown in Fig~\ref{fig:singlet}.

\begin{figure}[tbh]
    \centering
    \includegraphics[width=0.45\textwidth]{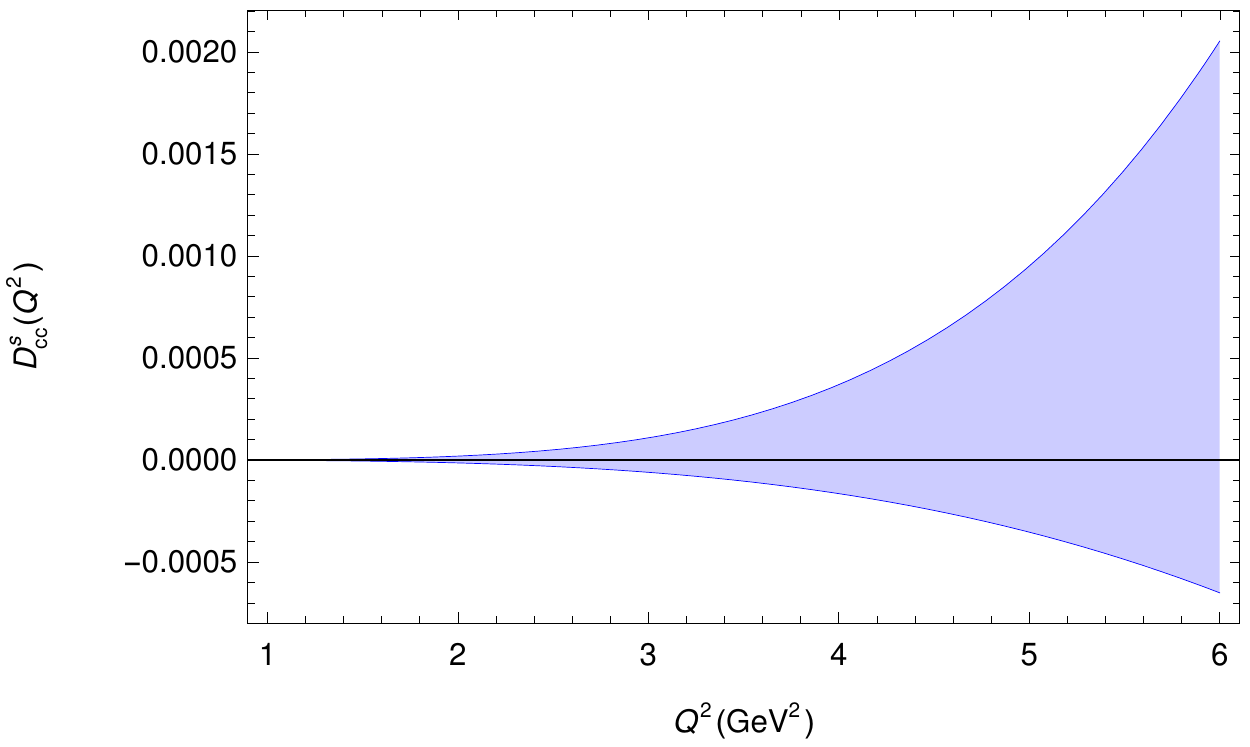}\hspace{0.5 cm}
        \includegraphics[width=0.45\textwidth]{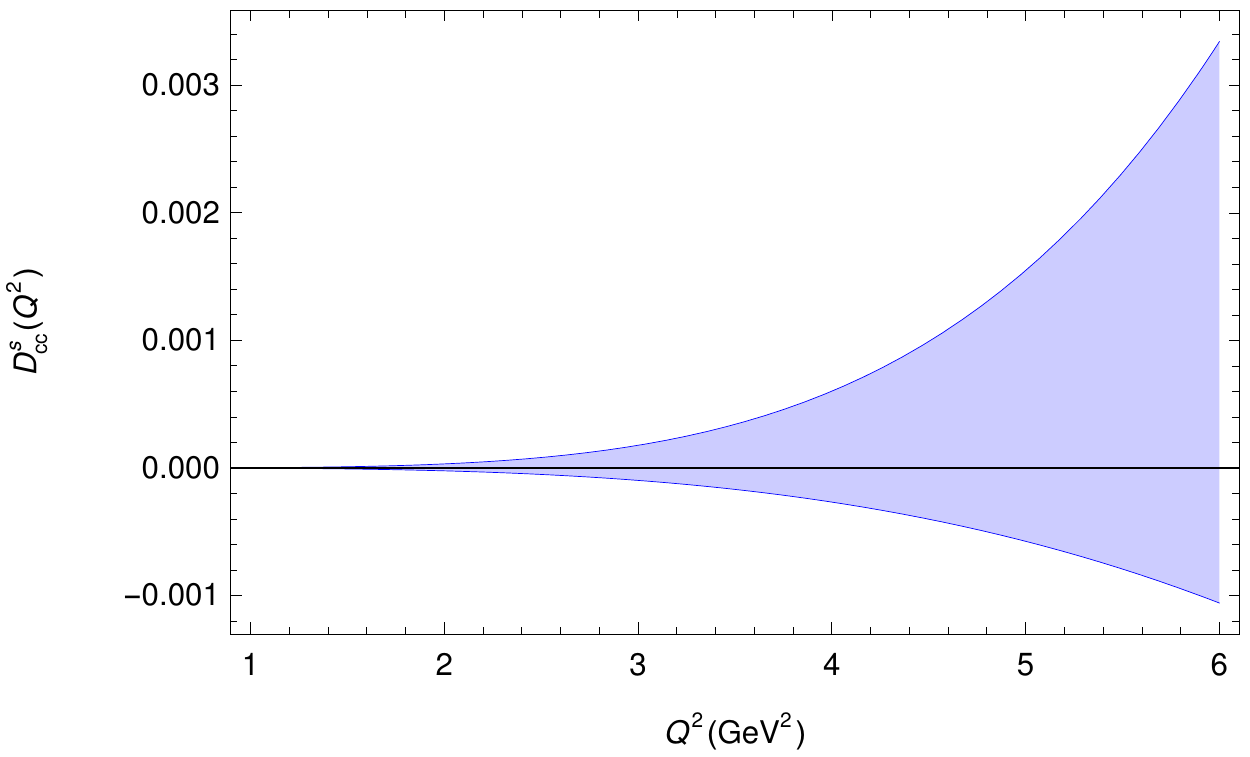}
    \caption{\label{fig:singlet}Estimated value of $D_{cc}^{s}(Q^2)$ for $\alpha_s^{(n_f=5)}(M_Z^2)=0.115$ (left) and  $0.120$ (right). $\mathrm{GeV}^2$ units.} 
\end{figure}

Adding everything up, one finds the contribution to the Adler function associated with the charm mass. The suppressed (both by $\frac{Q^2}{4m_b^2}$ and the electromagnetic charge factor $Q_b^2$) loop corrections
associated to the bottom quark are incorporated in a completely analogous way. The uncertainty of this contribution is dominated by the error on the input value of the bottom-quark mass. 

\subsection{QED corrections}
At the precision level that we have, it is worth assessing the size of the leading QED corrections. For large Euclidean momenta, QED corrections to the Adler function can be computed perturbatively just as one computes the QCD ones. 

At this level, there is a subtlety to be considered. Technically, in a full computation of the Adler function in QED plus QCD, one should incorporate the disconnected topology corresponding to a photon between two quark loops. However, this contribution is part of the vacuum polarization of the photon propagator, which, by definition, does not enter into $R(s)$. Then, consistently, at least with R-ratio data,\footnote{In the $Q^2$ region we are interested in, this effect is subleading compared to the quoted uncertainties of the lattice-based evaluation, and then can also be neglected.} we will not take that contribution into account. 
The remaining leading QED corrections are well known. Up to $\mathcal{O}(\alpha \alpha_s)$ \cite{Kataev:1992dg} and heavy-quark effects, the relevant contributions can be taken into account by the following shift
\begin{align}
D_{ii}^{\mathrm{L,(0)}}&\rightarrow (1+\delta D_{ii}^{\mathrm{QED}}) D_{ii}^{\mathrm{L,(0)}} \, ,
\end{align}
where
\begin{equation}
\delta D_{ii}^{\mathrm{QED}}=\frac{3}{4}Q_{i}^2\,\frac{\alpha}{\pi} \, .
\end{equation}
Taking as central value the average between the results for $D_{ii}^{L(0)}$ obtained at $Q=2\,\mathrm{GeV}$  with $\alpha_s=0.1184$ and $\alpha_s=0$ (both choices are correct up to $\mathcal{O}(\alpha \alpha_s)$ effects)  and half their difference as perturbative uncertainty, one finds,

\begin{equation}
\Delta D_{\mathrm{QED}}(Q^2)\equiv D(Q^2)-D^{\alpha=0}(Q^2)=0.0012\; (4) \, ,
\end{equation}
where the error includes a conservative estimate of missing QED corrections associated with heavy-quark loops. Perturbative QED corrections are then negligible,  at the current precision level.

\subsection{Compilation}
We have now all the needed ingredients to build up the full perturbative Euclidean Adler function below the charm threshold. Up to negligibly small corrections, one has\footnote{The only ones not yet discussed are the $\sum_{i=1}^3 Q_{i}D_{ic(b)}$ ones, which can only enter through disconnected diagrams. However, taking into account that $\sum_{i=1}^3 Q_{i}=0$, they vanish in the chiral limit and then they are suppressed by $\alpha_s^3\frac{m_s^2 Q^2}{m_{c(b)}^4}$, which makes them completely negligible.}
\begin{align}\nonumber
\label{eq:AdlerPert}
D(Q^2)=\sum_{i,j}Q_{i}Q_{j}D_{ij}(Q^2)&=\frac{2}{3}\, D_{ii}^{L,(0)}(Q^2)+\frac{1}{9}\,\Delta_{m_s}D_{33}^{L}(Q^2)+\frac{2}{3}\, D_{ii}^{L,m_c}(Q^2)\\&+\frac{4}{9}\, D_{cc}(Q^2)+\frac{4}{9}\, D_{cc}^{s}(Q^2)+\frac{1}{9}\, D_{bb}(Q^2) +\Delta D_{\mathrm{QED}}(Q^2)\, .
\end{align}
In Table~\ref{tab:together} we present our numerical results for $\alpha_s^{(n_f=5)}(M_Z^2)=(0.115,0.120)$ and the current lattice average $\alpha_s^{(n_f=5)}(M_Z^2)=0.1184 \pm 0.0008$ \cite{FlavourLatticeAveragingGroupFLAG:2021npn,Ayala:2020odx,Bazavov:2019qoo,Cali:2020hrj,Bruno:2017gxd,PACS-CS:2009zxm,Maltman:2008bx} for $Q^2=3,4,5\, \mathrm{GeV}^2$, including uncertainties.

\begin{sidewaystable}[ph!]\centering\setlength{\tabcolsep}{5pt}\renewcommand{\arraystretch}{1.2}
{\begin{tabular}{|c|c||c|c|c|c|c|c|c|c|}\hline 
$\alpha_{s}^{(n_f=5)}(M_Z^2)$ & $Q^2$  & $\frac{2}{3}D_{ii}^{L,(0)}$ & $\frac{1}{9}\Delta_{m_s}D_{33}^{L}$ & $\frac{2}{3}D_{ii}^{L,m_c}$ & $\frac{4}{9}D_{cc}$    & $\frac{4}{9}D_{cc}^{s}$  & $\frac{1}{9}D_{bb}$ & $\Delta D_{\mathrm{QED}}$ &  $D$  \\ \hline
\multirow{3}{*}{$0.115$}  &
$3$ & $2.2395(77)$ & $-0.0123(12)$ & $0.0039(10)$   & $0.4484(21)(24)$ &  $0.0000(00)$ & $0.0130(01)$ & $0.0012(04)$ & $2.694(09)$
\\
& $4$ & $2.2175(52)$ & $-0.0080(07)$ & $0.0045(12)$ & $0.5435(24)(26)$  & $0.0000(01)$ & $0.0171(02)$ & $0.0012(04)$ & $2.776(07)$
\\
& $5$ & $2.2033(39)$ & $-0.0058(04)$ & $0.0050(13)$ & $0.6197(31)(26)$  & $0.0001(03)$ & $0.0212(02)$ & $0.0012(04)$ & $2.845(06)$
\\ \hline
\multirow{3}{*}{$0.120$} &
$3$  & $2.2866(156)$ & $-0.0141(17)$  & $0.0053(21)$ & $0.4649(47)(24)$ &  $0.0000(01)$ & $0.0132(01)$ & $0.0012(04)$ & $2.757(17)$
\\
& $4$  & $2.2542(98)$& $-0.0089(09)$ & $0.0062(24)$ & $0.5629(53)(26)$ &  $0.0001(02)$ & $0.0174(02)$ & $0.0012(04)$ & $2.833(12)$
\\
& $5$  & $2.2343(70)$ & $-0.0063(06)$ & $0.0069(26)$ & $0.6429(65)(27)$ &  $0.0002(05)$ & $0.0215(02)$ & $0.0012(04)$ & $2.900(11)$
\\ \hline
\multirow{3}{*}{$0.1184(8)$} &
$3$  & $2.2699(124)$ & $-0.0134(15)$  & $0.0048(17)$ & $0.4591(36)(24)$ &  $0.0000(01)$ & $0.0131(01)$ & $0.0012(04)$ & $2.735(17)$
\\
& $4$  & $2.2414(79)$ & $-0.0086(08)$ & $0.0055(19)$ & $0.5561(41)(26)$ &  $0.0001(02)$ & $0.0173(02)$ & $0.0012(04)$ & $2.813(14)$
\\
& $5$  & $2.2236(58)$ & $-0.0061(05)$ & $0.0062(20)$ & $0.6348(51)(27)$ &  $0.0002(04)$ & $0.0214(02)$ & $0.0012(04)$ & $2.881(13)$
\\ \hline
\end{tabular}}
\caption{Values of the different contributions to the Euclidean Adler function for different input values of $\alpha_s^{(n_f=5)}(M_Z^2)$. The first and second uncertainties in $D_{cc}$ correspond, respectively, to the perturbative error and the one coming from the input charm mass. Uncertainties in $\alpha_s$ in the last three rows are only included for the final number, $D$.}
\label{tab:together}
\end{sidewaystable}

In Fig.~\ref{fig:TotalAdler} we plot our final result for the perturbative Adler function $D(Q^2)$, including all estimated uncertainties.\footnote{Let us note how the observed bending at $Q^2\sim 5.5-6\, \mathrm{GeV}^2$ is not a physical feature of the Euclidean Adler function, but a first signature of the breakdown of the series in powers of $\frac{Q}{2 m_c}$. As a consequence we will restrict the comparisons with other determinations to $Q^2< 5.5\, \mathrm{GeV}^2$.}
The relative size of the different contributions and their corresponding errors are shown in Fig.~\ref{fig:relativeAdler}.

\begin{figure}[tb]
    \centering
    \includegraphics[width=0.75\textwidth]{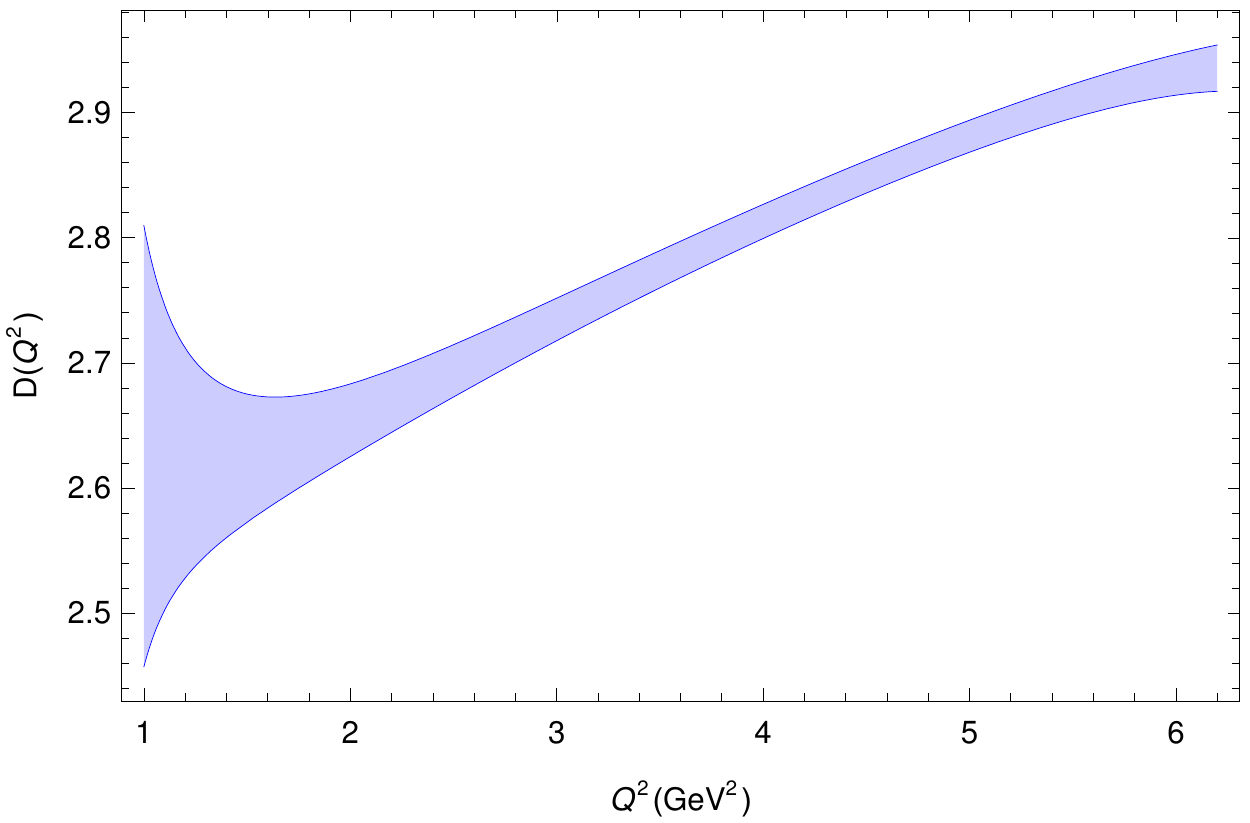}
    \caption{Final result for the perturbative $D(Q^2)$ with $\alpha_s^{(n_f=5)}(M_Z^2)=0.1184\pm0.0008$ along with its uncertainty.} 
    \label{fig:TotalAdler}
\end{figure}
\begin{figure}[tb]
    \centering
    \includegraphics[width=0.49\textwidth]{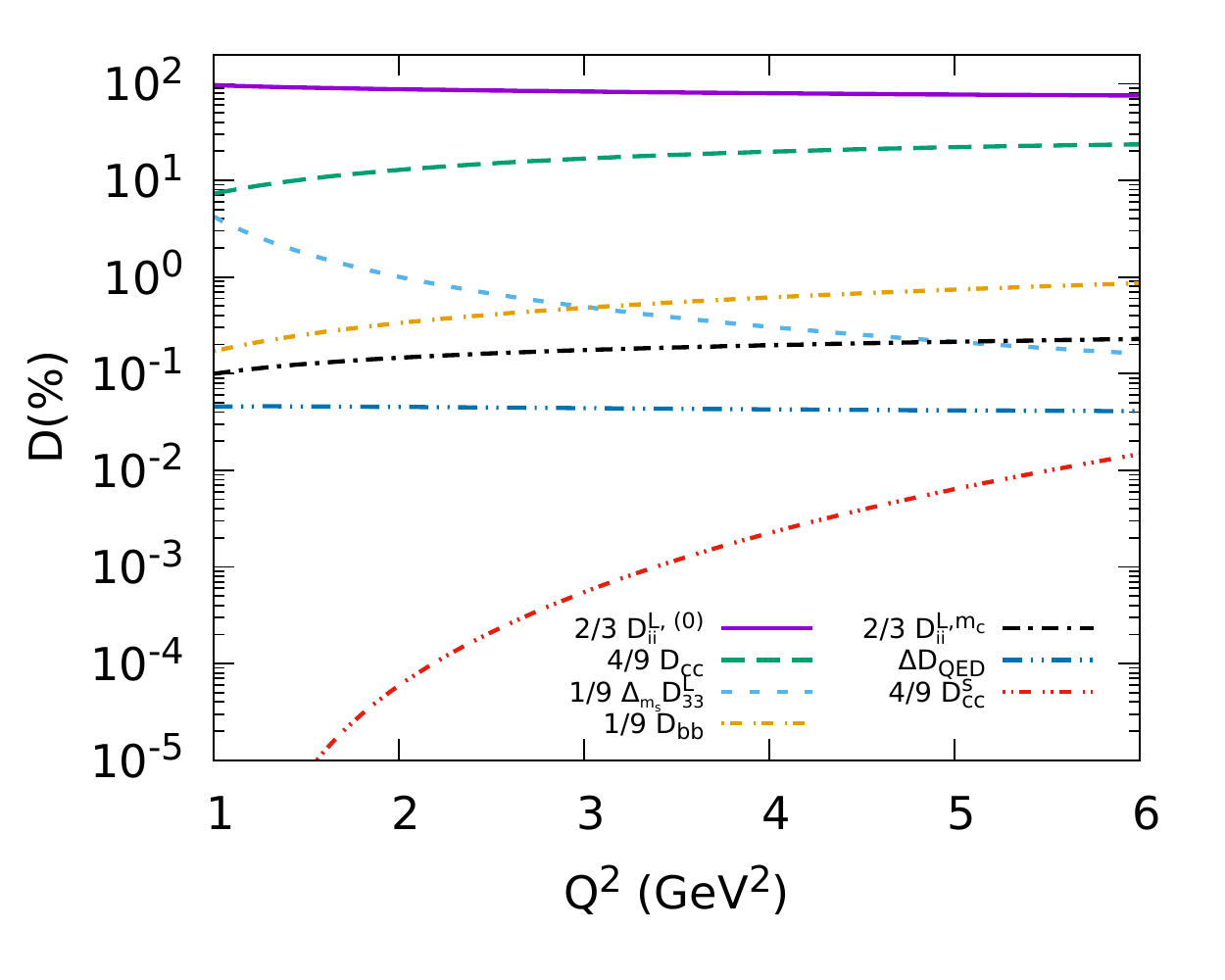}
     \includegraphics[width=0.49\textwidth]{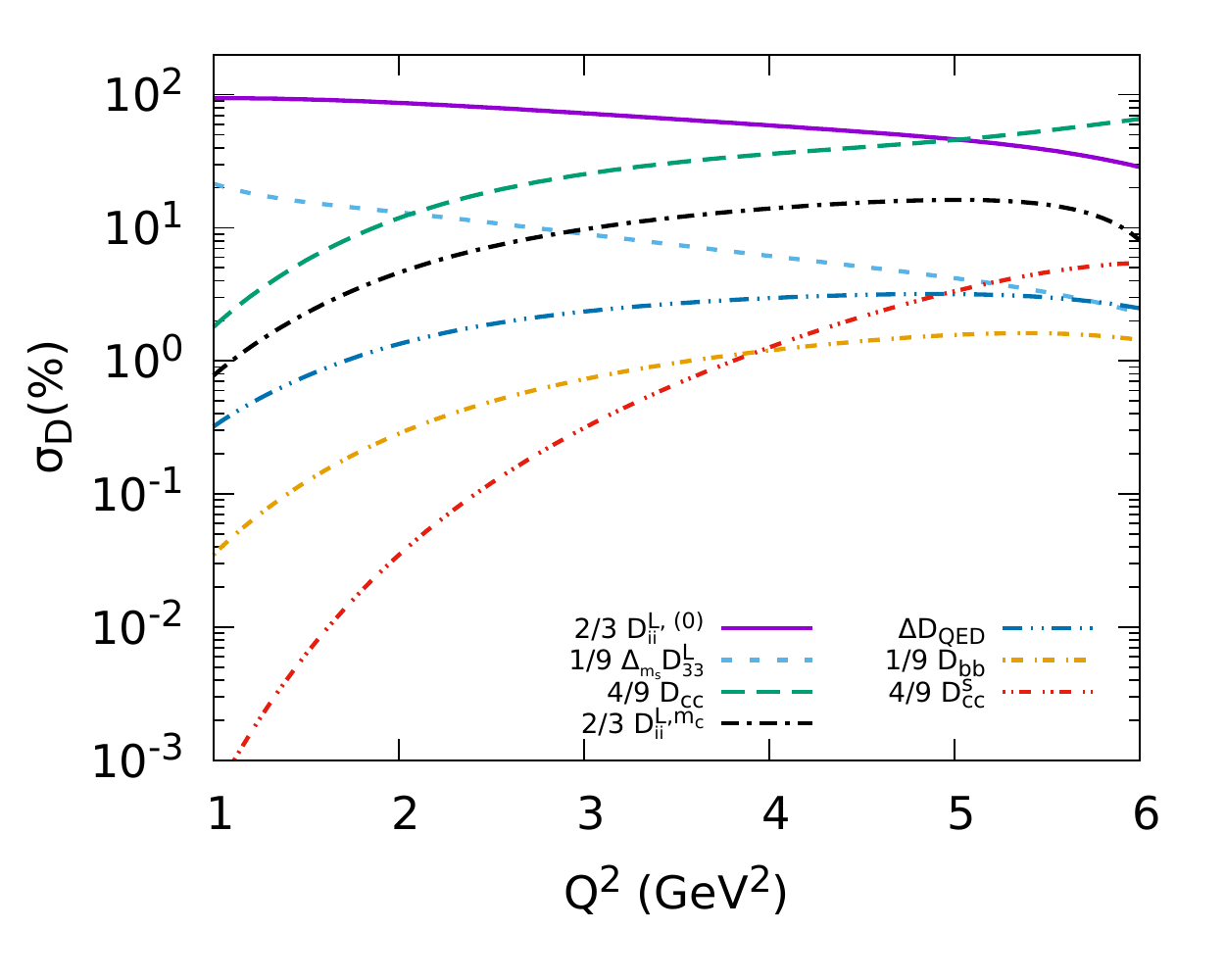}
    \caption{Relative contribution to the value (left) and the uncertainty (right) of $D(Q^2)$ as a function of $Q^2$ taking as input $\alpha_s^{(n_f=5)}(M_Z^2)=0.1184$.} 
    \label{fig:relativeAdler}
\end{figure}

\section{Adler function based on the experimental ratio {\boldmath $R(s)$} and lattice data}\label{sec:otheradlers}
In this section we discuss in some detail the experimental data that we use for $R(s)$ (based on the DHMZ compilation \cite{Davier:2017zfy,Davier:2019can}) and the lattice inputs (based on the Mainz results of Ref.~\cite{Ce:2022eix}).

\subsection[Experimental ratio $R(s)$]{Experimental ratio {\boldmath $R(s)$}}
The experimental results on $R(s)$ used in this work are based on the DHMZ data compilation made in Refs.~\cite{Davier:2017zfy,Davier:2019can}. 
More concretely, we make use of a previous study of the different contributions to $R(s)$, performed with a full treatment of the uncertainties and their correlations.
 This information allows us to evaluate the hadronic running of $\alpha(Q^2)$, $\Delta\alpha_{\mathrm{had}}(Q^2)$~(see the top panels of Fig.~\ref{fig:AdlerR}), which is then used in order to derive the Adler function with its various uncertainty components and its covariance matrix.

The Adler function emerging from the $R(s)$ data is displayed in the bottom panels of Fig.~\ref{fig:AdlerR}, together with the associated correlation matrix. 
The relative size of the various contributions to $D(Q^2)$ and of their corresponding uncertainties are shown in Fig.~\ref{fig:relativeD}. 
At low values of $Q^2$, the exclusive channels fully dominate. 
This contribution, corresponding to the region $\sqrt{s_{\mathrm{th}}}<\sqrt{s}<1.8\, \mathrm{GeV}$, is derived based on the measurements of $32$ exclusive channels. 
This compilation takes into account the statistical and systematic correlations between the different points/bins of a given measurement, between different experiments measuring a given channel, as well as between different channels~\cite{Davier:2010rnx, Davier:2010nc,Davier:2017zfy,Davier:2019can}.
In this procedure, possible tensions between different measurements of a given channel are also taken into account.
This is generally done in the combination, through an enhancement of the uncertainties by a factor $\sqrt{\chi^{2} / {\rm ndof}}$, applied in all the $\sqrt{s}$ bins where this factor is larger than unity.
In addition, an extra uncertainty accounting for the systematic deviations between the BaBar~\cite{BaBar:2009wpw,BaBar:2012bdw} and KLOE~\cite{KLOE:2008fmq,KLOE:2010qei,KLOE:2012anl} measurements in the $2\pi$ channel has been included for the first time in Ref.~\cite{Davier:2019can}, by comparing the combination results obtained when excluding either of the two experiments.\footnote{Note: Recently, a new precise measurement of the $2\pi$ channel performed by the CMD-3 collaboration has been made public~\cite{CMD-3:2023alj}. In the dominant $\rho$-resonance region, it features larger cross-section values compared to all previous experiments, in particular the most precise ones from BaBar~\cite{BaBar:2009wpw,BaBar:2012bdw} and KLOE~\cite{KLOE:2008fmq,KLOE:2010qei,KLOE:2012anl}. While dispersive integrals computed with these new inputs are certainly enhanced and closer to the ones obtained from Lattice QCD, it is of upmost importance to first achieve a better understanding of the tensions on the experimental side. In particular, one needs to understand the source of tension between the CMD-3 results and the former CMD-2 measurements~\cite{Aulchenko:2006dxz,CMD-2:2006gxt}, performed in somewhat similar conditions by the same group.}
This uncertainty turns out to be dominant in the case of the theoretical prediction for the anomalous magnetic moment of the muon, hence the importance of fully taking this systematic effect into account.
The remaining data-based contributions come from the $3.7\, \mathrm{GeV}<\sqrt{s}<5\,\mathrm{GeV}$ interval, the dispersive integrals being evaluated based on the inclusive measurements available in this range, and from the narrow $J/\Psi$ and $\Psi(2S)$ resonances.

Unfortunately, no precise enough data are yet available for the remaining regions, $1.8\, \mathrm{GeV}<\sqrt{s}<3.7 \, \mathrm{GeV}$~\footnote{The relatively precise BES III and KEDR results are in tension in the range $3.40\, \mathrm{GeV}<\sqrt{s}<3.67 \, \mathrm{GeV}$~\cite{BESIII:2021wib}.} and $\sqrt{s}>5\, \mathrm{GeV}$, and there one needs to rely on perturbation theory for $R(s)$.\footnote{Notice, however, that the perturbative uncertainties include also the estimated size of potential violations of quark-hadron duality in the region $1.8-2\, \mathrm{GeV}$~\cite{Davier:2019can}.} 
This happens to be a more critical and limiting factor for the associated Adler function, especially at large $Q^2$, where this contribution eventually dominates when $Q\gtrsim 2 \, \mathrm{GeV}$.

\begin{figure}
    \centering
    \includegraphics[height=0.32\textwidth]{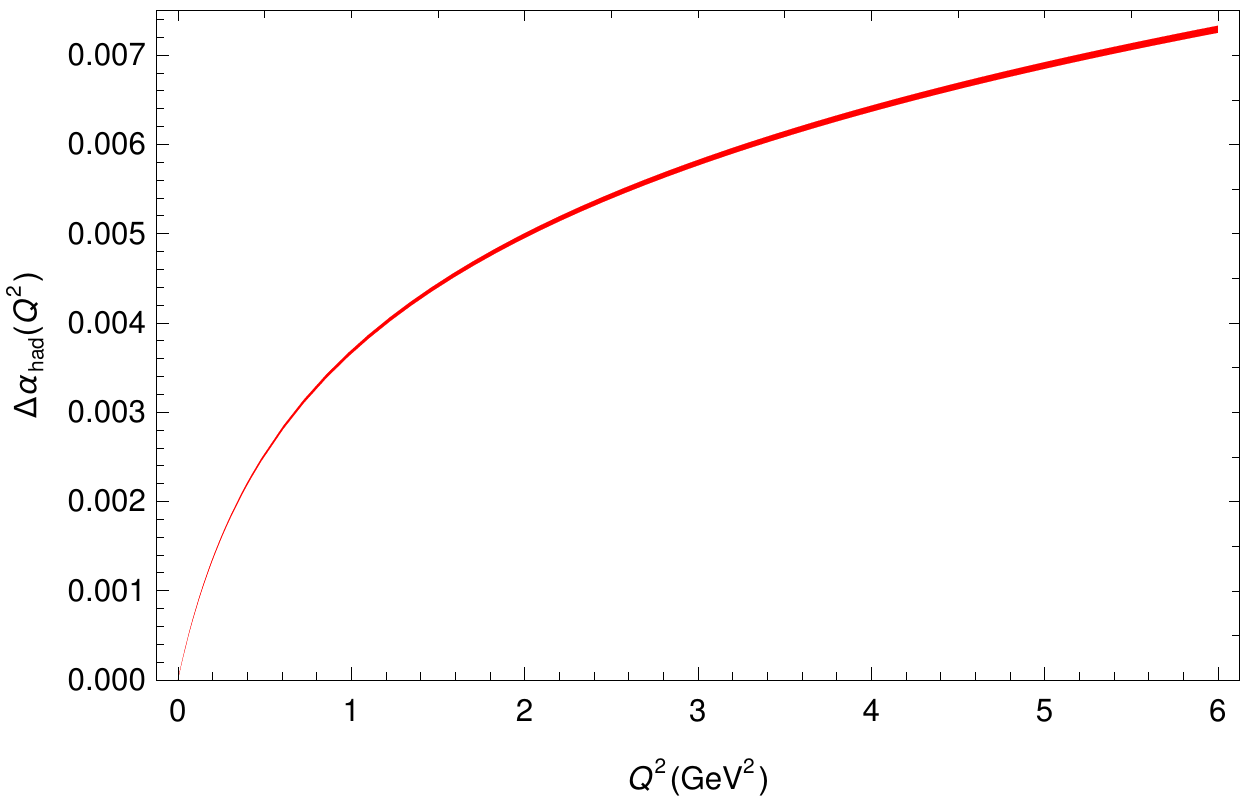}
    \includegraphics[height=0.36\textwidth]{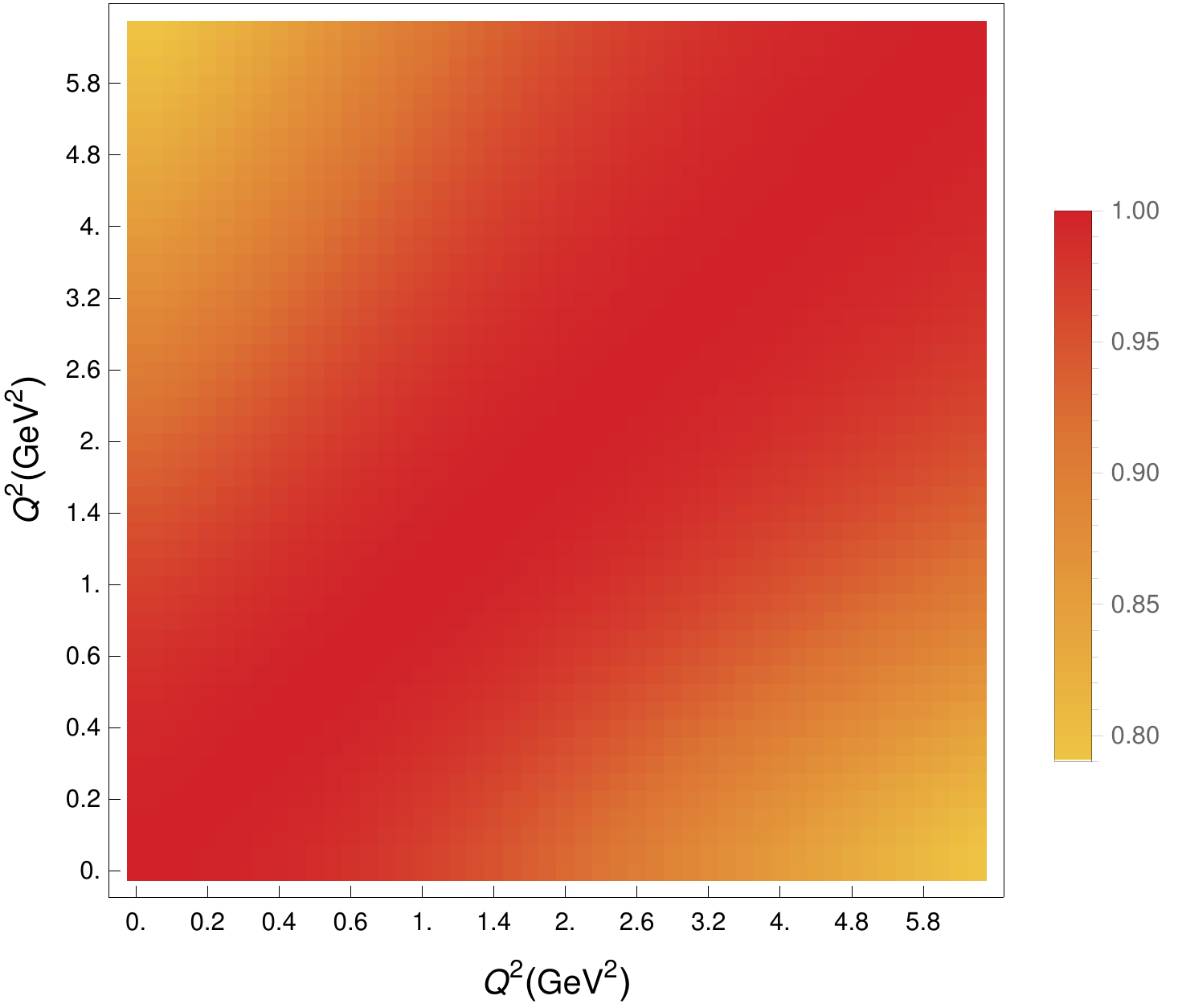}
    \includegraphics[height=0.32\textwidth]{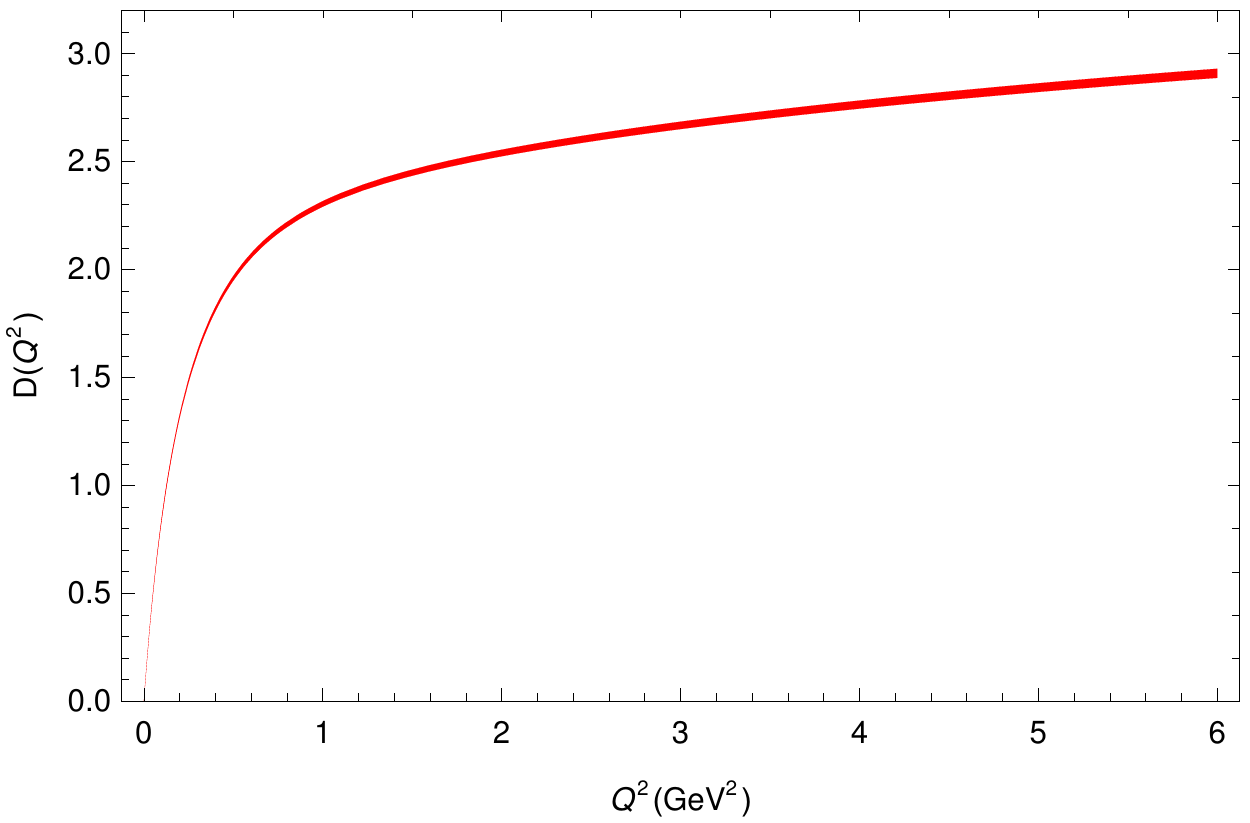}
    \includegraphics[height=0.36\textwidth]{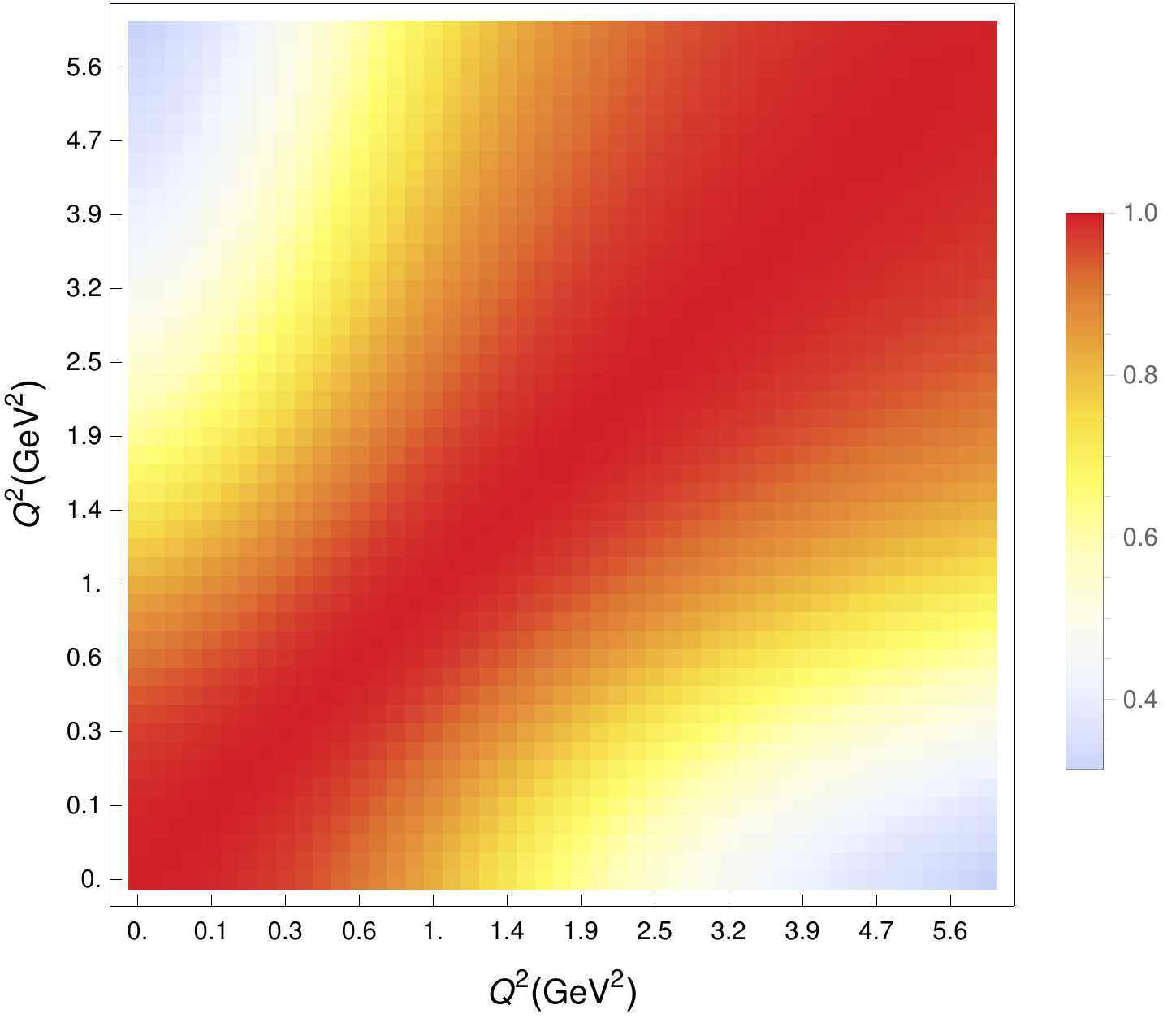}
\caption{Results for $\Delta \alpha_{\mathrm{had}}(Q^2)$ (top) and $D(Q^2)$ (bottom) from the data-driven dispersive evaluation based on the ratio $R(s)$, together with their correlations (right). The half-width of the red band indicates the total uncertainty.} 
    \label{fig:AdlerR}
\end{figure}

\begin{figure}
    \centering
    \includegraphics[width=0.48\textwidth]{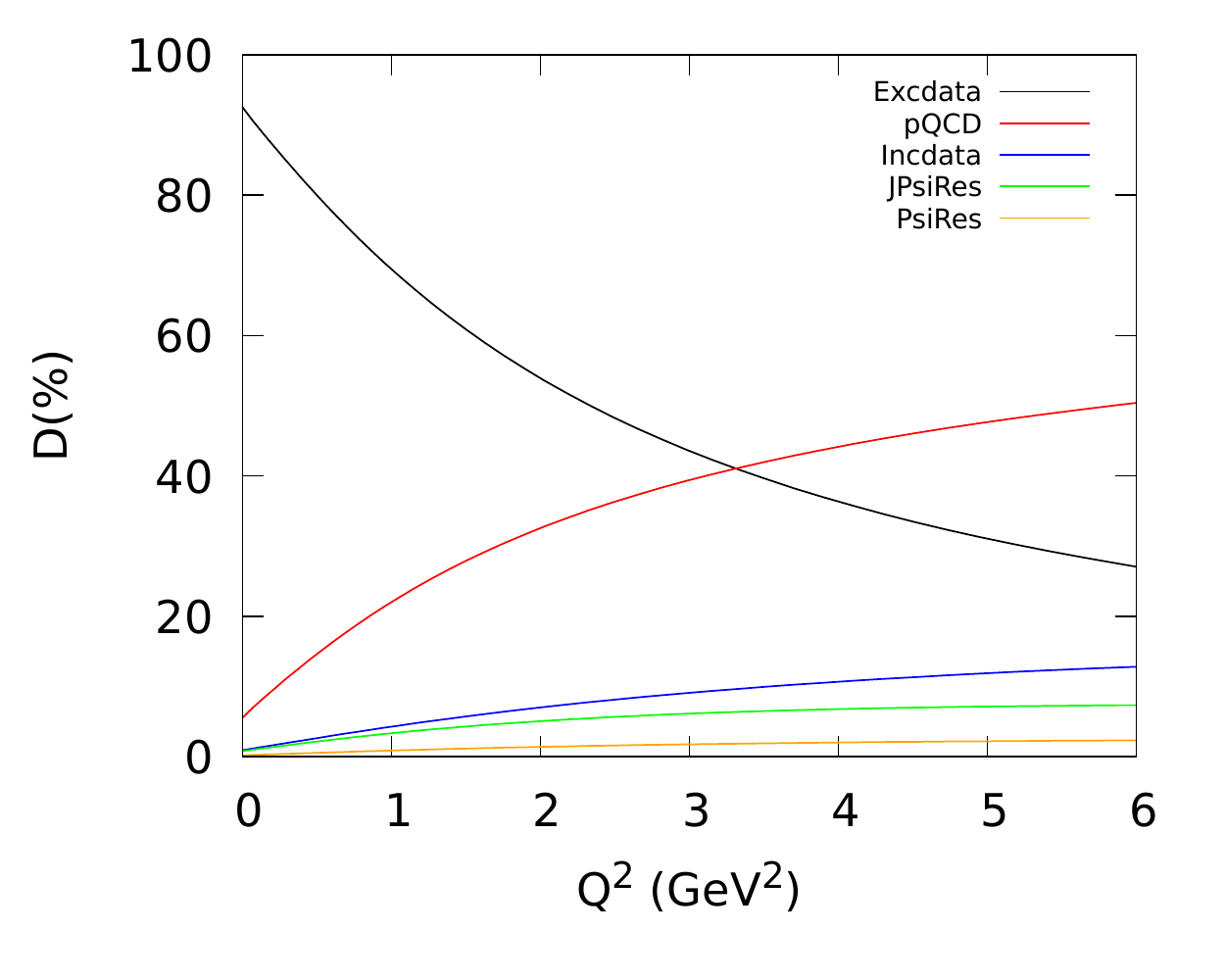}
        \includegraphics[width=0.48\textwidth]{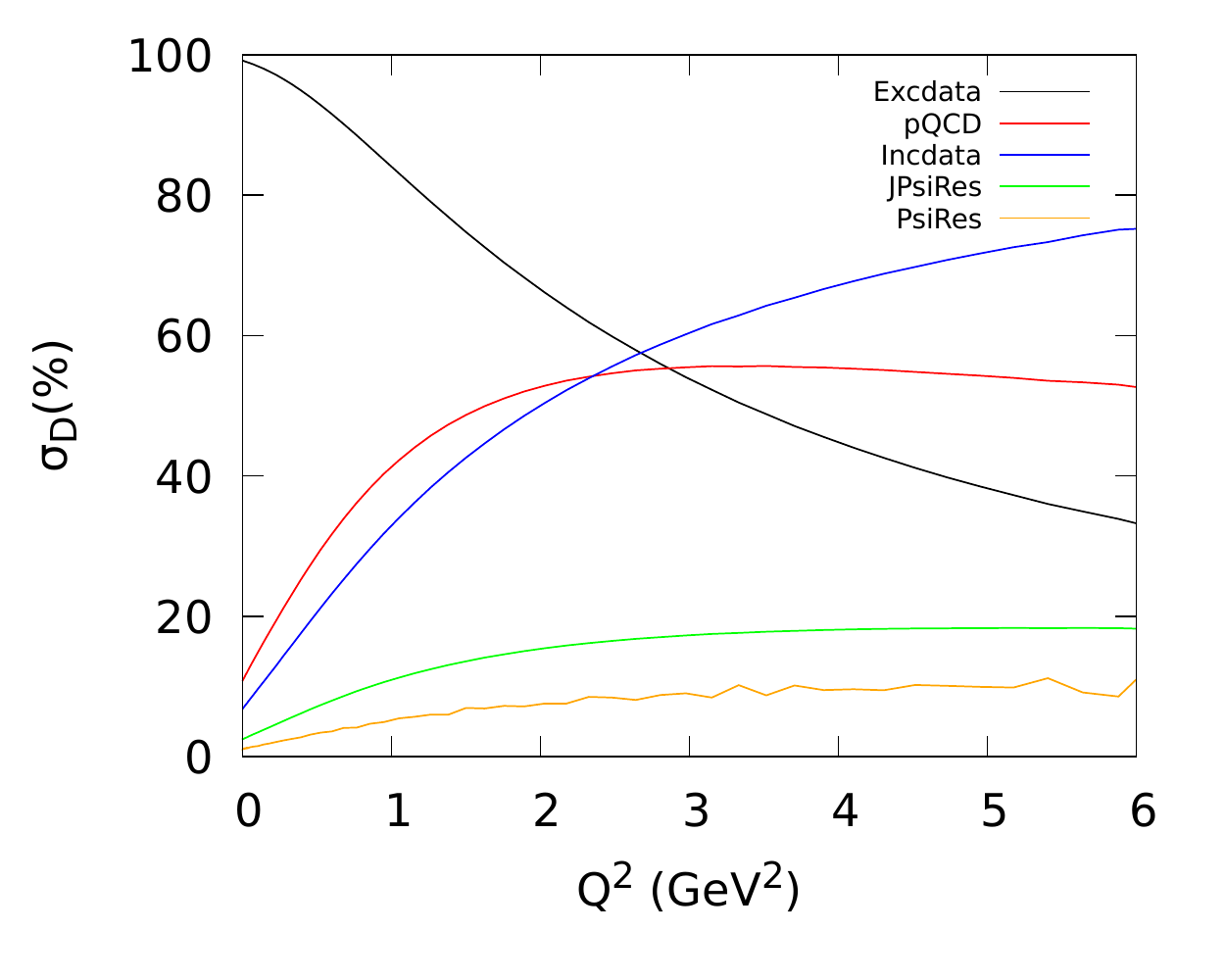}
    \caption{Relative size of the different contributions to $D(Q^2)$~(left) and to their uncertainties~(right) entering the data-driven dispersive evaluation based on the R-ratio.} 
    \label{fig:relativeD}
\end{figure}

\subsection{Lattice Adler function}
The study about the hadronic running of the electromagnetic coupling from lattice QCD presented in Ref.~\cite{Ce:2022eix} contains all the needed details to extract the corresponding Adler function together with estimated uncertainties and correlations. In order to be able to keep track of them, we use the rational approximation for $\bar{\Pi}(Q^2)$ presented in that reference:
\begin{equation}
\label{eq:approx_gg}
  \bar{\Pi}(Q^2) \,\approx\, 
\frac{\sum_{n=1}^3 a_n x^n}{1+\sum_{n=1}^3 b_n x^n}\, =\,
  \frac{0.1094\, (23)\, x + 0.093\, (15)\, x^2 + 0.0039\, (6)\, x^3}{1 + 2.85\, (22)\, x + 1.03\, (19)\, x^2 + 0.0166\, (12)\, x^3} \,, 
\end{equation}
where 
$x \equiv Q^2/\mathrm{GeV}^2$ and the correlation matrix of the expansion coefficients is given by
\begin{equation}
\label{eq:approx_gg_corr}
  \mathrm{corr}\begin{pmatrix}
    a_1 \\ a_2 \\ a_3 \\ b_1 \\ b_2 \\ b_3
  \end{pmatrix} = \begin{pmatrix}
    1     \\
    0.455 & 1     \\
    0.17  & 0.823 & 1     \\
    0.641 & 0.946 & 0.642 & 1     \\
    0.351 & 0.977 & 0.915 & 0.869 & 1     \\
    0.0489 & -0.0934 & 0.0667 & -0.044 & -0.115 & 1     \\ 
  \end{pmatrix} \, .
\end{equation}
It is worth noticing that, when comparing the rational approximation of $\bar{\Pi}(Q^2)$ with the tables of that reference, we observe that, while this approximation gives a very accurate description of the central value up to $Q^2\sim 7\, \mathrm{GeV}^2$, a significant reduction of the uncertainties~(up to $50\%$) starts appearing at $2\, \mathrm{GeV}^2$, due to a more conservative treatment of discretization effects and, to some extent, to the constraint due to the assumed rational approximation ansatz.

We obtain the corresponding lattice Adler function by simply using Eqs.~(\ref{eq:defalphad}) and (\ref{eq:defqedadler}). 
The result is displayed in Fig.~\ref{fig:latticeD}.
\begin{figure}
    \centering
    \includegraphics[height=0.32\textwidth]{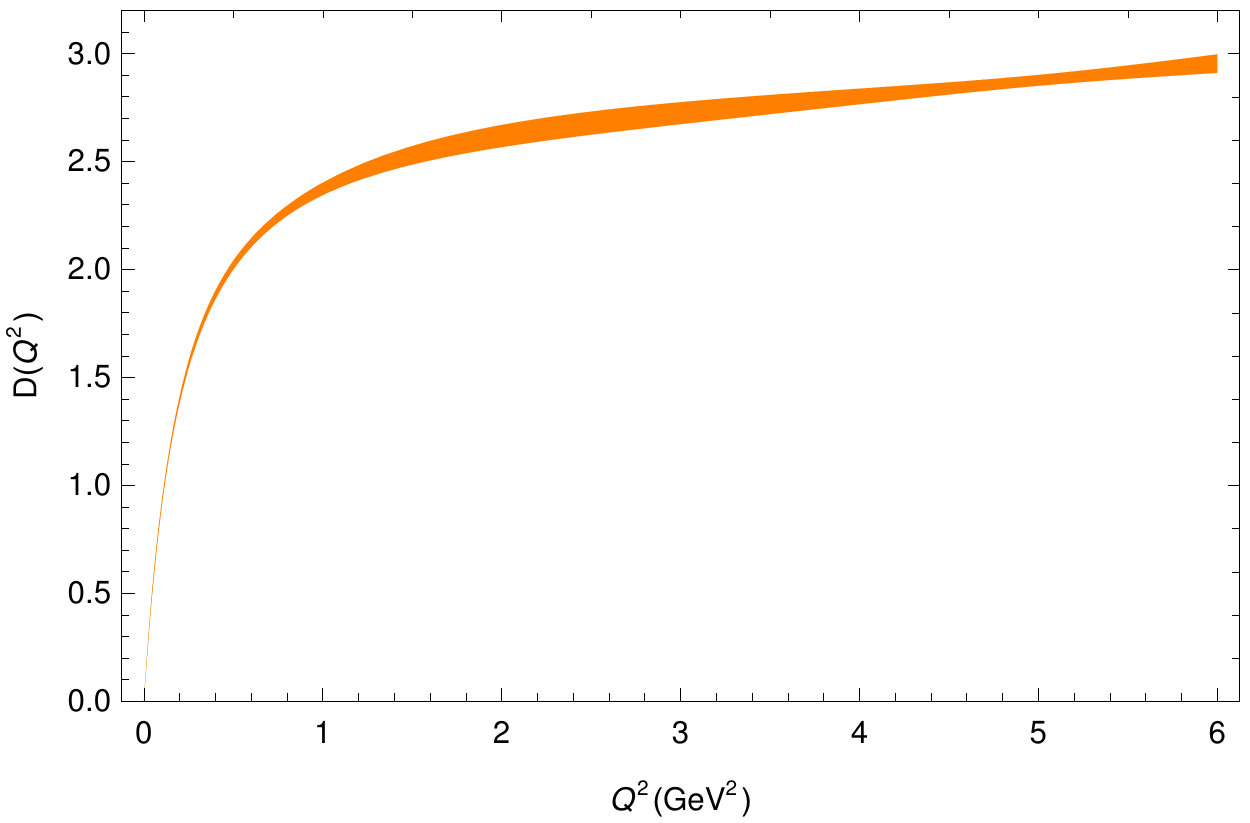}
    \includegraphics[height=0.36\textwidth]{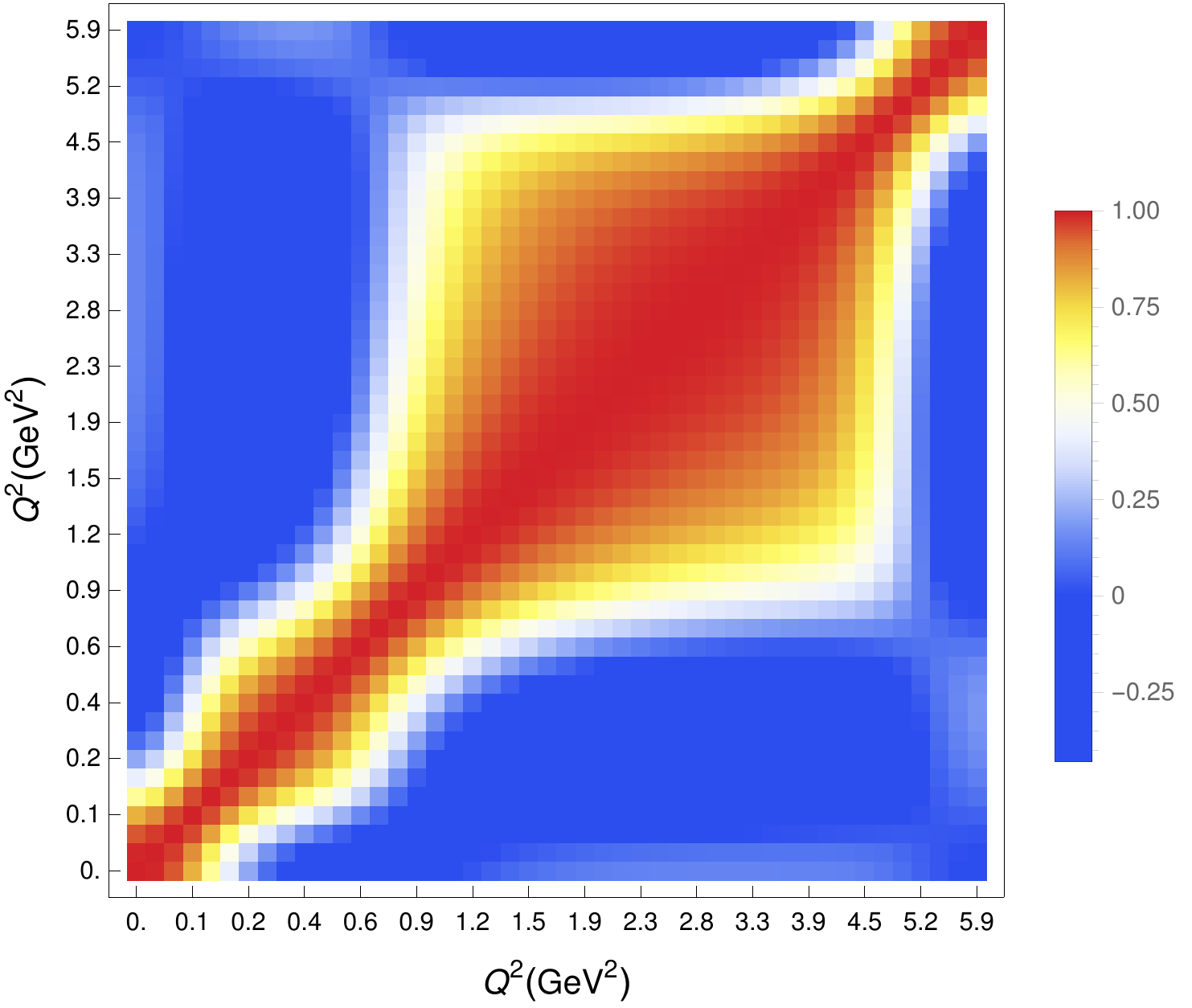}
    \caption{Adler function obtained from the lattice results of Ref.~\cite{Ce:2022eix} (left) together with its correlation matrix (right).}
    \label{fig:latticeD}
\end{figure}

\section{Comparison of the three different approaches to {\boldmath $D(Q^2)$}}\label{sec:comparison}
We can finally perform the comparison of the three descriptions of the Adler function. For the perturbative one we will take the same inputs as above \ie $\alpha_s^{(n_f=5)}(M_Z^2)=0.1184 \pm 0.0008$,
$m_s(\mu_0^2)=(92.03 \pm 0.88)\, \mathrm{MeV}$ at $\mu_0 = 2\, \mathrm{GeV}$,
$m_c(m_c^2)=1.275\, (5)\, \mathrm{GeV}$ and $m_b(m_b^2)=4.171\, (20)\, \mathrm{GeV}$ from the FLAG lattice review \cite{FlavourLatticeAveragingGroupFLAG:2021npn,Ayala:2020odx,Bazavov:2019qoo,Cali:2020hrj,Bruno:2017gxd,PACS-CS:2009zxm,Maltman:2008bx,MILC:2009ltw,Durr:2010vn,Durr:2010aw,McNeile:2010ji,RBC:2014ntl,FermilabLattice:2018est,Lytle:2018evc,Aoki:2021kgd,McNeile:2010ji,Yang:2014sea,Nakayama:2016atf,Petreczky:2019ozv,EuropeanTwistedMass:2014osg,Chakraborty:2014aca,Alexandrou:2014sha,Hatton:2020qhk,Hatton:2021syc,Colquhoun:2014ica,ETM:2016nbo,Gambino:2017vkx}.
The results are presented in Fig.~\ref{fig:fullcomparison}. In Fig.~\ref{fig:sigpert} we quantify the tension among the different descriptions of the Adler function 
in terms of the statistical significance of their differences, {\ie}
\begin{equation}
  S^{ij}(Q^2)\,\equiv\,  \frac{
  D^i (Q^2)-D^j (Q^2)
  }{
  \sigma_{[D^i(Q^2)-D^j(Q^2)]}  }\, ,
  \qquad\qquad (i,j = \mathrm{pQCD},\, e^+e^-\,\mathrm{data},\, \mathrm{latt}).
\end{equation}

We observe the following:
\begin{enumerate}
\item The estimate based on $e^+e^-$ data has smaller uncertainties than the lattice determination. 
The quoted pQCD precision becomes competitive at $2-3\, \mathrm{GeV}^2$.
\item In the whole energy range analysed, the lattice determination
of $D(Q^2)$ has a larger central value than the one inferred from $e^+e^-$ data.
This follows the same trend as in $g-2$ and $\Delta\alpha_{\mathrm{had}}$.
However, within the quoted uncertainties, these two estimates of the Adler function remain compatible at large values of $Q^2$,
the statistical significance of their difference being $\sim 1\sigma$.
\item The lattice determination is in excellent agreement with pQCD in the region where perturbation theory is reliable, \ie at $Q^2 \gtrsim 2-4 \, \mathrm{GeV}^2$.
The differences between the two determinations~(central values) is only $\sim 0.2 \sigma$.
Some tension, larger than $1\sigma$, is observed at lower values of $Q^2$.
This is expected, both because systematic perturbative uncertainties become less reliable and because power corrections are expected to emerge.
\item A significant tension between the determinations of $D(s)$ from $e^+e^-$ data and pQCD emerges below $5$ GeV$^2$. It is larger than $2\sigma$, and it even surpasses $3\sigma$ at $Q^2 \sim 1.5-3 \, \mathrm{GeV}^2$. Let us give three possible explanations for it. First, at low $Q^2$ values, power corrections can become sizeable and may explain the discrepancy. We analyze this possibility in more detail in the next section. The second possibility is that the strong coupling is lower than the value used as an input. From that perspective one can translate this tension into a tension between the value of $\alpha_s$ obtained from a fit to $e^+e^-$ data and the lattice average that we are using as input. We will study this perspective in more detail below. New precise results from novel methods agreeing with the lattice average~\cite{DallaBrida:2022eua} suggest that such a large disagreement is unlikely. The third possible
explanation for the tension at large $Q^2$ values, appears to be possible unaccounted systematic effects in the dispersive evaluation of the Adler function~(see Fig.~\ref{fig:relativeD} for the various contributions to this evaluation and to its uncertainty).

\end{enumerate}
\begin{figure}
    \centering
    \includegraphics[width=0.95\textwidth]{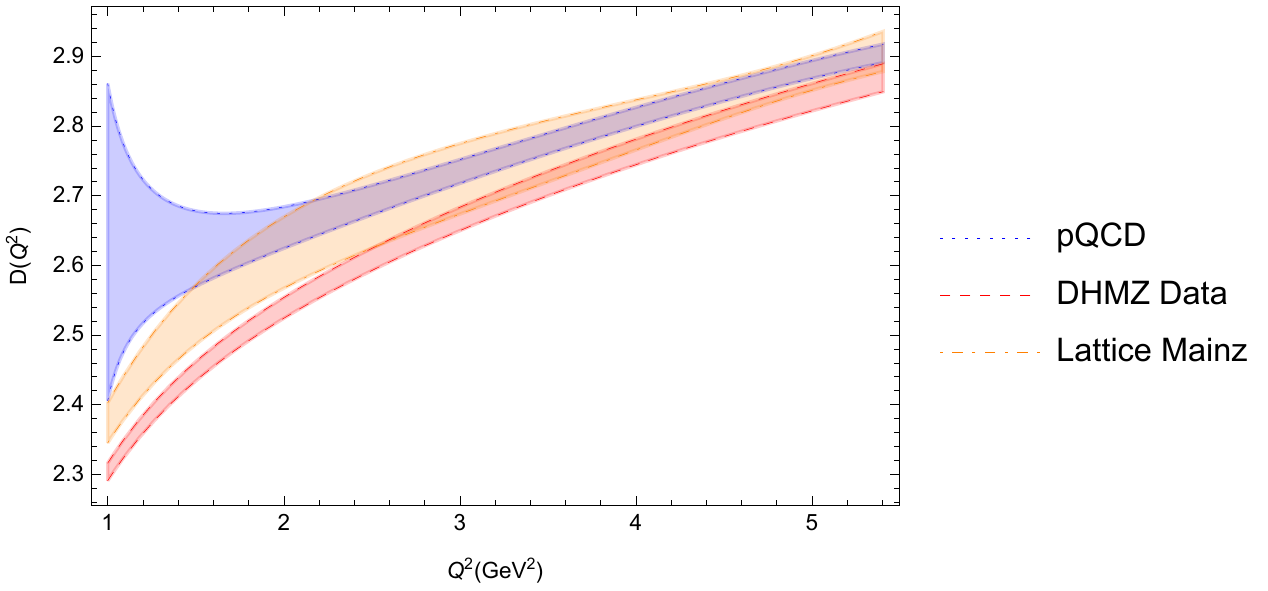}
    \caption{Comparison between the Adler functions obtained from pQCD, the DHMZ compilation of $e^+e^-$ data and the lattice results of Ref.~\cite{Ce:2022eix}.
    \label{fig:fullcomparison}}
\end{figure}

\begin{figure}
    \centering
    \includegraphics[width=0.95\textwidth]{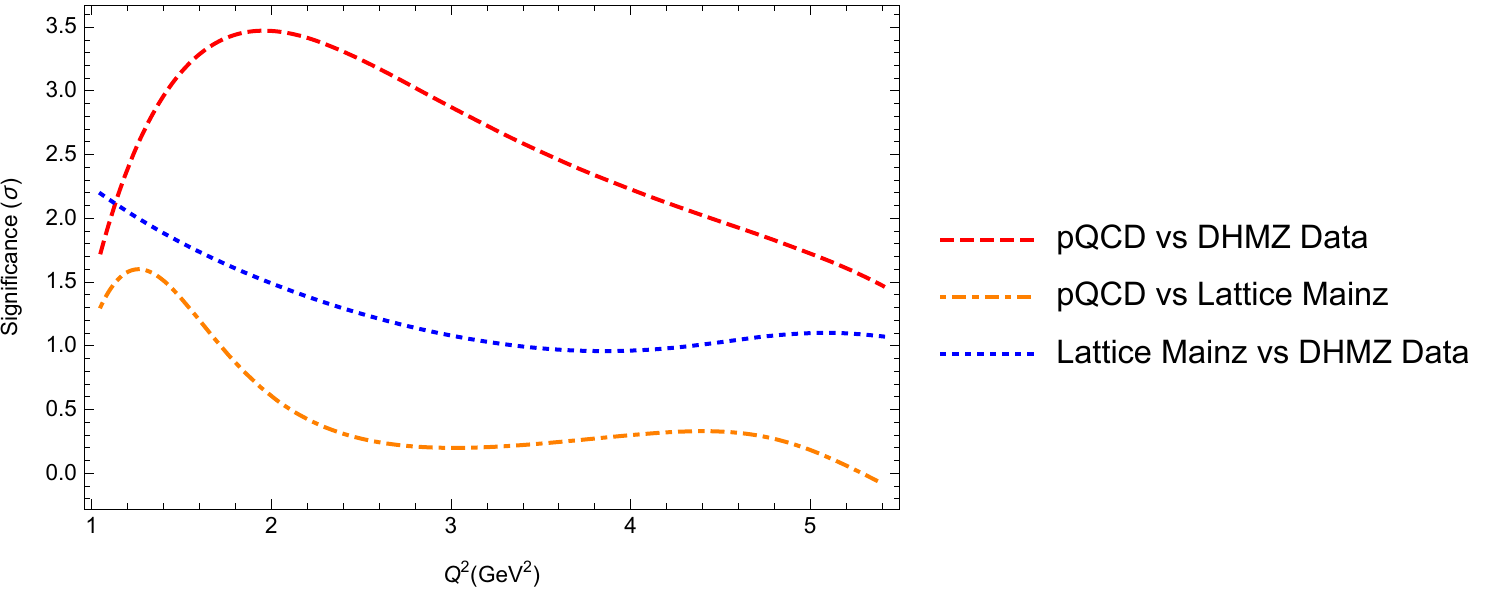}
    \caption{Statistical significance of the differences between the three determinations of the Adler function.
    \label{fig:sigpert}}
\end{figure}

\section{Nonperturbative corrections to the perturbative Adler function}\label{sec:nonpert}

We aim to assess up to which level power corrections can account for the deviations observed in the previous section between the Adler function emerging from $e^+e^-$ data and pQCD. This is not straightforward because the needed vacuum expectation values are not known from first principles and, in general, their numerical values can depend on the way one truncates the (asymptotic) perturbative series. 

In fact, the factorial growth of the perturbative  expansion at large orders
generates infrared ambiguities (when one tries to reconstruct the Adler function from its Borel sum) that scale as inverse powers of $Q^2$ and are expected to be reabsorbed into the nonperturbative terms of the OPE.
One may be tempted to state that those effects are already accounted for in the perturbative systematic uncertainties.
However, the existence of vacuum matrix elements is well established beyond perturbation theory \cite{Shifman:1978bx,Shifman:1978by}. The nonperturbative nature of the QCD vacuum generates non-zero vacuum expectation values for many composite operators such as the quark condensate $\langle 0|\bar q q|0\rangle$, responsible for the breaking of chiral symmetry, or the gluon condensate $\langle 0| a_s G_{\mu\nu} G^{\mu\nu}|0\rangle$, which breaks the scale invariance of massless QCD.

Pure nonperturbative observables, where perturbation theory vanishes, allow for cleaner determinations of the corresponding vacuum condensates because their numerical effects cannot be masked by perturbative uncertainties.
This is the case of the two-point function of a left-handed and a right-handed currents (the $VV-AA$ correlator), which is identically zero to all perturbative orders in $\alpha_s$ but receives non-zero contributions from $D\ge 6$ vacuum condensates that are order parameters of the chiral symmetry breaking. The sizes of the leading power corrections to this correlator are well known, since they can be directly extracted from the $\tau$ decay data \cite{Pich:2021yll,Gonzalez-Alonso:2016ndl}.
Since there is no reason to neglect them, neither in the vector correlator nor in the axial one, it is then a must to incorporate power corrections for a complete description of the OPE-based Adler function.

\subsection{Light-quark correlators}
The leading nonperturbative contribution to the Adler function is, up to negligible  up and down  quark-mass corrections~\cite{Vainshtein:1978wd,Pich:2020gzz},
\begin{equation}
\delta D_{ii}^{L,D=4}=\frac{2\pi^2}{Q^4}\left\{
\left(1+\frac{7}{6}\, a_s\right)\langle a_s GG \rangle 
+ 24\, m_s\,\langle \bar s s\rangle \left[ \delta_{i3} \left( 1 + \frac{a_s}{3} + \frac{47}{8}\, a_s^2\right) + (8\,\zeta_3-5)\,\frac{a_s^2}{12}
\right]\right\}  ,
\end{equation}
which implies for the total electromagnetic correlator:
\begin{equation}\label{eq:D4Adler}
\delta D_{\mathrm{em}}^{L,D=4}=\frac{4\pi^2}{3 Q^4}\left\{
\left(1+\frac{7}{6}\, a_s\right)\langle a_s GG \rangle 
+ 4 \left[ 1 + \frac{a_s}{3} + \left(\frac{27}{8} + 4\,\zeta_3\right)  a_s^2
\right]  m_s\,\langle\bar s s\rangle ]\right\}  .
\end{equation}

The numerical value of the gluon condensate is quite uncertain \cite{Narison:2018nbv,Gubler:2018ctz}, since it is difficult to separate its effect
from the ambiguity generated by the asymptotic tail of the perturbative series, which is supposed to be already included in the perturbative uncertainty. On the other hand, from general grounds we know that the gluon condensate is positively defined \cite{Shifman:1978bx}. This is an important point, because it actually means that its corresponding $D=4$ power correction goes into the wrong direction to explain the tension between pQCD and the experimental data on $R(s)$. To be on the conservative side, let us take the central value estimated in Ref.~\cite{Shifman:1978bx}, but with a $100\%$ of uncertainty, \ie 
\begin{equation}\label{eq:GGcond}
\langle \frac{\alpha_s}{\pi} GG \rangle=(0.012 \pm 0.012)\, \mathrm{GeV}^4 \, .
\end{equation}

The strange quark condensate is better known because it is related to the kaon mass and decay constant by chiral symmetry \cite{Pich:1995bw}. 
Since it is an order parameter of the chiral symmetry breaking (it vanishes to all orders in perturbation theory), the quark condensate does not suffer from the perturbative ambiguity mentioned before.
At lowest-order in chiral perturbation theory it gets determined by the old Gell-Mann--Oakes--Renner relation \cite{Gell-Mann:1968hlm}. However, it receives large higher-order corrections that enhance its final uncertainty \cite{Pich:1999hc,Jamin:2002ev,Gamiz:2002nu}:
\begin{equation}\label{eq:ms-sscond}
     m_s\,\langle\bar s s\rangle = - F_K^2 M_K^2 \left[ 1-\delta_{\mathcal{O}(p^4,m_{u,d})}\right] \approx - (1.3\pm 0.7)\cdot 10^{-3}\;\mathrm{GeV}^4\, .
\end{equation}

The combined dimension-four correction to the electromagnetic Adler correlator in Eq.~(\ref{eq:D4Adler}) takes then the value:
\begin{equation}
\delta D_{\mathrm{em}}^{L,D=4}\approx \frac{(0.10 \pm 0.18)\, \mathrm{GeV}^4}{Q^4} \, .
\end{equation}

Let us also account for the $D=6$ contribution. Up to residual pieces that vanish in the electromagnetic sum, one can write the $D=6$ contribution as
\begin{equation}
\Pi_{ii}^{L,D=6}=\frac{\mathcal{O}_{6,V}}{Q^6} \, .
\end{equation}
It is convenient to rewrite
\begin{equation}
\mathcal{O}_{6,V}=\frac{1}{2}\left(\mathcal{O}_{6,V-A}+\mathcal{O}_{6,V+A} \right) \, .
\end{equation}
The $\mathcal{O}_{6,V-A}$ contribution is a genuine vacuum condensate whose nonzero value, $\mathcal{O}_{6,V-A}\approx
-0.0035\; (9) \, \mathrm{GeV}^6$ \cite{Pich:2021yll}, is well established and understood beyond perturbation theory, and its effect is unrelated to the perturbative series, which is identical for the vector and axial channels. Nonperturbative effects in the observed spectrum (see for example~\cite{Davier:2013sfa,Pich:2022tca}) are known to be suppressed for the $V+A$ combination with respect to the $V-A$, which motivates to assume $|\mathcal{O}_{6,V+A}|<|\mathcal{O}_{6,V-A}|$ \cite{Cirigliano:2021yto}. This inequality
holds (by far) in the large-$N_C$ limit, which gives 
$\mathcal{O}_{6,V+A}^\infty = - \frac{2}{9}\, \mathcal{O}_{6,V-A}^\infty$,  reproducing the old
vacuum saturation approximation, which is also known to work well in predicting $\mathcal{O}_{6,V-A}$ and some rigorous inequalities~\cite{Shifman:1978bx}. Taking this into account, we will adopt
\begin{equation}\label{eq:O6V}
\mathcal{O}_{6,V}=(-0.0015 \pm 0.0015) \, \mathrm{GeV}^6 \, ,
\end{equation}
as
an estimate of this contribution. For the needed Adler function one finds
\begin{equation}
D_{\mathrm{em}}^{L,D=6}\approx 24 \pi^2\frac{\mathcal{O}_{6,V}}{Q^6}=\frac{-(0.36 \pm 0.36)\, \mathrm{GeV}^6}{Q^6} \, .
\end{equation}

Notice that the assigned uncertainties to those corrections potentially contaminated by perturbation theory (gluon condensate and $\mathcal{O}_{6,V+A}$) are above 100\% of their corresponding estimates, guaranteeing that any potential double-counting effect is consistently absorbed by our conservative errors.

\subsection{Charm correlator}
Much less relevant for our analysis is the contribution of power corrections to the charm correlator. The leading power correction is given by the gluon condensate contribution~\cite{Shifman:1978bx}:
\begin{equation}
D_{cc}(Q^2) =-12\pi^2 Q^2 \frac{d}{dQ^2}\left[\frac{\langle\frac{\alpha_s}{\pi}GG \rangle}{4 \cdot 12\, Q^4}\; F\!\left(1+\frac{4m_c^2}{Q^2}\right)\right] \, ,
\end{equation}
where
\begin{equation}
F(a)\equiv \frac{3(a+1)(a-1)^2}{a^2}\frac{1}{2\sqrt{a}}\ln{\left(\frac{\sqrt{a}+1}{\sqrt{a}-1}\right)}-\frac{3a^2-2a+3}{a^2}\, .
\end{equation}
Taking again $\langle a_s GG \rangle=(0.012 \pm 0.012)\, \mathrm{GeV}^2$, one can easily check that the contribution to $D(Q^2)$ remains below $10^{-3}$ and can then be neglected.

\subsection{Discussion}

The nonperturbative correction to the Adler function is then given by
\begin{equation}
\label{eq:deltaNP}
\delta D_{\mathrm{em}}^{\mathrm{NP}}\approx  \frac{4\pi^2}{3 Q^4}\left\{
\left(1+\frac{7}{6}\, a_s\right)\langle a_s GG \rangle 
+ 4 \left[ 1 + \frac{a_s}{3} + \left(\frac{27}{8} + 4\,\zeta_3\right)  a_s^2
\right]  m_s\,\langle\bar s s\rangle ]\right\} + 24 \pi^2\frac{\mathcal{O}_{6,V}}{Q^6} \, .
\end{equation}
\begin{figure}
    \centering
    \includegraphics[width=0.95\textwidth]{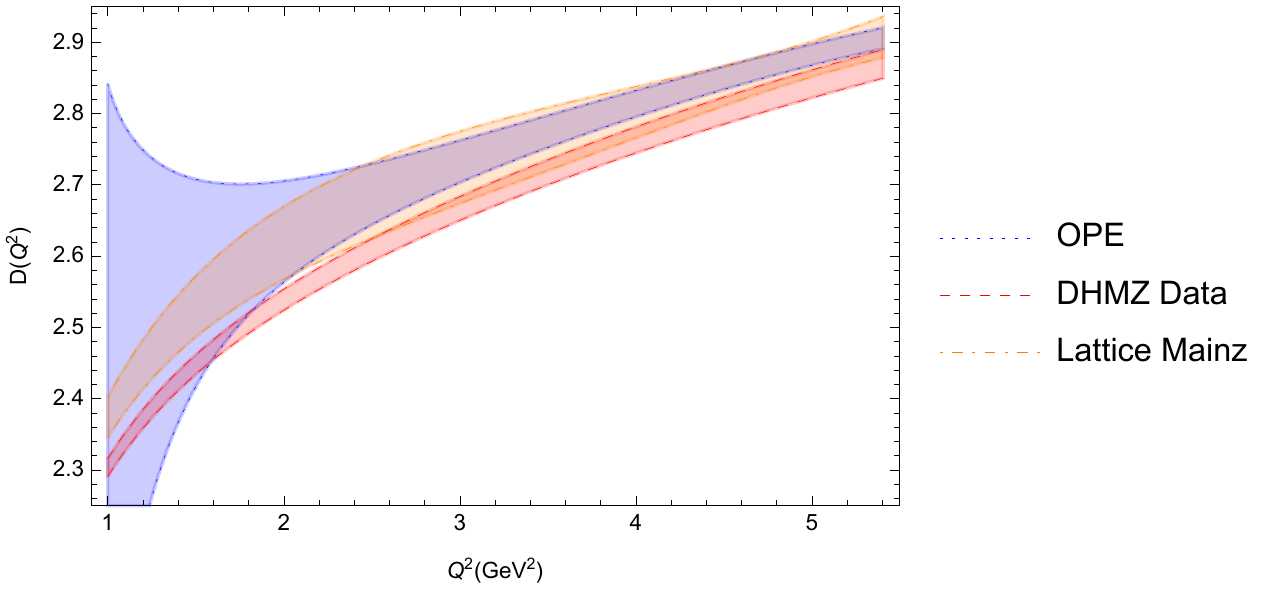}
    \caption{Same plot as Fig.~\ref{fig:fullcomparison} but including the nonperturbative contribution in Eq.~(\ref{eq:deltaNP}).
    \label{fig:fullcomparisonNP}}
\end{figure}
\begin{figure}
    \centering
    \includegraphics[width=0.95\textwidth]{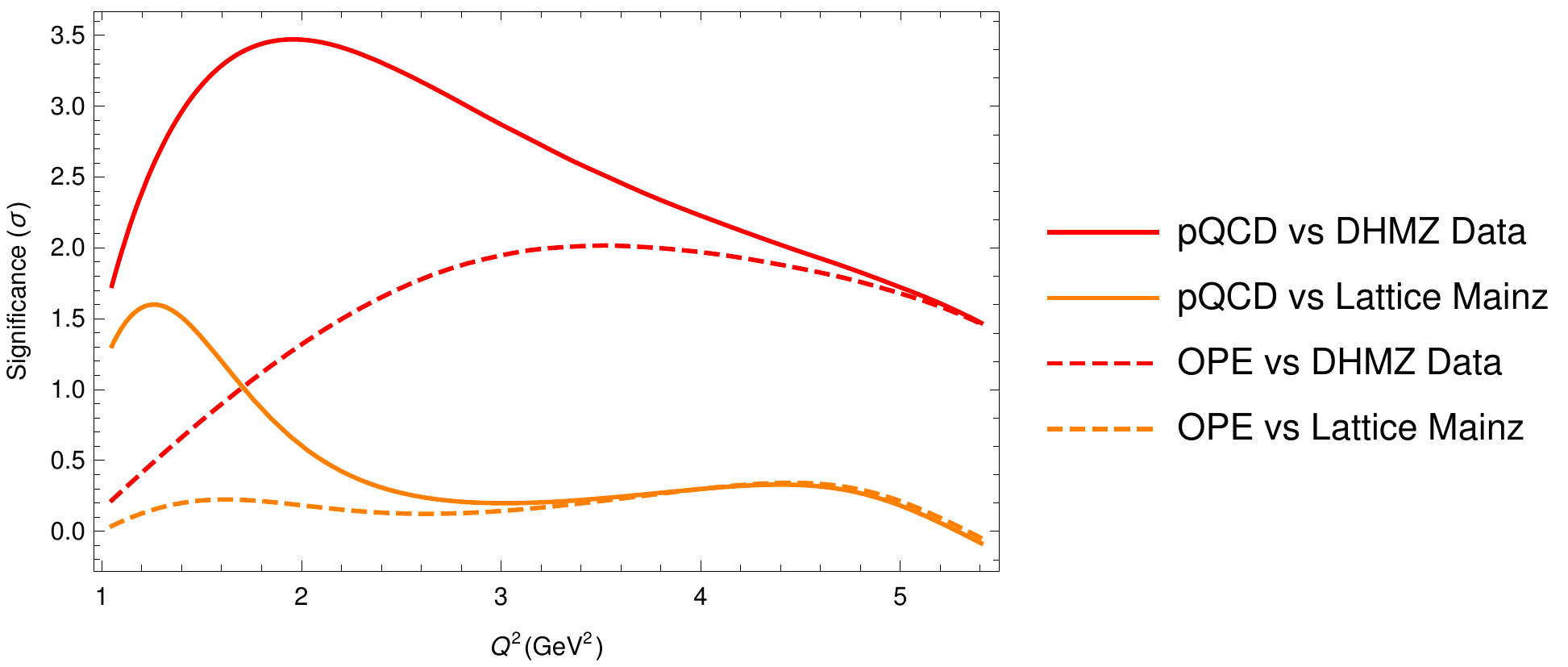}
    \caption{Statistical significance of the difference between the Adler functions extracted from $e^+e^-$ data and lattice results with respect to both the OPE and pQCD predictions.
    \label{fig:sigope}}
\end{figure}
Adopting the conservative numerical estimates in Eqs.~(\ref{eq:GGcond}), (\ref{eq:ms-sscond}) and (\ref{eq:O6V}),
we find that the nonperturbative uncertainty is actually larger than the perturbative one at low $Q^2$ values and, at the current precision level, it cannot be fully neglected even at $Q^2\approx 4\, \mathrm{GeV}^2$. We show in Fig.~\ref{fig:fullcomparisonNP} the comparison of the OPE (pQCD plus condensates) Adler function with the other approaches, and in Fig.~\ref{fig:sigope} the statistical significance of the differences with respect to the lattice and $e^+e^-$-data evaluations of the Adler function. We observe that the origin of the tension at $\sim 1\, \mathrm{GeV}^2$ between the perturbative prediction and both lattice and DHMZ results can indeed be explained by genuine nonperturbative effects. Incidentally, the input central values assumed for the vacuum condensates fit very well the shape of the distribution, although slightly different estimates cannot be discarded within the current experimental and lattice uncertainties, also depending on the size of higher-order power corrections. 
On the other hand, for $Q^2 \gtrsim\,  2\,  \mathrm{GeV^2} $, the observed tension between the analytic Adler function and the data-based one gets slightly reduced. This reduction decreases with $Q^2$, as the effect of the nonperturbative terms diminish, and a tension of up to $\sim 2 \,\sigma$ remains.

\section{\boldmath Determination of $\alpha_s$ from the Adler function}\label{sec:alphas}

Instead of using $\alpha_s$ as input to compare the perturbative $D(Q^2)$ with the other approaches, we can reconvert the comparison into an $\alpha_s$ extraction. The extraction can be done at each value of $Q^2$ by solving the equation:
\begin{equation}
\label{eq:pointbypointeq}
    D^{\mathrm{OPE}}(Q^2,\alpha_s)-D^{\mathrm{data}}(Q^2)=0\, ,
\end{equation}
for $\alpha_s$ while keeping $Q^2$ fixed. $D^{\mathrm{OPE}}(Q^2,\alpha_s)$ is the sum of the perturbative and nonperturbative contributions of the theoretical Adler function, Eqs.~(\ref{eq:AdlerPert}) and (\ref{eq:deltaNP}), keeping $\alpha_s$ as a variable. $D^{\mathrm{data}}(Q^2)$ is the Adler function obtained through either the experimental ratio $R(s)$, $D^{R(s)}(Q^2)$, or the lattice data, $D^{\mathrm{latt}}(Q^2)$, which were discussed in Sec.~\ref{sec:otheradlers}. In this section we illustrate the procedure by using the Adler function based on $R(s)$. The results for the analogous lattice fits are relegated to App.~\ref{app:lattalphasfits}.\footnote{Notice however how lattice data is not limited to the full EM correlator and then better strategies to extract $\alpha_s$ can in principle be pursued.}

As explained in Sec.~\ref{sec:otheradlers}, the $R(s)$ data relies on perturbation theory for the regions $1.8 \, \mathrm{GeV} < \sqrt{s} < 3.7\, \mathrm{GeV}$ and $\sqrt{s} > 5 \, \mathrm{GeV}$. This means that the $R(s)$-based Adler function also contains a residual dependence on $\alpha_s$, which we take into account in the extraction. We decompose $D^{R(s)}$ into the sum of two contributions: one coming from experimentally measured values, $D^{R(s)}_{\mathrm{exp}}(Q^2)$ (see Fig. \ref{fig:relativeD}), and one corresponding with the perturbation theory contribution, $D^{R(s)}_{P}(Q^2,\alpha_s)$. The resulting equation to solve for $\alpha_s$ at each $Q^2$ becomes:   
\begin{equation}
    D^{\mathrm{OPE}}(Q^2,\alpha_s)-(D^{R(s)}_{\mathrm{exp}}(Q^2)+D^{R(s)}_{P}(Q^2,\alpha_s))=0 \,.
\end{equation}

The values for $\alpha_s^{(n_f=5)}(M_Z^2)$ obtained from both lattice and $R(s)$ data, together with their uncertainties, including both all experimental and theoretical sources, are plotted in Fig.~\ref{fig:aspointbypoint}, as a function of the $Q^2$ value at which they are derived. 
As expected, the same observations made when comparing the different Adler functions are applicable here as well. That is,
\begin{enumerate}
    \item The central values of the strong coupling extracted from $e^+e^-$ data are smaller than the ones extracted using the lattice determination of $D(Q^2)$.
    \item The values of $\alpha_s$ extracted from lattice data are in agreement with the FLAG lattice average.
    \item The $\alpha_s$ extracted from $e^+e^-$ data is between  $1.5$ and $2\, \sigma$ below the FLAG lattice average in the $3\, \mathrm{GeV}^2<Q^2<5 \, \mathrm{GeV}^2$ region. Notice however how the different $Q^2$ points are strongly correlated among each other.
\end{enumerate}

In the following, we discuss how these different values for $\alpha_s$ can be compared quantitatively and combined.

\begin{figure}
    \centering
    \includegraphics[width=0.95\textwidth]{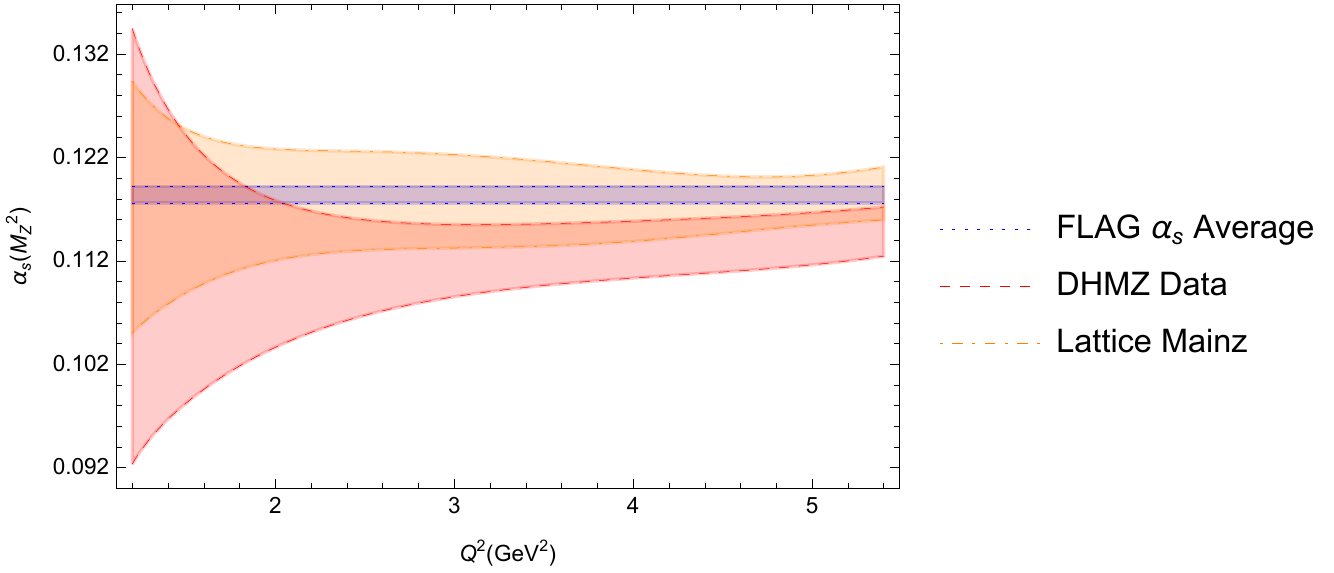}
    \caption{Comparison between the lattice FLAG average $\alpha_s^{(n_f= 5)}(M_Z^2)=0.1184\pm0.0008$ and the values of $\alpha_s$ obtained at each $Q^2$ for the $R(s)$ and lattice Adler functions.
    \label{fig:aspointbypoint}}
\end{figure}

\subsection{Averages}\label{sec:averages}
A potential improvement in terms of precision for the extracted $\alpha_s^{(n_f= 5)}(M_Z^2)$ could come from combining the values obtained at different $Q^2$ values, taking into account their correlations. 
One possibility would be to perform the combination through the minimisation of a $\chi^2$ function,
\begin{equation}
    \chi^2=\left( \bar{\alpha}_s^{\mathrm{extr}}-\bar{\alpha}_s^{\mathrm{av}} \right) \cdot C^{-1} \cdot \left( \bar{\alpha}_s^{\mathrm{extr}}-\bar{\alpha}_s^{\mathrm{av}} \right)^T \,,
    \label{Eq:Chi2Cov}
\end{equation}
with respect to $\alpha_s^{\mathrm{av}}$. 
In this function, $\bar{\alpha}_s^{\mathrm{extr}}$ is the vector containing the extracted values of $\alpha_s$ at each $Q^2$, $C$ is their covariance matrix~\footnote{In principle, for e.g. relative uncertainties, this covariance matrix can itself depend on $\alpha_s^{\mathrm{av}}$, which could be addressed through an iterative fitting approach~(see e.g. Refs.\cite{Lyons:1989gh,DAgostini:1993arp} and references therein) or using fitted nuisance parameters applied as (constrained)~scaling factors of the theoretical prediction~\cite{Blobel:2003wa,Pascaud:1995qs}. However, here the input $\alpha_s^{\mathrm{extr}}$ values are similar~(see e.g. Fig.~\ref{fig:aspointbypoint}) and the impact of this effect is small.} and $\bar{\alpha}_s^{\mathrm{av}}$ is a vector with the same dimension as $\bar{\alpha}_s^{\mathrm{extr}}$ containing the parameter $\alpha_s^{\mathrm{av}}$. However, one is faced with the following limitations:
\begin{enumerate}
\item As discussed in the previous section, the nonperturbative contributions to the Adler function are a large source of systematic uncertainty at low $Q^2$, which additionally is not fully controlled. This can clearly be seen in Fig.~\ref{fig:aspointbypoint}. Therefore, one should avoid using the extracted values in this energy region. 
\item The values extracted from $e^+e^-$ data display strong experimental correlations (see Fig.~\ref{fig:AdlerR}). These correlations are stronger between neighboring points and a larger precision in the estimation of the corresponding covariance matrix would be needed in order to find meaningful results for the combination, which may otherwise not be realistic. Indeed, if one includes too many consecutive points, eventually one is going to find a covariance matrix with null eigenvalues, which would imply absolute predictions~(\ie no uncertainties in certain linear combinations). Actually, this is a consequence of the approximations involved in finding the original covariance matrix.\footnote{An illustrative example of this (extreme) case, this time for fit inputs from lattice QCD, would consist in simply starting from Eq.~(\ref{eq:approx_gg}) and taking $7$ different points. While the rational approximation and their associated uncertainties are expected to work generally well both in $\bar\Pi(Q^2)$ and linear combinations of it, it would clearly fail in predicting the uncertainties of the linear combination corresponding to the zero eigenvalue~(corresponding to a null estimated uncertainty), which would directly dominate~(because of an apparently infinite precision) the determination of any theoretical parameter depending on it.}
Additionally this makes the experimental correlation matrix singular, which therefore cannot be inverted. As a result, a $\chi^2$ function cannot be constructed for the whole set of extracted values.

\item A similar problem arises for the theoretical correlation matrix. In general one expects that the perturbative uncertainties are dominated by the first unknown coefficient, in this case $K_5$, and the nonperturbative ones by the first unknown power correction, say $\mathcal{O}_6$. We have supplemented the perturbative uncertainty by renormalization-scale variations, which in general one expects to account for uncertainties due to higher-order effects. However this is going to fail if one artificially looks for linear combinations of data points such that either $K_5$, $\mathcal{O}_6$ and/or the scale variations cancel or they appear suppressed, for example by numerical prefactors, with respect to the contributions of higher-order coefficients, which then by construction are not going to be negligible with respect to the accounted effects. In a naive $\chi^2$ fit, the extracted value of $\alpha_s$ is going to be dominated by the most precise linear combination of data points, taking into account those theoretical uncertainties, which are precisely the directions where the estimators are prone to underestimate them, leading to very aggressive predictions. An explicit example of this kind of direction is the logarithmic derivative of the Adler function, implicit in fits to consecutive data points.\footnote{In fact we find that the eigenvectors associated to the lowest eigenvalues of the experimental covariance matrix correspond, in first approximation, to the highest-order derivatives in the discrete approximation. They can be related to different, more localized, weights when integrating $R(s)$.} Indeed by taking consecutive (adimensionalized) derivatives one is going to trigger the sooner breakdown of the OPE. Schematically, for the nonperturbative contributions to $D$, $D_{\mathrm{NP}}$, one has\footnote{An analogous issue occurs for the $\left(\frac{Q^2}{4m_{c}^2}\right)^n$ series, possibly inducing the breakdown of the associated expansion from lower energies for high-order derivatives.}
\begin{equation}
D_{\mathrm{NP}}\,\sim\, \sum_{D}c_D\;\frac{\Lambda^D}{Q^D}\qquad\longrightarrow\qquad \frac{d^n D_{\mathrm{NP}}}{d\ln Q^n}\, =\,\sum_D c_D\; (-D)^n\;\frac{\Lambda^D}{Q^D} \, ,
\end{equation}
and then higher-dimensional corrections become more and more important with respect to lower-dimensional ones at a fixed energy, eventually leading to potentially underestimated theoretical uncertainties in those directions. Analogously, the scale variation is going to be in general a good estimator of perturbative uncertainties, because its variation is in general of the same order as the neglected perturbative contributions. However, one would clearly fall into one version of the well-known look-elsewhere effect if one artificially looks for directions in which it happens to cancel: the fact that the scale-dependence accidentally cancels in some linear combination does not guarantee that the contributions from higher-order coefficients are also cancelling.
There is also the possibility that new topologies emerge at higher orders, inducing an uncertainty not well accounted by the scale variations.

\end{enumerate}

In summary, the $\chi^2$ function defined in Eq.~(\ref{Eq:Chi2Cov}) and the result of its minimisation are sensitive to uncertainties on the uncertainties and on the correlations~(\ie to uncertainties on the covariance matrix), present for both the experimental and the theoretical components.
Starting from remarks made in the context of ATLAS jet performance and cross-section studies~\cite{ATLAS:2014hvo,ATLAS:2017kux,ATLAS:2017ble}, the relevance of the uncertainties on the covariance matrices~(in particular for what concerns the implications for combination methods) has been pointed out in the context of the theoretical predictions for the anomalous magnetic of the muon~\cite{Davier:2019can,bogdan-Mainz-2018-DHMZ-UncOnUnc,Aoyama:2020ynm}. 
More recently, similar remarks about the uncertainties on uncertainties were made for what concerns the procedure of quantifying the significance of the data-theory tensions, in this same context~\cite{Cowan:2021sdy}.

Taking all these aspects into account, we will restrict to data sets with at most three points and $Q^2 \gtrsim 3 \, \mathrm{GeV}^2$. We will consider 
the two sets of three $Q^2$ points shown in Table~\ref{tab:sets}. 
The first set, set $1$, has a better behaviour in the expansion in powers of $Q^2/m_c^2$, whereas the second set, set 2, is less affected by potential nonperturbative effects. Additionally we will also consider two variations with only two $Q^2$ values, set $1^*$ and set $2^*$,
where the midpoint of the corresponding set has been removed.

\begin{table}[]
    \centering
    \begin{tabular}{|c|c|c|}
    \hline
    &   $Q^2$  & $\alpha_s(M_Z^2)$ \\ \hline \hline
    \multirow{3}{*}{Set 1} &
        $3.15$ & $0.1122\; (17)_{\mathrm{exp}}\, (11)_{\mathrm{pert}}\, (26)_{\mathrm{th}}$ \\ \cline{2-3}
    &    $4.10$ & $0.1132\; (19)_{\mathrm{exp}}\, (12)_{\mathrm{pert}}\, (16)_{\mathrm{th}}$ \\ \cline{2-3}
    &    $5.18$ & $0.1144\; (19)_{\mathrm{exp}}\, (11)_{\mathrm{pert}}\, (10)_{\mathrm{th}}$ \\ \hline \hline
    \multirow{3}{*}{Set 2} &
        $4.10$ & $0.1132\; (19)_{\mathrm{exp}}\, (12)_{\mathrm{pert}}\, (16)_{\mathrm{th}}$ \\ \cline{2-3}
    &    $4.63$ & $0.1138\; (20)_{\mathrm{exp}}\, (12)_{\mathrm{pert}}\, (12)_{\mathrm{th}}$ \\ \cline{2-3}
    &    $5.41$ & $0.1148\; (19)_{\mathrm{exp}}\, (11)_{\mathrm{pert}}\, (10)_{\mathrm{th}}$ \\ \hline
    \end{tabular}
    \caption{The two sets of three $Q^2$ points chosen for the averages with their experimental (exp), perturbative (pert) and theoretical (th) symmetrized uncertainties. By perturbative uncertainty we mean the uncertainty coming from the use of perturbation theory in the regions $1.8\, \mathrm{GeV}<\sqrt{s}<3.7\, \mathrm{GeV}$ and $\sqrt{s}>5\, \mathrm{GeV}$ for the ratio $R(s)$. }
    \label{tab:sets}
\end{table}

The only remaining input necessary to compute the $\chi^2$ is the covariance matrix $C$. 
In order to evaluate it, we can use linear error propagation, for both the experimental and theoretical components of this matrix.
In particular, for the theoretical component this is justified, since the Adler function is an approximately linear function of $\alpha_s$ in the energy region we are considering. 
However, there are different possible choices aimed to avoid the issues mentioned above when assigning correlations between the theoretical uncertainties at different points. 
We will explore the different possibilities in Sec.~\ref{sec:thunc}. 

When performing such combination, we must also take into account the possibility of the $\chi^2$ minimisation yielding a biased result, caused by the limited precision with which the covariance matrix is known. 
According to the Gauss-Markov Theorem, the minimisation of the $\chi^2$ in Eq.~(\ref{Eq:Chi2Cov}) is equivalent to performing a weighted average, while optimizing the weights~(constraint to have the sum equal unity) such that the uncertainty of the average is minimum~\cite{Cowan:1998ji,ParticleDataGroup:2010dbb}.
This yields
\begin{equation}
    \alpha_s^{\mathrm{av}}=\frac{\bar{1}\cdot C^{-1}\cdot (\bar{\alpha_s}^{\mathrm{extr}})^T}{\bar{1}\cdot C^{-1} \cdot \bar{1}^T} \, ,
\end{equation}
with $\bar{1}$ being a vector of the same dimension as $\bar{\alpha_s}^{\mathrm{extr}}$, with all its entries equal to 1. 
As a consequence, if the covariance matrix is poorly known, these weights can be biased~(in some cases they can e.g. take negative values or values larger than unity, which could make the average value to be outside the range of the extracted values). 
Therefore, we have to consider alternative averaging procedures that, although will yield results with somewhat larger uncertainties~(see Sec.~\ref{sec:results}), are free of any bias caused by implicit assumptions in the derivation of their weights.
We have considered the following:
\begin{itemize}
    \item[1.] A simple average, where all the inputs have the same weight. This is, the inverse of the number of input values considered.
    \item[2.] A weighted average where the weights are proportional to the inverse of the experimental uncertainty squared.\footnote{In Ref.~\cite{Malaescu:2012ts} one can find further discussions on unbiased combination procedures. These feature realistic uncertainty estimates, based on the propagation of the full information on the uncertainties of the inputs, with their correlations.}
\end{itemize}
The averaged values obtained using these different averaging procedures for the four sets considered can be found in Sec.~\ref{sec:results}.

\subsection{Theoretical uncertainties}\label{sec:thunc}
As discussed throughout the text, there are different sources of theoretical uncertainty to the Adler function and for some of them assigning a $100 \%$ 
correlation between different $Q^2$ values may not be the best choice. 
This is for example the case of the scale variations (it is probably not realistic to assume that there is a single renormalization-scale choice such that the perturbative uncertainties of all truncated series can be removed) or the nonperturbative one parameterized by $\mathcal{O}_6$ (knowledge of $\mathcal{O}_6$ would not completely remove the nonperturbative uncertainty, since we would need to account for the higher-dimensional contributions). We will then repeat the same fits under four assumptions on the correlations between the theoretical uncertainties at different $Q^2$:
\begin{enumerate}
\item All correlations for the same ``theoretical source'', renormalization scales or $\mathcal{O}_6$, are $100 \%$.
\item All correlations for the same ``theoretical source'' are $100 \%$ except for the scale variations, that are assumed to be uncorrelated among points with large separations in $Q^2$.
\item All correlations for the same ``theoretical source'' are $100 \%$ except for the $\mathcal{O}_6$ one, which is assumed to be uncorrelated among points with large separations in $Q^2$.
\item All correlations for the same ``theoretical source'' are $100 \%$ except for the scale uncertainties and the $\mathcal{O}_6$ ones, which are assumed to be uncorrelated among points with large separations in $Q^2$.
\end{enumerate}

\subsection{Results}\label{sec:results}
The numerical results obtained with the four different assumptions on the theoretical correlations are displayed in Tables~\ref{tab:all-corr},~\ref{tab:scale-uncor},~\ref{tab:O6-uncor} and~\ref{tab:scales-O6-uncor}, respectively. Each table shows the average values of $\alpha_s^{(n_f=5)}(M_Z^2)$ extracted from the four different choices of $Q^2$ points (sets 1, 2, 1* and 2*) and with the three different averaging procedures: $\chi^2$ minimisation, simple average and weighted average. 
Clearly, the results in Table~\ref{tab:all-corr} have too large $\chi^2$ values associated to them, both for sets 1 and 2. As explained above, this is not necessarily a signature of inconsistent experimental data sets, but can also be due to the limitations explained in Sec.~\ref{sec:averages}. Much more reasonable values are obtained by taking points with broader separation in $Q^2$, {\ie}the set $1^*$, or with the alternative choices for the theoretical correlations proposed in Sec.~\ref{sec:thunc}.

All in all we find that there is not much gain in combining several $Q^2$ values. We observe that the minimisation procedure yields a bias towards larger values of $\alpha_s$, which are preferred by the (potentially dangerous) directions associated to small eigenvalues of the original covariance matrix. Once these are treated more conservatively ({\ie}Tables \ref{tab:scale-uncor},~\ref{tab:O6-uncor} and~\ref{tab:scales-O6-uncor}), one suppresses the associated bias and obtains values much more compatible with the ones from the other averaging procedures. As expected, the associated $\alpha_s$ values cluster around
\begin{equation}
\alpha_s^{(n_f=5)}(M_Z^2) = 0.1136 \pm 0.0025 \, ,
\end{equation}
which is approximately $2\, \sigma$ below the FLAG lattice average. 
The corresponding fit to lattice data (see App.~\ref{app:lattalphasfits}) returns larger values clustering around\footnote{Note that the uncertainties in these two combinations (with a different hierarchy compared to Fig.~\ref{fig:aspointbypoint}) strongly depend on the estimated correlations between the combined points, see Figs.~\ref{fig:AdlerR} and \ref{fig:latticeD}, which in the lattice case is based on Eqs.~(\ref{eq:approx_gg}) and (\ref{eq:approx_gg_corr}).}  \begin{equation}
\alpha_s^{(n_f=5)}(M_Z^2) = 0.1179 \pm 0.0025\, ,
\end{equation}
exhibiting again the discrepancy between the dispersive and the lattice-based results. Nevertheless, this
also shows that, once the situation with respect to the different tensions related to $R(s)$ is clarified, a determination of $\alpha_s^{(n_f=5)}(M_Z^2)$ with a precision of $\mathcal{O} (1\%)$ could be achievable from the Euclidean Adler function.

\begin{table}[]
    \centering
    \begin{tabular}{|c||c|c||c|c||c|c|}
    \hline
       & $\chi^2$ & Minimisation & $\chi^2$ & Simple Average & $\chi^2$ & Weighted Average \\ \hline \hline
       Set 1  & $11.2
       $ & $0.1190(16)(7)(12)$  & $18.6
       $ & $0.1133(18)(12)(17)$ & $18.8
       $  &$0.1132(18)(12)(18)$\\ \hline
       Set 2  &  $8.8
       $ & $0.1198(8)(3)(13)$ & $22.4
       $ &$0.1139(19)(11)(12)$ & $22.4
       $ & $0.1140(19)(11)(12)$\\ \hline
       Set 1* & $1.5
       $ & $0.1149(19)(13)(8)$ & $1.8
       $ & $0.1133(18)(12)(18)$ & $1.9
       $ & $0.1132(19)(12)(19)$\\ \hline
       Set 2* & $4.3
       $ & $0.1163(19)(10)(7)$ & $5.3
       $ & $0.1139(19)(11)(12)$ & $5.3
       $ & $0.1140(19)(11)(12)$\\ \hline 
    \end{tabular}
    \caption{Averaged results for $\alpha_s^{(n_f=5)}(M_Z^2)$ obtained when the theory uncertainties originating from the same source are assumed to be fully correlated.}
    \label{tab:all-corr}
\end{table}

\begin{table}[]
    \centering
    \begin{tabular}{|c||c|c||c|c||c|c|}
    \hline
       & $\chi^2$ & Minimisation & $\chi^2$ & Simple Average & $\chi^2$ & Weighted Average \\ \hline \hline
       Set 1  & $1.9
       $ & $0.1148(20)(11)(9)$  & $2.2
       $ & $0.1133(18)(12)(17)$ & $2.3
       $  &$0.1132(18)(12)(18)$\\ \hline
       Set 2  &  $2.8
       $ & $0.1155(19)(11)(10)$ & $3.2
       $ &$0.1139(19)(11)(12)$ & $3.2
       $ & $0.1140(19)(11)(12)$\\ \hline
       Set 1* & $1.3
       $ & $0.1147(20)(11)(9)$ & $1.7
       $ & $0.1133(18)(12)(19)$ & $1.7
       $ & $0.1132(19)(12)(19)$\\ \hline
       Set 2* & $2.7
       $ & $0.1155(19)(11)(9)$ & $3.1
       $ & $0.1139(19)(11)(12)$ & $3.1
       $ & $0.1140(19)(11)(12)$\\ \hline 
    \end{tabular}
    \caption{Averaged results for $\alpha_s^{(n_f=5)}(M_Z^2)$ obtained  when the theory uncertainties from scale variations are assumed to be uncorrelated.}
    \label{tab:scale-uncor}
\end{table}

\begin{table}[]
    \centering
    \begin{tabular}{|c||c|c||c|c||c|c|}
    \hline
       & $\chi^2$ & Minimisation & $\chi^2$ & Simple Average & $\chi^2$ & Weighted Average \\ \hline \hline
       Set 1  & $2.0
       $ & $0.1148(20)(11)(9)$  & $2.4
       $ & $0.1133(18)(12)(16)$ & $2.5
       $  &$0.1132(18)(12)(17)$\\ \hline
       Set 2  &  $3.9
       $ & $0.1160(18)(10)(9)$ & $4.7
       $ &$0.1139(19)(11)(12)$ & $4.7
       $ & $0.1140(19)(11)(12)$\\ \hline
       Set 1* & $1.2
       $ & $0.1146(20)(11)(10)$ & $1.4
       $ & $0.1133(18)(12)(17)$ & $1.5
       $ & $0.1132(19)(12)(18)$\\ \hline
       Set 2* & $2.9
       $ & $0.1156(19)(11)(9)$ & $3.3
       $ & $0.1139(19)(11)(12)$ & $3.3
       $ & $0.1140(19)(11)(12)$\\ \hline 
    \end{tabular}
    \caption{Averaged results for $\alpha_s^{(n_f=5)}(M_Z^2)$ obtained when the theory uncertainty from $\mathcal{O}_6$ is assumed to be uncorrelated.}
    \label{tab:O6-uncor}
\end{table}

\begin{table}[]
    \centering
    \begin{tabular}{|c||c|c||c|c||c|c|}
    \hline
       & $\chi^2$ & Minimisation & $\chi^2$ & Simple Average & $\chi^2$ & Weighted Average \\ \hline \hline
       Set 1  & $1.6
       $ & $0.1145(19)(11)(10)$  & $1.8
       $ & $0.1133(18)(12)(16)$ & $1.9
       $  &$0.1132(18)(12)(17)$\\ \hline
       Set 2  &  $2.3
       $ & $0.1152(19)(11)(10)$ & $2.6
       $ &$0.1139(19)(11)(11)$ & $2.6
       $ & $0.1140(19)(11)(11)$\\ \hline
       Set 1* & $1.1
       $ & $0.1145(19)(11)(10)$ & $1.3
       $ & $0.1133(18)(12)(17)$ & $1.4
       $ & $0.1132(19)(12)(18)$\\ \hline
       Set 2* & $2.1
       $ & $0.1151(19)(11)(10)$ & $2.3
       $ & $0.1139(19)(11)(12)$ & $2.3
       $ & $0.1140(19)(11)(12)$\\ \hline 
    \end{tabular}
    \caption{Averaged results for $\alpha_s^{(n_f=5)}(M_Z^2)$ obtained when the theory uncertainties from scale variations and $\mathcal{O}_6$ are assumed to be uncorrelated.}
    \label{tab:scales-O6-uncor}
\end{table}

In the combinations of pairs of $\alpha_s^{(n_f=5)}(M_Z^2)$ values evaluated at different $Q^2$ points, the derived $\chi^2$ values also provide an implicit test of the RGE~(used for evolving $\alpha_s^{(n_f=5)}$ from each $Q^2$ point to $M_Z^2$), within the assumptions for the treatment of the uncertainties and of their correlations discussed above.
We implement an alternative approach of testing the RGE.
It consists in computing the differences between the extracted $\alpha_s^{(n_f=5)}(M_Z^2)$ values at some reference $Q^2$ point~(in this case $5.41\, \mathrm{GeV}^2$, yielding the most precise $\alpha_s^{(n_f=5)}(M_Z^2)$ value among the $Q^2$ points considered here) and each of the other considered $Q^2$ points respectively.
We also compute the corresponding relative differences normalised with respect to the former~(reference) $\alpha_s^{(n_f=5)}(M_Z^2)$ value~(see Fig.~\ref{fig:RGtest}).
The uncertainties on these differences are evaluated through a linear error propagation, taking into account the full information on the correlations among the uncertainties of the two corresponding $\alpha_s^{(n_f=5)}(M_Z^2)$ input values.
The significance of the deviation from zero for each of these differences, computed as the difference divided by its uncertainty, represents a test of the RGE within the $Q^2$ range of the two points that are being considered.\footnote{ The significance squared for the simple differences corresponds to the minimum of the $\chi^{2}$ in Eq.~(\ref{Eq:Chi2Cov})~(see e.g. Ref.~\cite{Nisius:2014wua}). Very similar values are obtained for the significance of the relative differences squared~(up to small non-linear effects in the uncertainty propagation). We have also checked this correspondence numerically in the current study. }
In addition, the uncertainty of each such difference provides a measure of the precision within which the RGE test has been performed.
This represents an important information, in addition to the $Q^2$ range and to the outcome of the test itself.
Indeed, we observe that the RGE test here is performed within a precision between 1.4 permil and about two percent, depending on the $Q^2$ range and the correlation assumptions employed for the theory uncertainties. 
This can be seen in the plots at the bottom of Fig.~\ref{fig:RGtest}, where we display the differences between the extracted $\alpha_s^{(n_f=5)}(M_Z^2)$ value at $Q^2=5.41\, \mathrm{GeV}^2$ and the rest of extracted values of Table~\ref{tab:sets}, normalised by the former.\footnote{The analogous plots using instead lattice data are once again relegated to App.~\ref{app:lattalphasfits}.}

\begin{figure}
    \centering
    
        \includegraphics[width=0.48\textwidth]{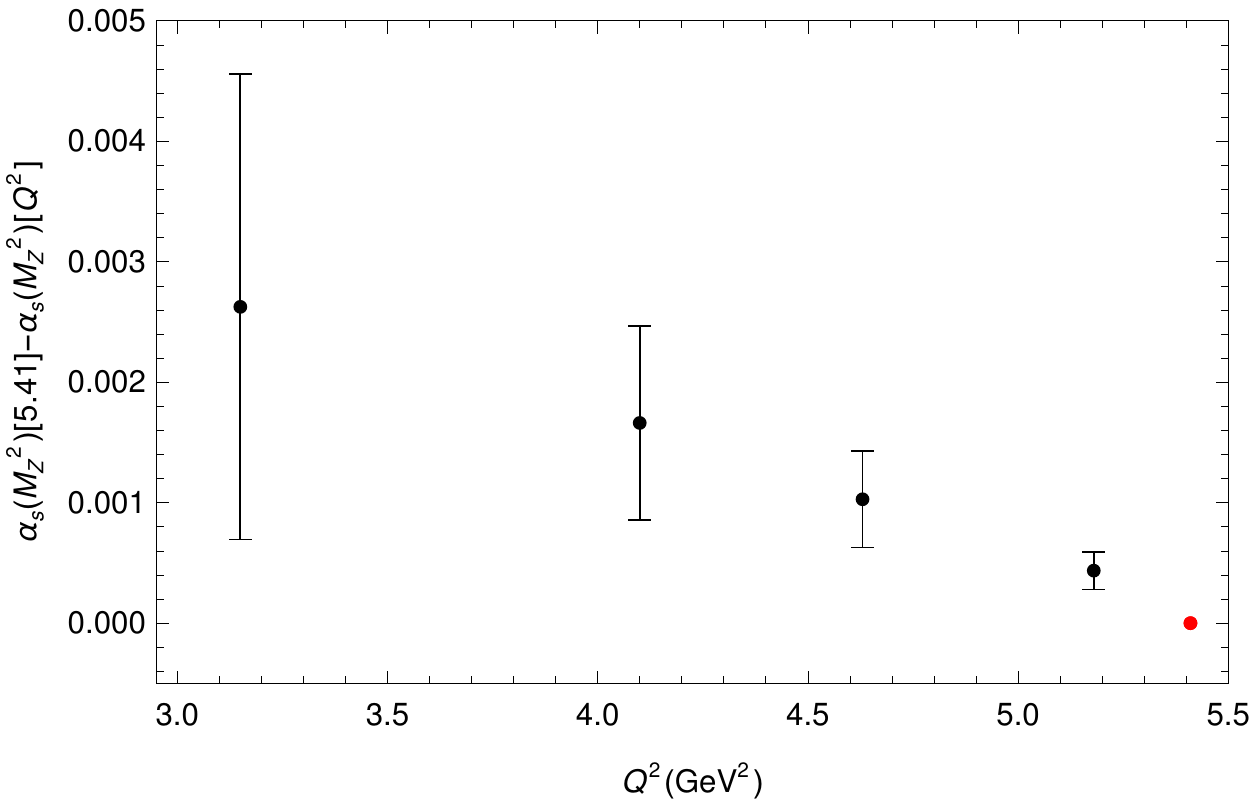} \quad
        \includegraphics[width=0.48\textwidth]{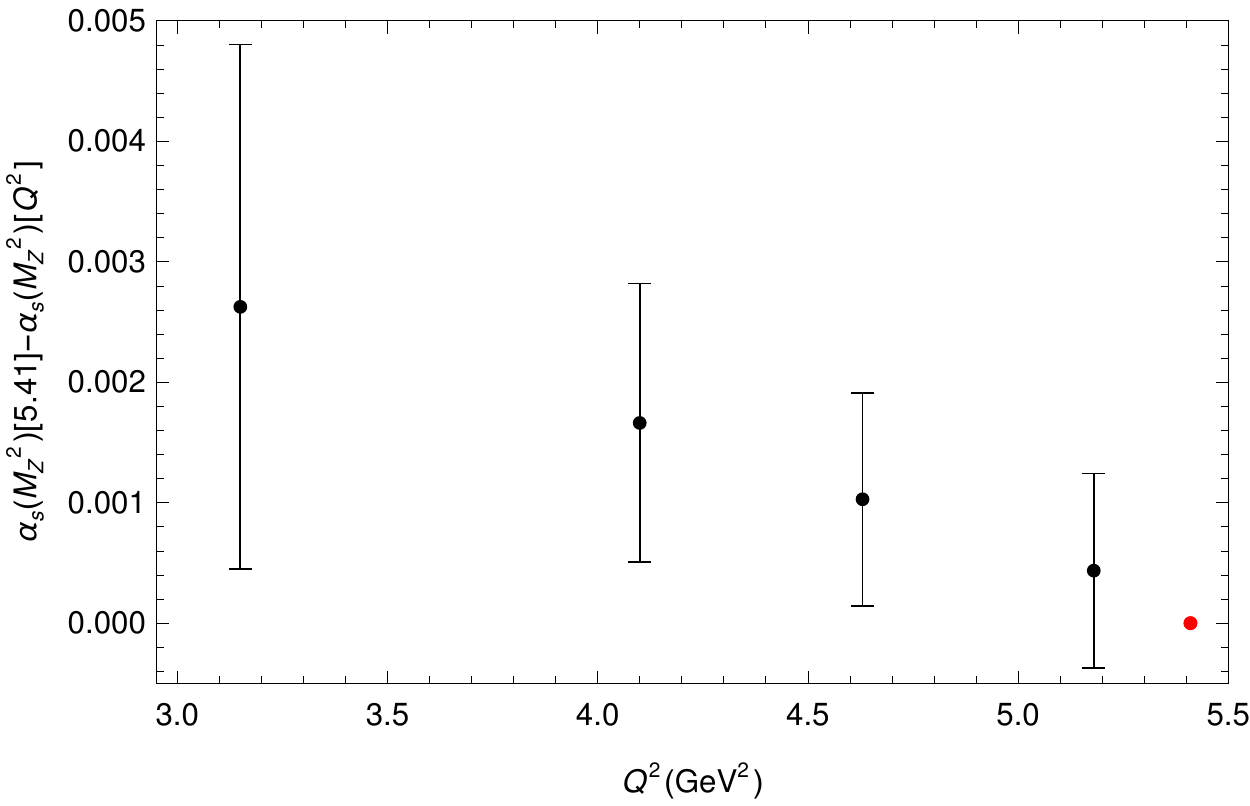}
        \includegraphics[width=0.48\textwidth]{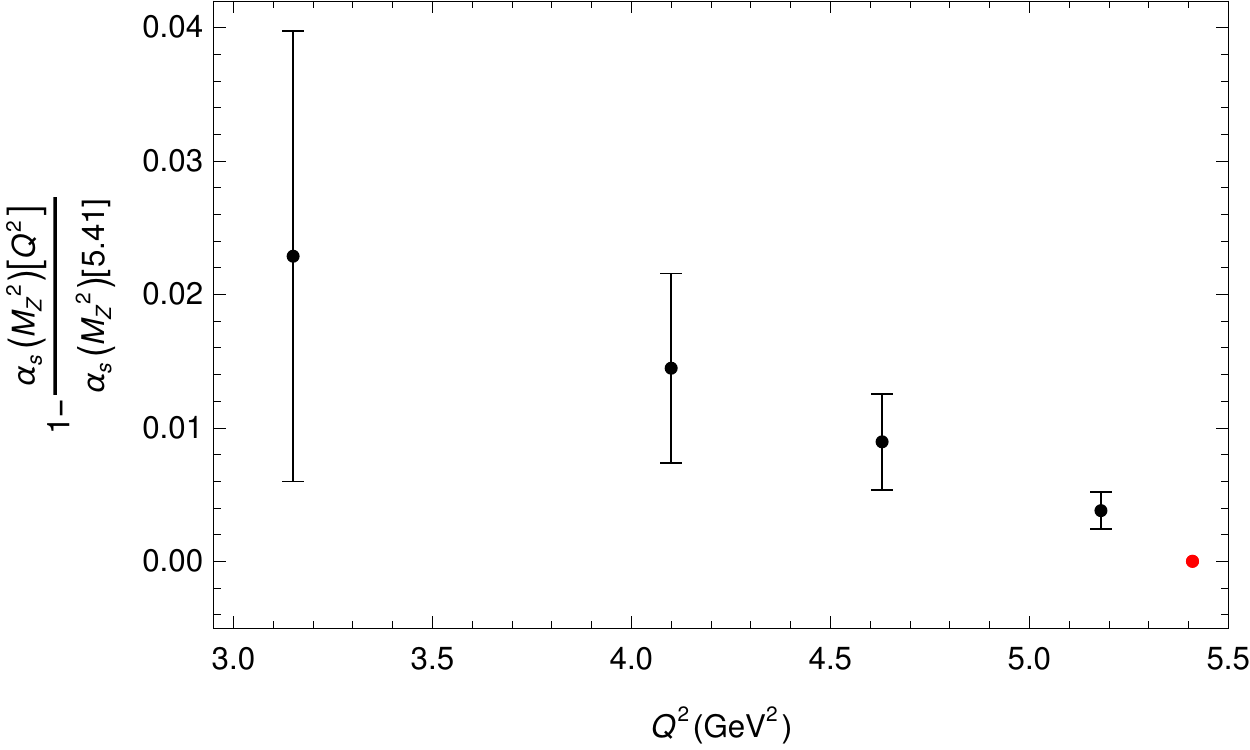} \quad
        \includegraphics[width=0.48\textwidth]{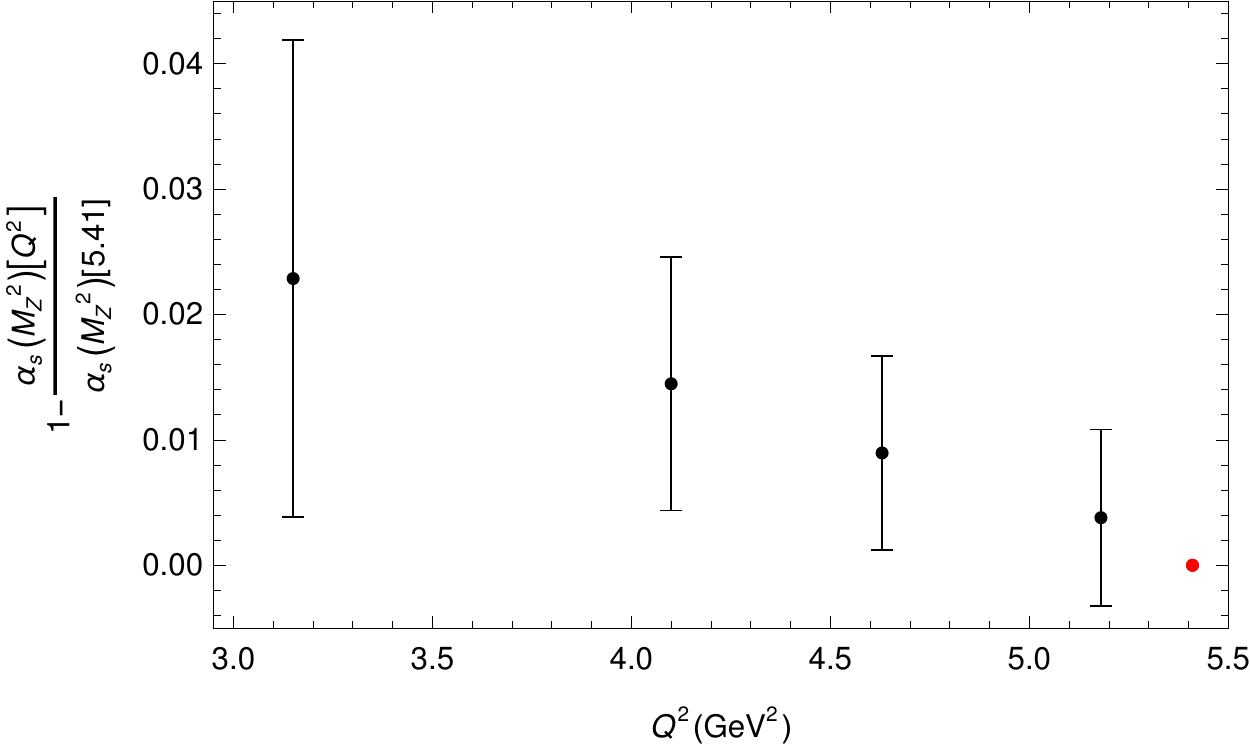}
     \caption{The top plots show the differences between the extracted $\alpha_s^{(n_f=5)}(M_Z^2)$ value at $Q^2=5.41\, \mathrm{GeV}^2$ and the rest of extracted values in Table~\ref{tab:sets}, together with their uncertainties computed taking into account the correlations. The bottom plots show the same differences normalised with respect to the $\alpha_s^{(n_f=5)}(M_Z^2)$ value extracted at $Q^2=5.41\, \mathrm{GeV}^2$. The red circle indicates the reference point at $Q^2=5.41\, \mathrm{GeV}^2$. For the left plots we assume the theory uncertainties originating from the same source to be fully correlated, whereas for the right plots we assume the theory uncertainties from scale variations and $\mathcal{O}_6$ to be uncorrelated.} 
    \label{fig:RGtest}
\end{figure}

\section{Conclusions}\label{sec:conclusions}

In this work we have carefully analysed the QCD predictions for the electromagnetic Euclidean Adler function, below the charm threshold. It was already known that this observable could be studied within pQCD from relatively low energies. From the phenomenological point of view,  this is relevant because 
one can independently determine it at relatively low energies from recent precise $e^+e^-$ and lattice data, which are known to produce results in clear tension with each other. A comparison between 
these three different determinations of the Adler function is
then compelling because 1. pQCD may discriminate, up to a certain extent, between the $e^+e^-$-based and the latticed-based Adler functions if they do not agree. 2. Assuming the validity of one of them, one can test the validity domain of pQCD, just at the edge of the perturbative breakdown. 3. One can study the sensitivity to the QCD coupling of a direct comparison between pQCD and the ratio $R(s)$.

In order to unlock the full potential from known pQCD results at relatively low energies, a consistent analysis beyond the available $\mathcal{O}(\alpha_s^2)$ precision  in mass-dependent (decoupling) schemes such as MOM was preferred. Most of this work has consisted in assembling all the needed pieces in an EFT-based $\overline{\mathrm{MS}}$ set-up, to perform such a comparison at $\mathcal{O}(\alpha_s^4)$, with a careful treatment of all associated expansions and uncertainties. This has been done in Sec.~\ref{sec:pqcdadler}, where many details are given in order to simplify the reproducibility of our results and facilitate future applications and improvements. The extraction of alternative Adler functions, based on $R(s)$ data and lattice results, has been studied in Sec.~\ref{sec:otheradlers}.

We have first compared in Sec.~\ref{sec:comparison} the pure perturbative predictions with the results obtained from both  $e^+e^-$ data and lattice QCD, confirming (although at a somewhat reduced level) the tension observed in $g-2$ between these two different approaches. Taking as input the value of $\alpha_s$ from the FLAG compilation, $\alpha_s^{(n_f=5)}(M_Z^2)=0.1184\pm 0.0008$,
the pQCD prediction of the Adler function turns out to be in excellent agreement with the lattice determination at $Q^2\gtrsim 2\;\mathrm{GeV}^2$, while the determination from $e^+e^-$ data lies systematically below it in the energy range $Q^2\in[1,5.5]~\mathrm{GeV}^2 $.

In Sec.~\ref{sec:nonpert} we have incorporated the leading nonperturbative power corrections, in order to have a better assessment of the theoretical uncertainties in the lowest range of $Q^2$ values. Once these corrections are taken into account, the agreement between the QCD OPE prediction and the lattice result extends to the whole analysed range of $Q^2$, although the theoretical uncertainties turn out to be large below $2~\mathrm{GeV}^2$.

Assuming the validity of the OPE at $Q\approx 2\, \mathrm{GeV}$, the Adler function extracted from the $e^+e^-$ data
lies approximately $2\, \sigma$ below the OPE predictions. This appears to follow the same trend as the deficit observed in the muon $g-2$ integral, when comparing the dispersive $e^+e^-$ estimate
to both the experimental measurement of the muon anomalous magnetic moment and the lattice results, as well as the deficit observed  by different lattice groups for the window integral and the deficit in the hadronic running of $\alpha$ when compared to the lattice result from Ref.~\cite{Ce:2022eix}.

In order to fit the Adler function extracted from $e^+e^-$ data, one would need values of the strong coupling significantly below the current lattice (and PDG) average and/or much larger nonperturbative corrections that would not scale with the expected power corrections. Both possibilities are disfavoured by the lattice data that match beautifully the pQCD predictions, at the achieved precision, even at energies as low as $Q \sim 1.25 \, \mathrm{GeV}$.

On the other hand, several possible applications can be expected from our study. 
Combining the results of this work with future lattice studies could help in assessing the validity domain of pQCD for the different involved correlators at higher precision, while at large $Q^2$ values, where pQCD is more reliable, the corresponding results can help in understanding discretization effects in the lattice. We also leave for future work exploring an alternative implementation of the Euclidean running of $\alpha$, based on EFTs (using the $\overline{\mathrm{MS}}$ scheme) supplemented by appropriate energy expansions and possibly interpolations, just starting from the massless descriptions at the different number of flavors, as we have done in this first work at $n_f=3$. This can also be combined with lattice data, as shown in~\cite{Jegerlehner:2019lxt,Ce:2022eix}.

Once the current tensions between the lattice QCD and data-driven approaches~(which are also very relevant in the context of the anomalous magnetic moment of the muon) are resolved, the comparison between data and QCD could provide a determination of $\alpha_s^{(n_f=5)}(M_Z^2)$ at the per-cent level.
With this respect, forthcoming precise measurements of the hadronic production cross-sections, as well as independent lattice calculations of similar precision, will play a major role.

\section*{Acknowledgements}

MD, BM and ZZ would like to acknowledge their fruitful collaboration with Andreas Hoecker. We are also grateful to Andrei Kataev, Eduardo de Rafael and Harmut Wittig for useful comments on the manuscript
ARS was supported by European Union’s Horizon 2020 research and innovation programme under grant agreement No 101002846, ERC CoG ``CosmoChart'' and funded in part by MIUR contract number 2017L5W2PT. ARS and BM acknowledge the support from LPNHE, CNRS/IN2P3,
Sorbonne Université. BM is also supported by Université de Paris. DDC and AP are supported in part by Generalitat Valenciana, Grant No. Prometeo/2021/071, and
MCIN/AEI/10.13039/501100011033, Grant No. PID2020-114473GB-I00. DDC was funded in part by CIDEGENT/2018/014.

\appendix

\section{Compilation of perturbative coefficients}\label{app:pertcoef}
In this section we compile the different coefficients used to evaluate the perturbative Adler function, in the $\overline{\mathrm{MS}}$ renormalization scheme.

\subsection[Running of $\alpha_s$, quark masses and decoupling relations]{\boldmath Running of $\alpha_s$, quark masses and decoupling relations}

The known $\beta$-function coefficients associated with the running of $\alpha_s$,
\begin{equation}
\mu\,\frac{d\alpha_s}{d\mu}\; =\; \alpha_s\;\beta(\alpha_s)\, ,
\qquad\qquad\qquad\qquad
\beta(\alpha_s)\; =\;\sum_{n=1}\, \beta_n\, 
\left(\frac{\alpha_s}{\pi}\right)^n\, ,
\end{equation}
are  \cite{Tarasov:1980au,vanRitbergen:1997va,Czakon:2004bu,Baikov:2016tgj,Luthe:2016ima,Herzog:2017ohr,Luthe:2017ttc,Luthe:2017ttg,Chetyrkin:2017bjc}
\be
\beta_1\, =\,\frac{1}{3}\, n_f - \frac{11}{2}\, ,
\qquad\qquad\qquad
\beta_2\, =\, -\frac{51}{4} + \frac{19}{12} \, n_f \, ,
\ee
\begin{eqnarray}
\label{eq:beta3}
\beta_3 & =& {1\over 64}\left[ -2857 + {5033\over 9} \, n_f
- {325\over 27} \, n_f^2 \right]  ,
\\[5pt]
\label{eq:beta4}
\beta_4 & =&
\frac{-1}{128}\, \left[
\frac{149753}{6}+3564 \,\zeta_3 
-\left( \frac{1078361}{162}+\frac{6508}{27}\,\zeta_3 \right)\, n_f
+ \left( \frac{50065}{162}+\frac{6472}{81}\,\zeta_3 \right)\, n_f^2
+\frac{1093}{729}\, n_f^3 \right] ,
\nonumber\\ \\[5pt]
\beta_5 & = &
-\frac{1}{512}\,\Biggl\{
\frac{8157455}{16} +\frac{621885}{2} \,\zeta_{3} -\frac{88209}{2} \,\zeta_{4}
-288090 \,\zeta_{5}
\nonumber\\
&& \hskip 0.94cm \mbox{} +\, n_f\:\left[
-\frac{336460813}{1944} 
-\frac{4811164}{81}  \, \zeta_{3}
+\frac{33935}{6}  \, \zeta_{4}
+\frac{1358995}{27}  \, \zeta_{5}
\right]
\nonumber\\
&& \hskip 0.94cm \mbox{} +\, n_f^2\:\left[
\frac{25960913}{1944} 
+\frac{698531}{81}  \, \zeta_{3}
-\frac{10526}{9}  \, \zeta_{4}
-\frac{381760}{81}  \, \zeta_{5}
\right]
\nonumber\\
&& \hskip 0.94cm \mbox{} +\, n_f^3\:\left[
-\frac{630559}{5832} 
-\frac{48722}{243}  \, \zeta_{3}
+\frac{1618}{27}  \, \zeta_{4}
+\frac{460}{9}  \, \zeta_{5} \right]
\, +\, n_f^4\:\left[ \frac{1205}{2916} -\frac{152}{81}  \, \zeta_{3} \right]\,
\Biggr\} .
\end{eqnarray}
The RGE for the running masses is
\be
\mu\,\frac{d m_q}{d\mu}\; =\; -m_q \;\gamma(\alpha_s)\, ,
\qquad\qquad\qquad\qquad
\gamma(\alpha_s)\; =\;\sum_{n=1}\, \gamma_n \,
\left(\frac{\alpha_s}{\pi}\right)^n\, .
\ee
The known coefficients are \cite{Tarasov:1982plg,Larin:1993tq,Chetyrkin:1997dh,Vermaseren:1997fq,Baikov:2014qja,Luthe:2016xec,Baikov:2017ujl}
\be
\gamma_1\, =\, 2\, ,
\qquad\qquad\qquad
\gamma_2\, =\, \frac{101}{12} - \frac{5}{18} \, n_f \, ,
\ee
\be\label{eq:gamma3}
\gamma_3 &\!\! = &\!\!
\frac{1}{24}\left[ \frac{3747}{4} - \left(\frac{554}{9} + 40\, \zeta_3\right) n_f - \frac{35}{27}\, n_f^2\right]\, ,
\\[5pt]
\gamma_4 &\!\! = &\!\!
\frac{1}{128}\left\{ \frac{4603055}{162} + \frac{135680}{27}\,\zeta_3 -8800\,\zeta_5
\, +\, n_f\:
\left[-\frac{91723}{27}-\frac{34192}{9}\,\zeta_3 + 880\,\zeta_4 + \frac{18400}{9}\,\zeta_5\right]
\right.\nonumber\\
&& \hskip .4cm \mbox{} \left. +\, n_f^2\:
\left[\frac{5242}{243} + \frac{800}{9}\,\zeta_3 - \frac{160}{3}\,\zeta_4\right]
+ n_f^3\left[-\frac{332}{243}+ \frac{64}{27}\,\zeta_3\right]\right\}\, ,
\\[5pt]
\gamma_5 &\!\! = &\!\!
\frac{1}{512}\Biggl\{
\frac{99512327}{162} + \frac{46402466}{243}\,\zeta_3 + 96800\,\zeta_3^2 -\frac{698126}{9}\,\zeta_4
-\frac{231757160}{243}\,\zeta_5 + 242000\,\zeta_6 + 412720\,\zeta_7
\nonumber\\
&& \hskip .45cm \mbox{}
+\, n_f\:\left[ -\frac{150736283}{1458} -\frac{12538016}{81}\,\zeta_3 -\frac{75680}{9}\,\zeta_3^2 + \frac{2038742}{27}\,\zeta_4 + \frac{49876180}{243}\,\zeta_5 -\frac{638000}{9}\,\zeta_6
\right.\nonumber\\
&& \hskip 1.3cm\left. \mbox{}
-\frac{1820000}{27}\,\zeta_7\right]
\nonumber\\
&& \hskip .45cm \mbox{}
+\, n_f^2\:\left[ \frac{1320742}{729} + \frac{2010824}{243}\,\zeta_3 +\frac{46400}{27}\,\zeta_3^2 -\frac{166300}{27}\,\zeta_4 -\frac{264040}{81}\,\zeta_5 + \frac{92000}{27}\,\zeta_6\right]
\nonumber\\
&& \hskip .45cm \mbox{}
+\, n_f^3\:\left[ \frac{91865}{1458} +\frac{12848}{81}\,\zeta_3 + \frac{448}{9}\,\zeta_4 -\frac{5120}{27}\,\zeta_5\right]
\, +\, n_f^4\:\left[-\frac{260}{243} -\frac{320}{243}\,\zeta_3 + \frac{64}{27}\,\zeta_4\right]
\Biggr\} .
\ee

\subsection{Light-quark loop coefficients}
In the limit of $n_f$ massless quarks and no extra (or infinitely massive) heavy quarks the Adler function is determined by the $K_{n,0}$, coefficients\footnote{Again we do not display here the singlet contributions (see main text).}
\be
D^{L,(0)}_{ii}(Q^2)&\!\! = &\!\! 
N_C\;\left\{
1 + \sum_{n=1} K_{n,0}\,
\left( {\alpha_s(Q^2)\over \pi}\right)^n\,\right\} \, .
\ee
They are \cite{Appelquist:1973uz,Zee:1973sr,Chetyrkin:1979bj,Dine:1979qh,Gorishnii:1990vf,Surguladze:1990tg,Chetyrkin:1996ez,Baikov:2008jh,Baikov:2010je,Herzog:2017dtz}
\begin{eqnarray}
K_{1,0} &=& 1 \, ,
\qquad\qquad\qquad\qquad\qquad
K_{2,0} \, =\,
\frac{365}{24} - 11\,\zeta_3 +  \left( \frac{2}{3}\,\zeta_3 - \frac{11}{12}\right) n_f
\, , 
\nonumber\\
K_{3,0} &=&
\frac{87029}{288} - \frac{1103}{4}\, \zeta_3 + \frac{275}{6}\, \zeta_5 + 
      \left(-\frac{7847}{216} + \frac{262}{9}\, \zeta_3 - \frac{25}{9}\, \zeta_5\right) n_f + 
     \left(\frac{151}{162} - \frac{19}{27}\, \zeta_3\right) n_f^2 
\, ,
\nonumber\\
K_{4,0}  &=&
\frac{144939499}{20736} - \frac{5693495}{864}\,\zeta_3 + \frac{5445}{8}\,
\zeta^2_3 + \frac{65945}{288}\,\zeta_5 - \frac{7315}{48}\,\zeta_7
\nonumber\\
&+& \left(  -\frac{13044007}{10368} + \frac{12205}{12}\,\zeta_3 - 55\,\zeta^2_3 + \frac{29675}{432}\,\zeta_5 + \frac{665}{72}\,\zeta_7\right) n_f
\nonumber\\
&+& \left( \frac{1045381}{15552} - \frac{40655}{864}\,\zeta_3 + \frac{5}{6}\,\zeta^2_3 - \frac{260}{27}\,\zeta_5\right) n_f^2
+ \left( -\frac{6131}{5832} + \frac{203}{324}\,\zeta_3 + \frac{5}{18}\,\zeta_5
\right) n_f^3 \, .
\nonumber\\
\end{eqnarray}

Following the notation of Ref.~\cite{Pich:2020gzz}, the strange mass corrections are given by a linear combination of three coefficients
\be
\Delta_{m_s} D^{L}_{33}(Q^2)=-3 N_C \frac{m_s^2(Q)}{Q^2}\sum_{n} (2c_n^{L+T}+e_n^{L+T}+f_{n}^{L+T})\left( {\alpha_s(Q^2)\over \pi}\right)^n +\mathcal{O}\left( \frac{m_s^4}{Q^4}\right) \, .
\ee
In general they are all known up to three loops \cite{Generalis:1989hf,Chetyrkin:1993hi,Gorishnii:1986pz,Bernreuther:1981sp},
\begin{equation}
    c_0^{L+T}=1\, , \qquad c_1^{L+T}=\frac{13}{3}\, , \qquad c_2^{L+T}=\frac{25291}{432}+\frac{215}{54}\zeta_3-\frac{520}{27}\zeta_5-n_f\left(\frac{41}{24}+\frac{2}{9}\zeta_3 \right),
\end{equation}
\begin{equation}
    e_0^{L+T}=0\, , \qquad e_1^{L+T}=\frac{2}{3}\, , \qquad e_2^{L+T}=\frac{877}{54}-\frac{91}{27}\zeta_3-\frac{5}{27}\zeta_5-n_f\left(\frac{2}{3}-\frac{4}{9}\zeta_3 \right),
\end{equation}

\begin{equation}
    f_0^{L+T}=0\, , \qquad f_1^{L+T}=0\, , \qquad f_2^{L+T}=-\frac{32}{9}+\frac{8}{3}\zeta_3\, ,
\end{equation}

but in fact the needed linear combination is known up to four loops~\cite{Baikov:2004ku}
\begin{align}\nonumber
2\, c^{L+T}_3 + e^{L+T}_3 + f^{L+T}_3\, &=\,
\frac{16828967}{7776} -\frac{12295}{81}\,\zeta_3 + \frac{7225}{108}\,\zeta_3^2 -\frac{93860}{81}\,\zeta_5 + \frac{1027019}{2592}\,\zeta_7\\&-\, n_f \left(\frac{33887}{216} +\frac{721}{486}\,\zeta_3 +\frac{106}{27}\,\zeta_3^2 +\frac{5}{3}\,\zeta_4 -\frac{10355}{243}\, \zeta_5 \right) \nonumber
\\&+ n_f^2 \left( \frac{9661}{5832} +\frac{2}{27}\,\zeta_3
\right) \, .
\end{align}

\subsection{Heavy-quark loop coefficients}
The contribution to the Adler function from heavy-quark loops depends on the $\overline{C}_{j}^{(n)}(\mu)$ coefficients defined in Eqs.~(\ref{eq:PiHeavy}) and (\ref{eq:CjCoefficients}).
In Table \ref{tab:coefheavy} we compile the known (non-singlet) coefficients $\overline{C}_{j}^{(n)}(m_c(m_c))$, up to $j=10$
and $n=3$, taken from the references referred in the main text. Let us remark that for $n=3,j>4$ they are only approximated values. The $n=3$ singlet contribution is given separately in Eq.~(\ref{eq:SingletHeavy}).
\begin{table}[tbh]\centering\renewcommand{\arraystretch}{1.2}
\begin{tabular}{|c|c|c|c|c|}\hline
$j$ & $\overline{C}_{j}^{(0)}(m_c(m_c))$ & $\overline{C}_{j}^{(1)}(m_c(m_c))$ & $\overline{C}_{j}^{(2)}(m_c(m_c))$ & $\overline{C}_{j}^{(3)}(m_c(m_c))$ \\ \hline
$1$ &$1.0667  $&$   \phantom{-}2.5547  $&$   \phantom{-}2.4967 $ & $-5.6404$ \\ \hline 
$2$ &$0.4571  $&$   \phantom{-}1.1096  $&$   \phantom{-}2.7770  $ &$-3.4937$\\ \hline 
$3$ &$0.2709  $&$   \phantom{-}0.5194  $&$   \phantom{-}1.6389 $ & $-2.8395$\\ \hline 
$4$ &$0.1847  $&$   \phantom{-}0.2031  $&$   \phantom{-}0.7956 $ & $-3.349$\\ \hline 
$5$ &$0.1364 $&$  \phantom{-}0.0106  $&$   \phantom{-}0.2781 $ & $-3.737$\\ \hline 
$6$ &$0.1061  $&$  -0.1158  $&$   \phantom{-}0.0070$  & $-3.735$\\ \hline 
$7$ &$0.0856 $&$  -0.2033  $&$  -0.0859$  & $-3.39$\\ \hline 
$8$ &$0.0709 $&$  -0.2660  $&$  -0.0496$ & $-2.85$\\ \hline 
$9$ &$0.0601  $&$  -0.3122 $&$   \phantom{-}0.0817$ & $-2.22$\\ \hline 
$10$ &$0.0517  $&$  -0.3470 $&$   \phantom{-}0.2838$ & $-1.65$\\ \hline
\end{tabular}
\caption{Non-singlet heavy-quark coefficients $\overline{C}_{j}^{(n)}(m_c(m_c))$. The values given for $\overline{C}_{j>4}^{(3)}(m_c(m_c))$ are only approximate estimates.
}
\label{tab:coefheavy}
\end{table}

Using the scale-invariance of the Adler function and the RGE of the running coupling and masses, it is straightforward to find the values of these coefficients at any other renormalization scale. For example, for the four-loop coefficients one finds:
\begin{equation*}
    \bar{C}^{(3)}_1(\mu)=\bar{C}^{(3)}_1(m_c(m_c))+\frac{26 }{405}L^3-\frac{42001 }{10935}L^2+\left(-\frac{21640907 }{233280}\zeta (3) +\frac{144646921 }{1049760}\right)
    L\, ,
\end{equation*}
\begin{equation*}
    \bar{C}^{(3)}_2(\mu)=\bar{C}^{(3)}_2(m_c(m_c))-\frac{92 }{945}L^3+\frac{1236401 }{127575}L^2+\left(-\frac{160906453 }{3483648} \zeta (3)+\frac{1514929311547
   }{17635968000}\right)L\, ,
\end{equation*}
\begin{equation*}
    \bar{C}^{(3)}_3(\mu)=\bar{C}^{(3)}_3(m_c(m_c))+\frac{16544 }{8505}L^3+\frac{301549372 }{13395375}L^2+\left(\frac{588425644445059}{240045120000}\zeta_3-\frac{2101159030799659 }{720135360000}\right)L\, ,
\end{equation*}
\begin{equation*}
    \bar{C}^{(3)}_4(\mu)=\bar{C}^{(3)}_4(m_c(m_c))\frac{104512 }{18711}L^3+\frac{1207474918 }{37889775}L^2+\left(\frac{3882485996940952 }{139244923125}\zeta_3-\frac{4663762048907267 }{139244923125}\right)L\, ,
\end{equation*}
\begin{align*}\nonumber
    \bar{C}^{(3)}_5(\mu)=&\bar{C}^{(3)}_5(m_c(m_c))+\frac{121600}{11583} L^3+\frac{419559137 }{11188131}L^2+\left(\frac{11493749549904284922937344 }{51676194450456575903}\zeta_3 \right.\\
   &\left.-\frac{13815337731941240844320768 }{51676194450456575903}\right)L\, ,
\end{align*}
\begin{align*}\nonumber
    \bar{C}^{(3)}_6(\mu)=&\bar{C}^{(3)}_6(m_c(m_c))+\frac{2863616 }{173745}L^3+\frac{917112400 }{23175603}L^2+\left(\frac{651547528689568126172154298368 }{429356162878789378794041}\zeta_3\right.\\
   &\left.-\frac{783193764656552535030069460992 }{429356162878789378794041}\right)L\, ,
\end{align*}
\begin{align*}\nonumber
    \bar{C}^{(3)}_7(\mu)&=\bar{C}^{(3)}_7(m_c(m_c))+\frac{483328 }{20655}L^3+\frac{584955462 }{15264583}L^2+\left(\frac{535501006096859255438313967560163328 }{56757449883272928029520032067}\zeta_3 \right.\\
   &\left.-\frac{1931107816850828622369214941259366400 }{170272349649818784088560096201}\right)L\, ,
\end{align*}
\begin{align*}\nonumber
   \bar{C}^{(3)}_8(\mu)&=\bar{C}^{(3)}_8(m_c(m_c))+ \frac{194265088 }{6235515}L^3+\frac{488784115 }{14385877}L^2\\ & \nonumber +\left(\frac{3604714064840321574346924622791084364267520 }{65447583034900845993238160708606879}\zeta_3 \right.\\
   &\left.-\frac{4333071686738680398413127461273056938295296
   }{65447583034900845993238160708606879}\right)L\, ,
\end{align*}
\begin{align*}\nonumber
    \bar{C}^{(3)}_9(\mu)&=\bar{C}^{(3)}_9(m_c(m_c))+\frac{1038942208 }{26189163}L^3+\frac{499493070 }{18676903}L^2\\ & \nonumber+\left(\frac{159592894419467095119221552398948945854700978176 }{519784704463182451617731521986046572497}\zeta_3\right.\\
   &\left.-\frac{191839744405509297915091679432644835963148697600
   }{519784704463182451617731521986046572497}\right)L\, ,
\end{align*}
\begin{align*}\nonumber
    \bar{C}^{(3)}_{10}(\mu)&=\bar{C}^{(3)}_{10}(m_c(m_c))+\frac{140902400}{2882061} L^3+\frac{178198601 }{10606396}L^2\\
    & \nonumber +\left(\frac{21882965916951934885062063776238518244057546752 }{13246590806263931598330852967289655819}\zeta_3 \right.\\
   &\left.-\frac{26304570367320737503829762957552484175252029440
   }{13246590806263931598330852967289655819}\right)L\, ,
\end{align*}
where $L \equiv \log{\left(\frac{m_c^2(m_c)}{\mu^2}\right)}$.

\section{\boldmath Interplay of $\bar{\Pi}^{08}$ with perturbative QCD}

In Ref.~\cite{Ce:2022eix}, the rational approximation of another correlator, obtained using lattice data, is presented, $\Pi^{08}$, which is defined as,
\begin{equation}
\bar{\Pi}^{08}(Q^2)\equiv\frac{1}{4\sqrt{3}}\sum_{i,j}Q_{8}^i\;\bar{\Pi}_{ij}(Q^2) \, , \quad \vec{Q}_{8}\equiv\{1,1,-2\} \, .
\end{equation}
Let us check what can we learn using analytic methods. First, it can be rewritten as
\begin{equation}\label{eq:08}
\bar{\Pi}^{08}(Q^2)=\frac{1}{4\sqrt{3}}\left(2(\bar{\Pi}_{ll}-\bar{\Pi}_{33})+\sum_{i,j\neq i}Q_{8}^{i}\bar{\Pi}_{ij}\right) \, .
\end{equation}
We have defined
\begin{equation}
\bar{\Pi}_{ll}\equiv\frac{\bar{\Pi}_{11}+\bar{\Pi}_{22}}{2}\, .
\end{equation}
From Eq.~(\ref{eq:08}) it is straightforward to see that the correlator vanishes at all orders in massless perturbative QCD, since in that limit $\bar{\Pi}_{ll}=\bar{\Pi}_{33}$ and the disconnected topology associated to the second term, which in general starts at $\mathcal{O}(\alpha_s^3)$, vanishes for the sum. 

The leading OPE contribution to $\bar{\Pi}^{08}(Q^2)$ is then given by the perturbative mass correction, scaling as $\sim \frac{m_s^2}{Q^2}$, so that $\bar{\Pi}^{08}(Q^{2}\rightarrow\infty)= \Pi^{08}(Q^{2}\rightarrow 0)$.
In fact, let us note that this asymptotic scaling implies that the correlator satisfies an 
unsubtracted dispersion relation,
\begin{equation}
\Pi^{08}(Q^2)=\,\frac{1}{\pi}\int^{\infty}_{|Q_{\mathrm{th}}|^2}dQ'^2\;\frac{\mathrm{Im}\Pi^{08}(Q'^2 e^{-i\pi})}{Q'^2+Q^2} \, .
\end{equation}

The associated (massive) perturbative Adler function is, safely neglecting $m_{u,d}$, simply given by
\begin{equation}
D^{08}(Q^2)=-\frac{1}{2\sqrt{3}}\left\{\Delta_{m_s}D_{33}^{L}(Q^2)
+\, \Delta_{m_s}D_{3l}(Q^2)\right\} \, ,
\end{equation}
where $\Delta_{m_s}D_{33}^{L}(Q^2)$ and $\Delta_{m_s}D_{3l}(Q^2)$ refer to the strange mass correction of the associated correlators, being $\Delta_{m_s}D_{3l}(Q^2)=\frac{1}{2}\, [\Delta_{m_s}D_{3u}(Q^2)+\Delta_{m_s}D_{3d}(Q^2)]$. Within the large perturbative uncertainties, the second term, which only starts at $\mathcal{O}(\alpha_s^3)$ and comes with an extra $\mathcal{O}\left(\frac{1}{N_c}\right)$ suppression, can be safely neglected. We can then recall Eq.~(\ref{eq:strangecorr}) to arrive at the perturbative result, valid 
at large $Q^2$:\footnote{Once again, the very poor behaviour of this perturbative series at small $Q^2$ can be seen in Table \ref{tab:strange}.}
\be
D^{08}_{\mathrm{pQCD}}(Q^2)\,
=\, \frac{\sqrt{3}}{2} N_C\, \frac{m_s^2(Q^2)}{Q^2}\,\sum_{n} (2c_n^{L+T}+e_n^{L+T}+f_{n}^{L+T})\left( {\alpha_s(Q^2)\over \pi}\right)^n   .
\ee
Since the perturbative QCD contributions are suppressed by two powers of the energy, one may expect nonperturbative effects to be numerically more relevant. They only enter suppressed by two extra powers of the energy and, additionally, they are linear instead of quadratic in the small strange quark mass, since the chirality-conserving nature of the vector current insertions can be recovered by combining a chirality-flipping insertion of the strange quark mass with a second one from the quark condensate. The associated contribution is
\begin{equation}
D^{08}(Q^2)=-\frac{24 \pi^2}{\sqrt{3} Q^4}m_s\langle \bar{s} s \rangle\left(1+\frac{\alpha_s}{3\pi} \right) \, .
\end{equation}

Let us then take the corresponding rational approximation given in Ref.~\cite{Ce:2022eix},
\begin{equation}
\label{eq:approx_08}
  \bar{\Pi}^{08}(Q^2) = \frac{0.0217\, (11)\, x + 0.0151\, (12)\, x^2}{1 + 2.93\, (8)\, x + 2.15\, (12)\, x^2} \,, \qquad\qquad x = \frac{Q^2}{\mathrm{GeV}^2} \,,
\end{equation}
where the numerator $a_i$ and denominator $b_j$ parameters are
strongly correlated according to 
\begin{equation}
\label{eq:approx_08_corr}
  \mathrm{corr}\begin{pmatrix}
    a_1 \\
    a_2 \\
    b_1 \\
    b_2
  \end{pmatrix} = \begin{pmatrix}
    1     \\
    0.97  & 1     \\
    0.97  & 0.984 & 1     \\
    0.944 & 0.994 & 0.98  & 1     \\
  \end{pmatrix} ,
\end{equation}
and compare it to the perturbative and OPE descriptions. The result is displayed in Fig.~\ref{fig:D08}. In spite of the extremely bad behaviour of the associated pQCD series, which fortunately plays a very marginal quantitative role in the EM Adler function, an excellent agreement to pQCD (with a somewhat arbitrary truncation criteria) appears to emerge up to $Q^2\sim 5\, \mathrm{GeV}^2$. However, when incorporating the $D=4$ power correction, a slight tension emerges. Given the different scaling both in energy and in strange quark mass of the pQCD series, $\sim \frac{m_s^2}{Q^2}$, with respect to the leading power corrections $\sim \frac{m_s \Lambda_{\mathrm{QCD}}^3}{Q^4}$ and with respect to higher power corrections, most likely dominated by a quark-gluon condensate scaling as
$\sim \frac{m_s \Lambda_{\mathrm{QCD}}^5}{Q^6}$, comparisons to further lattice simulations at different strange quark masses and at different (relatively large) $Q^2$ values could shed some further light on where the rigorous OPE limits are fulfilled for the corresponding correlator.

\begin{figure}
\centering
\includegraphics[width=0.8\textwidth]{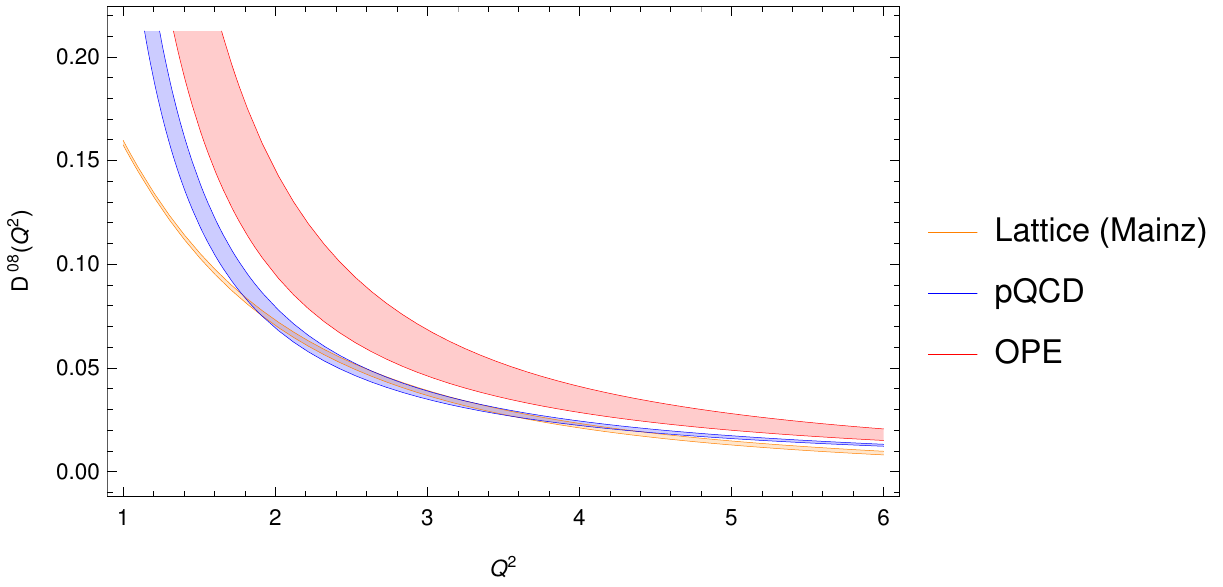}
\caption{Comparison between the Adler function $D^{08}(Q^2)$ obtained from the lattice results of Ref.~\cite{Ce:2022eix} and the one from the OPE of the associated correlator.\label{fig:D08}} 
\end{figure}

\section{\boldmath Fit results for $\alpha_s$ using lattice data}\label{app:lattalphasfits}
For completeness we provide in this appendix the corresponding lattice data results for Tables \ref{tab:sets}-\ref{tab:scales-O6-uncor} in Tables \ref{tab:setslatt}-\ref{tab:scales-O6-uncorlatt}. 
The lattice version of Fig.~\ref{fig:RGtest} is shown in Fig.~\ref{fig:RGtestlatt}.

\begin{table}[]
    \centering
    \begin{tabular}{|c|c|c|}
    \hline
    &   $Q^2$  & $\alpha_s(M_Z^2)$ \\ \hline \hline
    \multirow{3}{*}{Set 1} &
        $3.15$ & $0.1177(40)_{\mathrm{Latt}}(19)_{\mathrm{th}}$ \\ \cline{2-3}
    &    $4.10$ & $0.1174(31)_{\mathrm{Latt}}(13)_{\mathrm{th}}\; $ \\ \cline{2-3}
    &    $5.18$ & $0.1181(22)_{\mathrm{Latt}}(10)_{\mathrm{th}}\; $ \\ \hline \hline
    \multirow{3}{*}{Set 2} &
        $4.10$ & $0.1174(31)_{\mathrm{Latt}}(13)_{\mathrm{th}}\; $ \\ \cline{2-3}
    &    $4.63$ & $0.1176(23)_{\mathrm{Latt}}(11)_{\mathrm{th}}\;$ \\ \cline{2-3}
    &    $5.41$ & $0.1185(24)_{\mathrm{Latt}}(9)_{\mathrm{th}}\;$ \\ \hline
    \end{tabular}
    \caption{The two sets of three $Q^2$ points chosen for the averages with their uncertainties using lattice data.}
    \label{tab:setslatt}
\end{table}

\begin{table}[]
    \centering
    \begin{tabular}{|c||c|c||c|c||c|c|}
    \hline
       & $\chi^2$ & Minimisation & $\chi^2$ & Simple Average & $\chi^2$ & Weighted Average \\ \hline \hline
       Set 1  & $1.1
       $ & $0.1185(23)$  & $1.2
       $ & $0.1178(30)$ & $1.1
       $  &$0.1178(26)$\\ \hline
       Set 2  &  $4.5
       $ & $0.1205(21)$ & $6.4
       $ &$0.1178(25)$ & $6.3
       $ & $0.1179(24)$\\ \hline
       Set 1* & $0.01
       $ & $0.1181(24)$ & $0.01
       $ & $0.1179(28)$ & $0.01
       $ & $0.1180(24)$\\ \hline
       Set 2* & $0.1
       $ & $0.1182(23)$ & $0.1
       $ & $0.1179(24)$ & $0.1
       $ & $0.1181(24)$\\ \hline 
    \end{tabular}
    \caption{Averaged results for $\alpha_s^{(n_f=5)}(M_Z^2)$ obtained when the theory uncertainties coming from the same source are assumed to be fully correlated.}
    \label{tab:all-corrlatt}
\end{table}

\begin{table}[]
    \centering
    \begin{tabular}{|c||c|c||c|c||c|c|}
    \hline
       & $\chi^2$ & Minimisation & $\chi^2$ & Simple Average & $\chi^2$ & Weighted Average \\ \hline \hline
              Set 1  & $0.4
       $ & $0.1181(23)$  & $0.4
       $ & $0.1178(30)$ & $0.4
       $  &$0.1178(25)$\\ \hline
       Set 2  &  $0.3
       $ & $0.1182(23)$ & $0.4
       $ &$0.1178(24)$ & $0.3
       $ & $0.1179(24)$\\ \hline
       Set 1* & $0.01
       $ & $0.1181(24)$ & $0.01
       $ & $0.1179(28)$ & $0.01
       $ & $0.1180(24)$\\ \hline
       Set 2* & $0.1
       $ & $0.1182(23)$ & $0.1
       $ & $0.1179(24)$ & $0.1
       $ & $0.1181(23)$\\
       \hline 
    \end{tabular}
    \caption{Averaged results for $\alpha_s^{(n_f=5)}(M_Z^2)$ obtained  when the theory uncertainties from scale variations are assumed to be uncorrelated.}
    \label{tab:scale-uncorlatt}
\end{table}

\begin{table}[]
    \centering
    \begin{tabular}{|c||c|c||c|c||c|c|}
    \hline
       & $\chi^2$ & Minimisation & $\chi^2$ & Simple Average & $\chi^2$ & Weighted Average \\  \hline \hline
              Set 1  & $0.3
       $ & $0.1182(24)$  & $0.4
       $ & $0.1178(30)$ & $0.4
       $  &$0.1178(26)$\\ \hline
       Set 2  &  $0.7
       $ & $0.1184(23)$ & $0.8
       $ &$0.1178(25)$ & $0.8
       $ & $0.1179(24)$\\ \hline
       Set 1* & $0.01
       $ & $0.1181(24)$ & $0.01
       $ & $0.1179(28)$ & $0.01
       $ & $0.1180(24)$\\ \hline
       Set 2* & $0.1
       $ & $0.1182(23)$ & $0.1
       $ & $0.1179(24)$ & $0.1
       $ & $0.1181(23)$\\
       \hline 
    \end{tabular}
    \caption{Averaged results for $\alpha_s^{(n_f=5)}(M_Z^2)$ obtained when the theory uncertainty from $\mathcal{O}_6$ is assumed to be uncorrelated.}
    \label{tab:O6-uncorlatt}
\end{table}

\begin{table}[]
    \centering
    \begin{tabular}{|c||c|c||c|c||c|c|}
    \hline
       & $\chi^2$ & Minimisation & $\chi^2$ & Simple Average & $\chi^2$ & Weighted Average \\ \hline     \hline
       Set 1  & $0.2
       $ & $0.1181(23)$  & $0.2
       $ & $0.1178(29)$ & $0.2
       $  &$0.1178(25)$\\ \hline
       Set 2  &  $0.3
       $ & $0.1181(24)$ & $0.3
       $ &$0.1178(24)$ & $0.3
       $ & $0.1179(24)$\\ \hline
       Set 1* & $0.01
       $ & $0.1181(23)$ & $0.01
       $ & $0.1179(28)$ & $0.01
       $ & $0.1180(24)$\\ \hline
       Set 2* & $0.1
       $ & $0.1182(23)$ & $0.1
       $ & $0.1179(24)$ & $0.1
       $ & $0.1181(23)$\\
       \hline 
    \end{tabular}
    \caption{Averaged results for $\alpha_s^{(n_f=5)}(M_Z^2)$ obtained when the theory uncertainties from scale variations and $\mathcal{O}_6$ are assumed to be uncorrelated.}
    \label{tab:scales-O6-uncorlatt}
\end{table}

\begin{figure}
    \centering
    
        \includegraphics[width=0.48\textwidth]{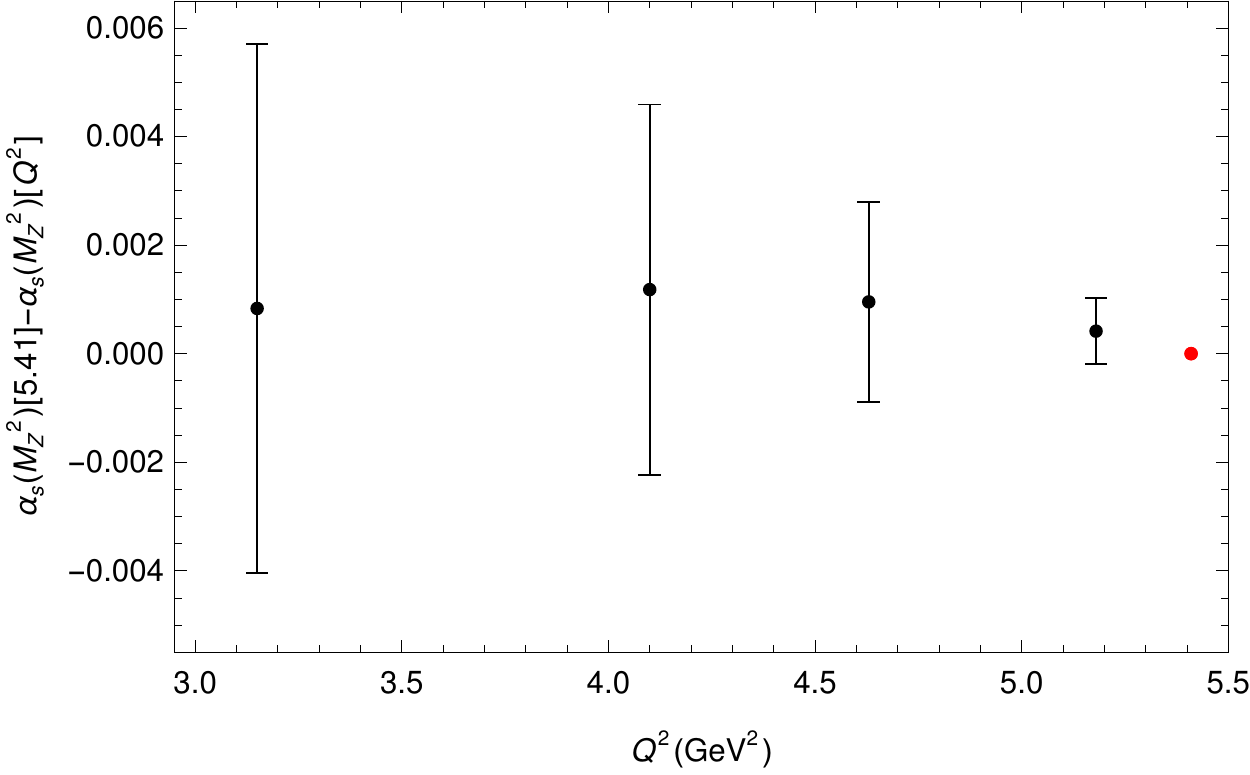} \quad
        \includegraphics[width=0.48\textwidth]{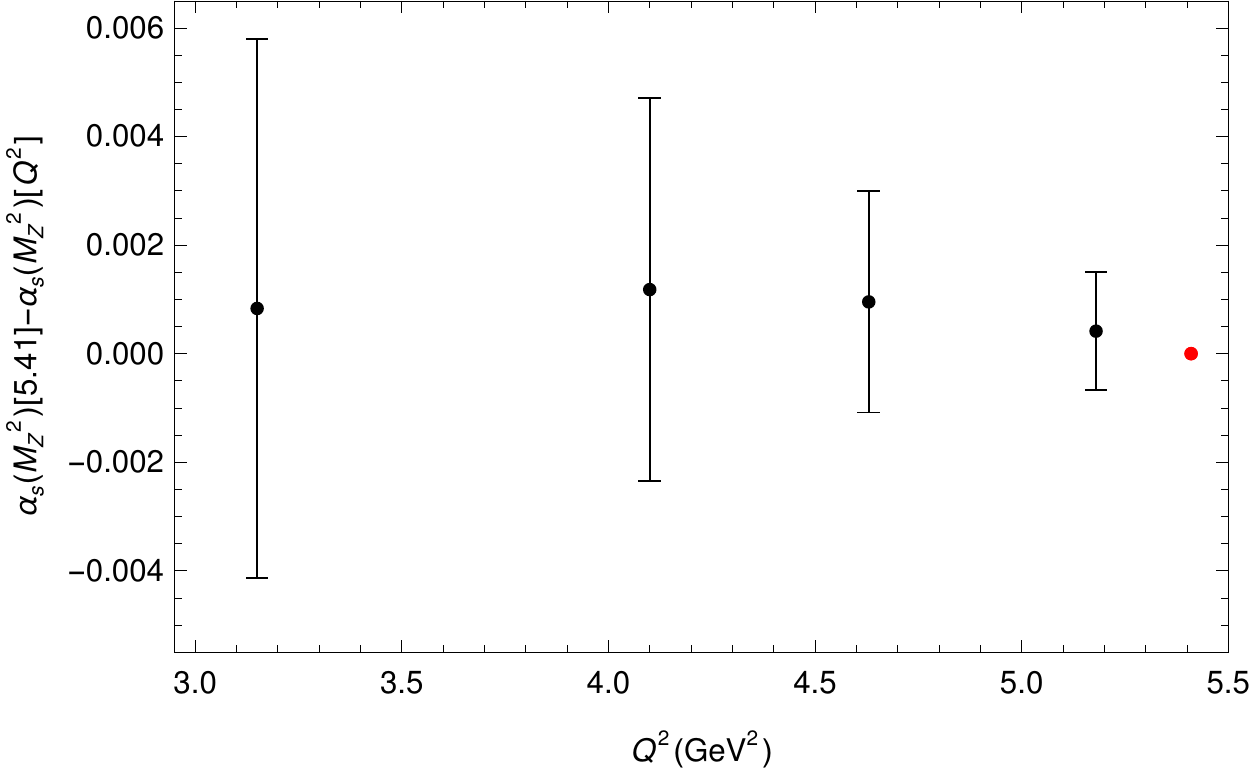}
        \includegraphics[width=0.48\textwidth]{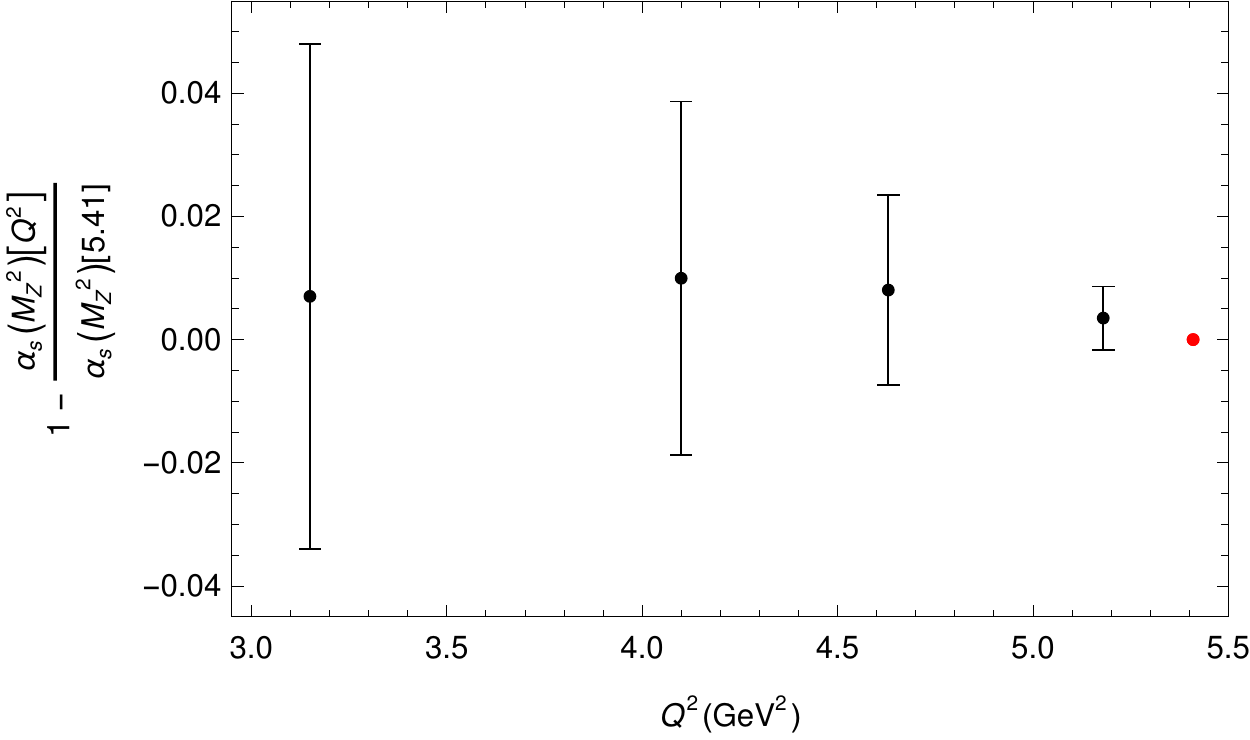} \quad
        \includegraphics[width=0.48\textwidth]{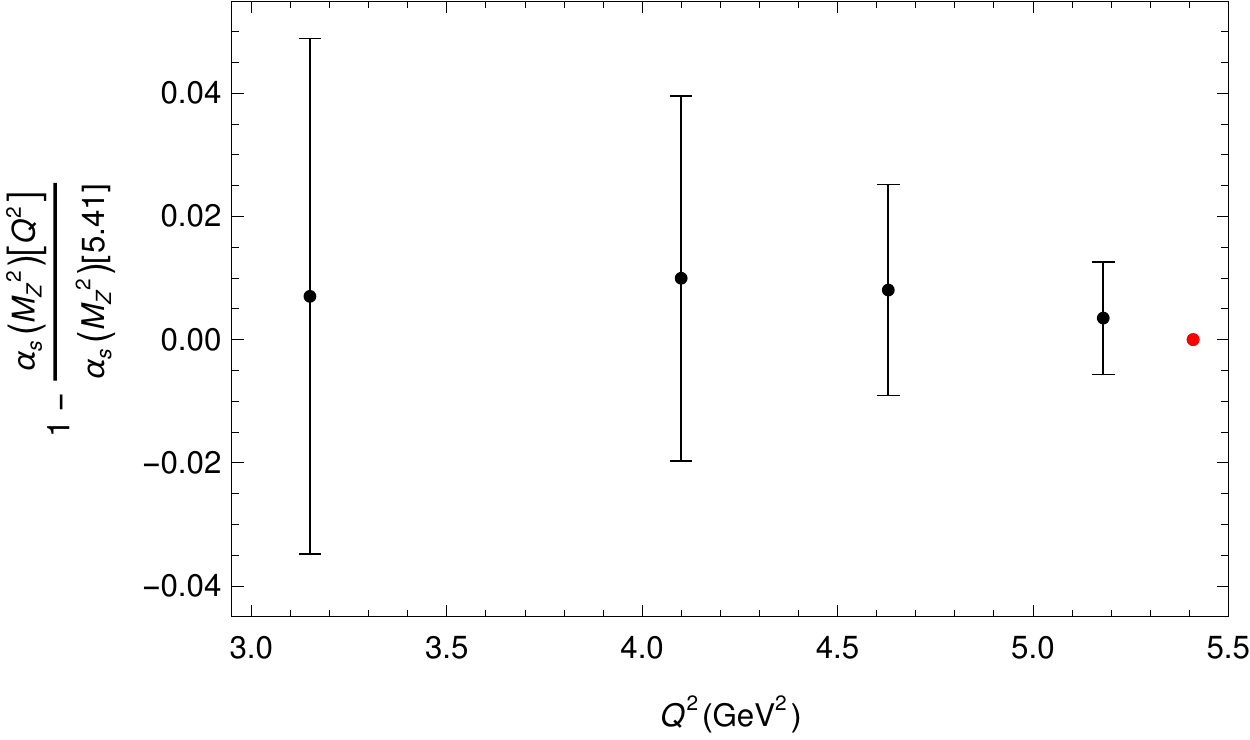}
        
    \caption{The same as Fig.~\ref{fig:RGtest} but using the results of Table~\ref{tab:setslatt}.} 
    \label{fig:RGtestlatt}
\end{figure}

\section{\boldmath Full correlation matrices for the extracted $\alpha_s^{(n_f=5)}(M_Z^2)$ values}
Full correlation matrix when the theory uncertainties coming from the same source are assumed to be fully correlated, Set 1:
\begin{equation}
\left(
\begin{array}{ccc}
 1. &  &  \\
 0.942 
 & 1. &  \\
 0.838
 & 0.971
 & 1. \\
\end{array}
\right)\,.
\end{equation}
Full correlation matrix when the theory uncertainties coming from the same source are assumed to be fully correlated, Set 2:
\begin{equation}
\left(
\begin{array}{ccc}
 1. &  &  \\
 0.986
 & 
 1. &  \\
 0.951
 & 0.988
 & 1. \\
\end{array}
\right)\,.
\end{equation}
Full correlation matrix when the theory uncertainties from scale variations and $\mathcal{O}_6$ are assumed to be uncorrelated, Set 1:
\begin{equation}
    \left(
\begin{array}{ccc}
 1. &  &  \\
 0.829
 & 1. &  \\
 0.768 
 & 0.917
 & 1. \\
\end{array}
\right)\,.
\end{equation}

Full correlation matrix when the theory uncertainties from scale variations and $\mathcal{O}_6$ are assumed to be uncorrelated, Set 2:
\begin{equation}
\left(
\begin{array}{ccc}
 1. &  &  \\
 0.912
 & 1. &  \\
 0.884
 & 0.925
 & 1. \\
\end{array}
\right)\,.
\end{equation}
Full correlation matrix when the theory uncertainties from $\mathcal{O}_6$ are assumed to be uncorrelated, Set 1:
\begin{equation}
    \left(
\begin{array}{ccc}
 1. &  &  \\
 0.829 
 & 1. &  \\
 0.762
 & 0.930 
 & 1. \\
\end{array}
\right)\,.
\end{equation}
Full correlation matrix when the theory uncertainties from $\mathcal{O}_6$ are assumed to be uncorrelated, Set 2:
\begin{equation}
    \left(
\begin{array}{ccc}
 1. &  &  \\
 0.941
 & 1. &  \\
 0.921
 & 0.968
 & 1. \\
\end{array}
\right)\,.
\end{equation}
Full correlation matrix when the theory uncertainties from scale variations are assumed to be uncorrelated, Set 1:
\begin{equation}
    \left(
\begin{array}{ccc}
 1. &  &  \\
 0.943
 & 1. &  \\
 0.882
 & 0.940
 & 1. \\
\end{array}
\right)\,.
\end{equation}
Full correlation matrix when the theory uncertainties from scale variations are assumed to be uncorrelated, Set 2:
\begin{equation}
    \left(
\begin{array}{ccc}
 1. &  &  \\
 0.956
 & 1. &  \\
 0.914
 & 0.945
 & 1. \\
\end{array}
\right)\,.
\end{equation}
\bibliographystyle{JHEP}
\bibliography{biblio}

\end{document}